\documentclass{aa}  
\usepackage{graphicx}
\usepackage{txfonts}
\begin{document}

   \title{Observability of characteristic binary-induced structures in circumbinary disks}
   \titlerunning{Observability of structures in circumbinary disks}

   \author{R.~Avramenko
          \inst{1}
          \and
          S.~Wolf\inst{1}
          \and
          T.F.~Illenseer\inst{1}
          }
   \authorrunning{Avramenko,~Wolf,~Illenseer}

   \institute{University of Kiel, Institute of Theoretical Physics and Astrophysics, Leibnizstrasse 15, 24118 Kiel, Germany\\
              \email{[ravramenko;wolf;tillense]@astrophysik.uni-kiel.de}}

   \date{}

% \abstract{}{}{}{}{} 
% 5 {} token are mandatory
 
  \abstract
   {A substantial fraction of protoplanetary disks form around stellar binaries. The binary system generates a time-dependent non-axisymmetric gravitational potential, inducing strong tidal forces on the circumbinary disk. This leads to a change in basic physical properties of the circumbinary disk, which should in turn result in unique structures that are potentially observable with the current generation of instruments.}
   {The goal of this study is to identify these characteristic structures, constrain the physical conditions that cause them, and evaluate the feasibility of observing them in circumbinary disks.}
   {To achieve this, first we perform 2D hydrodynamic simulations. The resulting density distributions are post-processed with a 3D radiative transfer code to generate re-emission and  scattered light maps. Based on these distributions, we study the influence of various parameters, such as the mass of the stellar components, mass of the disk, and binary separation on observable features in circumbinary disks.}
   {We find that the Atacama Large (sub-)Millimetre Array (ALMA) as well as the European Extremely Large Telescope (E-ELT) are capable of tracing asymmetries in the inner region of circumbinary disks, which are affected most by the binary-disk interaction. Observations at submillimetre/millimetre wavelengths allow the detection of the density waves at the inner rim of the disk and inner cavity. With the E-ELT one can partially resolve the innermost parts of the disk in the infrared wavelength range, including the disk's rim, accretion arms, and potentially the expected circumstellar disks around each of the binary components.}
   {}

   \keywords{Accretion, accretion disks -- 
                Binaries: general --
                Hydrodynamics --
                Radiative transfer
               }

   \maketitle
%
%-------------------------------------------------------------------

\section{Introduction}%%%%%%%%%%%%%%%%%%%%%%%%%%%%%%%%%%%%%%%%%%%%%%%%%%%%%%%%%%%%%%%%%%%%%%%%%%%%%%%%%%%%%%%%%%%%%%%%%%%%%%%%%%%%%%%%%%%%%%%%%%%%%%%%%%%%%%%

The development of circumstellar disks around single stars and their formation during the 
collapse of molecular clouds were studied extensively in the past three  decades
\citep{McKee_2007, Andrews_2010}. However, at the same time it was found that a 
substantial fraction of protoplanetary disks form around binaries 
\citep{Duquennoy_Mayor_1991, Kraus_2009}. 

The assumed mechanism by which a interstellar 
cloud evolves into a binary is either by multiple fragmentation or stellar capture \citep{Bate_2000, Bate_2002, Wolf_2001}. 
During the collapse both components can develop accretion disks in the same way as a single star system \citep{Regaly_2011, Muller_2012}.
In addition to that, the dust and gas from outside of the binary orbit form a third disk, the so-called cirumbinary disk \citep{Rodriguez_2010, Romero_Schreiber_2012,
Lines_Leinhardt_2015}, which revolves 
around the centre of mass of the binary. Furthermore, it is 
possible for a newly developed single star system to capture a transiting star to form a binary. 
In this case the primary could retain a part of its original disk and the captured star would accrete a disk of its own.
The remaining matter would form the circumbinary disk.  

Regardless of the exact formation mechanism the major difference from single star systems is that the orbiting binary 
generates a time-dependent non-axisymmetric gravitational potential, inducing strong 
tidal forces on the surrounding gas and dust \citep{Artymowicz_Lubow_1994}. This leads 
to a change in basic physical properties of the circumbinary disk, which should in turn 
result in unique structures such as a cavity in the disk centre, accretion arms, density waves, and spiral arms
\citep{Gunther_2002, Gunther_2004}. Previously it was shown that some of these structures, such as
the cavity and parts of the accretion arms, are 
observable with the current generation of instruments  \citep{Ruge_2015}. 
There are also examples of more complex systems, such as GG Tau, which is a triple star system \citep{Pietu_2011, Di_Folco_2014}. 
Some of the disk structures, for example spiral arms potentially associated with embedded planets, have also been observed 
\citep{Muto_2012, Garufi_2013, Stolker_2016}.

The presence of a planet further complicates the case \citep{Pierens_Nelson_2013}. A binary can also 
feature two types of planetary orbits: planets that revolve around a single binary component and  planets 
that orbit around both stars \citep{Welsh_2012, Orosz_2012, Schwamb_2013}.
Furthermore, the growth of planetesimals from smaller objects might be hindered \citep{Meschiari_2012, Marzari_2013}. 
The migration behaviour  of planets in a binary system can also be problematic since the 
planet can be ejected if it enters a 4:1 resonance with the binary \citep{Kley_Haghighipour_2014, Pierens_Nelson_2008}.

As mentioned above, the binary exerts a tidal force on the surrounding disk, which should result in characteristic 
structures. However, specific questions regarding the exact nature of these structures are still open: How do the characteristic spatial scales and timescales 
depend on parameters such as the mass and binary separation? How does the overall density and temperature structure 
compare to single star systems? Can structures induced by the 
binary be observed with the current or the next generation of instruments/observatories?

To answer the above questions, we simulate the density distribution of a circumbinary disk, using the 2D 
hydrodynamic code \texttt{Fosite} \citep{Illenseer_2009}. Subsequently, radiative transfer simulations are performed to obtain the corresponding temperature structure
of these disks with the code \texttt{Mol3D} \citep{Ober_2015}. Furthermore, scattered light and re-emission maps of these systems are generated. In each step the unique structures and quantities 
generated by the asymmetrical gravitational potential of the binary are discussed. Finally, a feasibility 
study to observe these quantities with the Atacama Large Millimeter/submillimeter Array (ALMA) and the future 
European Extremely Large Telescope (E-ELT) operating at optical/infrared wavelengths is performed. 

The paper is divided into three parts. First, we provide a general overview of the underlying 2D hydrodynamic 
simulations and 3D  radiative transfer calculations (Sect.~\ref{sec:sim}). Subsequently, the structures caused by the binary-disk 
interaction are analysed (Sect.~\ref{subsec:res_fos}). In the third part (Sect.~\ref{subsec:res_mol} and~\ref{subsec:obs}) we discuss the feasibility of observing 
distinctive structures resulting from the binary-disk interaction.

\section{Hydrodynamic and radiative transfer simulations}%%%%%%%%%%%%%%%%%%%%%%%%%%%%%%%%%%%%%%%%%%%%%%%%%%%%%%%%%%%%%%%%%%%%%%%%%%%%%%%%%%%%%%%%%%%%%%%%%%%%%%%%%%%%%%%%%%%%%%%%%%%%%%%%%%%%%%%%%%%%%%
\label{sec:sim}

In this chapter the applied hydrodynamic and radiative transfer simulation software is introduced. 
Furthermore, the  general simulation set-up is presented and discussed.

\subsection{Two-dimensional hydrodynamic simulation}%%%%%%%%%%%%%%%%%%%%%%%%%%%%%%%%%%%%%%%%%%%%%%%%%%%%%%%%%%%%%%%%%%%%%%%%%%%%%%%%%%%%%%%%%%%%%%%%%%%%%%%%%%%%%%%%%%%%%%%%%%%%%
\label{subsec:sim_fos}

We simulate the density distribution of the gas and dust 
around a binary system for a fixed parameter set. We consider the heating by the star 
by calculating the temperature for an initial density distribution after the system has reached a quasi-stationary stage.
This density distribution is calculated 
using a time- and spatial-constant temperature profile. With the new time-independent temperature profile a second density distribution, which is used in our study, 
is calculated. With this iteration a more realistic hydrodynamic 
simulation can be performed compared to a simulation with a constant temperature distribution. 
 Similar work was performed by 
\cite{Gunther_2002} and \cite{Gunther_2004}. The major difference between these studies and our approach
 is the calculation of the temperature distribution. In case of  \cite{Gunther_2002} and \cite{Gunther_2004}, 
the heating via the viscosity and the stellar irradiation were considered in each 
time step of the hydrodynamic simulation. 
By avoiding the calculation of the temperature in each time step of the hydrodynamic simulation we can  
decrease the required simulation time and achieve a higher spatial resolution. 
The viscous heating of the disk can be neglected, since the stellar irradiation is the dominant form of heating in circumbinary disks as was shown by \cite{Gunther_2004}.

The hydrodynamic simulations are performed with \texttt{Fosite} \citep{Illenseer_2009}. The code solves the 2D continuity and 
Navier-Stokes equations 

 \begin{eqnarray} 
    \frac{\partial \Sigma}{\partial t} + \nabla \cdot (\Sigma \mathbf{v}) &=& 0 \\
    \frac{\partial \Sigma \mathbf{v}}{\partial t} + \nabla \cdot (\Sigma \mathbf{v} \otimes \mathbf{v} + \mathbf{\Pi}  \mathbb{I} ) &=& \nabla \cdot \mathbb{T} - \Sigma\nabla\Phi
 \end{eqnarray}

\noindent
for the vertically integrated surface density $\Sigma$, pressure $\mathbf{\Pi}$, and  gas velocity $\mathbf{v}$. The quantity $\mathbb{T}$ is the 
viscous stress tensor and $\Phi$ the gravitational potential of the binary, since the self-gravitation 
of the disk is neglected. The hydrostatic scale height $H$ of the disk depends on the midplane temperature $T_c$ 
and  the gravitational potential $\phi(\mathbf{r}) = \sum_{i=1,2} \frac{GM_i}{ \sqrt{\mid \mathbf{r} - \mathbf{r_i} \mid}}$ and
 can be written as follows \citep[see][]{Gunther_2002}:

\begin{eqnarray} 
   H(\mathbf{r}) = \left( \sum_{i=1,2} \frac{GM_i}{c_s^2 \sqrt{\mid \mathbf{r} - \mathbf{r_i} \mid^3}} \right)^{-\frac{1}{2}}  ,\end{eqnarray}

\noindent
where $c_s$ is the speed of sound and $M_i$ the masses of the components.

\paragraph{Binary system:}
The masses of the primary and secondary star are chosen to be equal to reduce the parameter space and inner 
radius $r_{\rm in}$ of the computational domain (see below). A small inner radius is chosen to trace the disk 
structures up to the binary orbit. Furthermore, with equally massive components the system 
obtains a point symmetry, making it easy to spot possible non-physical behaviour in the results. 
As was shown by \cite{Bate_2002}, the mass tends to be accreted by the lower mass component, as 
it sweeps a larger area of the disk. This mechanism could lead to binary systems with  mass ratio $q \approx 1$.
Long period systems with mass ratios $\sim$ 1 have also been observed \citep{Raghavan_2010}. 
The mass ratio of the primary and secondary is $q = M_{\rm sec}/M_{\rm prim}$ and we denote from now on 
$M_{\rm prim} = M_{\rm sec} = M_{\rm B}$.

Three different semi-major axes 
$a$ are applied to study their influence on the structures generated by the binaries, 
such as the inner cavity, accretion arms, and radial inhomogeneities in the density distribution ($a =$ 10 AU, 20 AU, and 30 AU). Furthermore, 
for a separation of 20 AU three different stellar masses $M_{\rm B}$ are applied to examine the influence 
of the stellar mass on the disk structures. 
The full set of parameters is summarised in Table ~\ref{fig:hyd_param}.

The eccentricity of the orbit is set to zero, which reduces the parameter space and 
simplifies the system. Simultaneously, it allows one to reduce the inner radius $r_{\rm in}$ (see below), since the grid centre is located at the centre of mass of the binary.
At this point it is important to note that although there are some examples of binary systems with 
eccentricity $\varepsilon < 0.1$, the majority of them have a short orbital period $P<1$ yr. Systems with larger values of $P$
tend to have smaller eccentricities \citep{Raghavan_2010}. 

\paragraph{Temperature:}
All calculations are performed in a locally 
isothermal mode based on a temperature profile $T_c$. We derive the temperature profile  
by performing a low-resolution hydrodynamical simulation for each disk set-up with a constant 
temperature profile as a starting condition. 
Subsequently, the dust temperature distribution is calculated with the 3D radiative transfer code 
\texttt{Mol3D} \citep[see. Sect.~\ref{subsec:sim_mol}]{Ober_2015}. The  
resulting azimutally averaged temperature profile is then used as default temperature $T_c$ for 
the high resolution hydrodynamic calculation.

\paragraph{Initial and boundary conditions:}
The initial disk mass is concentrated in a Gaussian surface density profile with its maximum at 150 AU. The mass 
is fixed at $M_{\rm disk} = 0.02 \,\, {\rm M_{\odot}}$ for all hydrodynamic simulations. 
The gas is set in a near Keplerian motion that accounts for the potential asymmetry with a higher
velocity to counter additional gravitational attraction.

In order to reduce the impact
of boundary conditions on the density distribution at the relevant region (inner ~300 AU), 
the outer radius is set to 1000 AU for all 
simulations. The density as well as the gas velocities are sufficiently small at this boundary so as not to influence the relevant areas of the disk. 
Furthermore, the choice of the inner radius has an impact 
on the disk structure \citep{Pierens_Nelson_2013}. It cannot be chosen to be zero because the
sources of gravitational potential have to be outside of the simulated region. An inner radius of 
$r_{\rm in} = a\cdot0.6$ AU is used, where $a$ is the semi-major axis of the binary. This ensures 
that no singularities occur inside the grid. 
It also permits us to trace the accretion flow nearly up to the binary orbits. This is 
important, as the biggest differences to a circumstellar disk are expected to be in the inner disk region.
The inner boundary condition allows for a mass flux from inside the computational domain through the 
boundary, but prevents any flow in the opposite direction. This has 
a distinct disadvantage of preventing the circumprimary or circumsecondary disks from forming.
The lack of matter in the direct vicinity of both stars should result in short wavelength radiation 
deficiency in the radiative transfer simulations. The accretion rate at the inner boundary amounts to 
$\sim10^{-8} \,\, {\rm M_{\odot}/yr}$ and hence does not significantly change the mass of the binary or the disk for the duration of the simulation.

\paragraph{Grid:}
In order to achieve a sufficient resolution for the subsequent radiative transfer simulation, especially in the inner regions, a 
polar grid $(r, \phi)$ with logarithmic spacing for the $r$ coordinate and a grid resolution of $(N_r = 508, \,\, N_{\phi} = 508)$ is chosen.

\paragraph{Simulation time:}
The evolution of the system is calculated over a period of $t_{\rm sim} = 5\cdot 10^{4}$ yr for all simulations. For the subsequent
temperature calculations, density distributions with time steps between $2.2\cdot 10^{4}$ yr and 
$2.7\cdot 10^{4}$ yr are used. This decision is based on two constraints. First, the gas has to be 
sufficiently mixed to ensure that the initial state does not influence the resulting distribution. 
The timescale related to the advective mixing processes can be estimated using the dynamical timescale, 

 \begin{eqnarray}
  t_{\rm dyn} \sim \frac{r}{v_{\phi}} \sim \Omega^{-1}  \, .
 \end{eqnarray}
 
\noindent
Here, $v_{\phi}$ is the $\phi$ component of the velocity and $\Omega$ is the angular velocity.
By choosing a simulation time step corresponding to over 100 binary periods this condition is met.
The second constraint is to avoid losing too much of the initial mass. The timescale for this process is the 
viscous timescale, which can be estimated using the radial drift velocity $v_r$,

\begin{eqnarray}
  t_{\rm visc} \sim \frac{r}{v_{r}}  
 .\end{eqnarray}
 
\noindent
This timescale is only reached in the simulation in the innermost regions of the disk. The viscous timescale at 200 AU, which is 
the outer radius of the disk in radiative transfer simulations, is of the order of $10^{6}$ years.
The averaged accretion rate is of the order of $5\cdot 10^{-10}{\rm M_{\odot}/{\rm yr}}$, which is consistent with  
observations \citep{Bary_2014}.

 \begin{table*}
  \centering
  \caption{Hydrodynamic simulation parameters and ranges.}
\begin{tabular}{lc|c}
        \multicolumn{2}{c}{}                                                                                \\ \hline \hline
       Parameter                            &                                      & Parameter value/range  \\ \hline
       Mass of individual binary components &   $M_{\rm B}$ [${\rm M_{\odot}}$]    & $0.5, 1, 1,5 $         \\ \hline
       Binary mass ratio                    &   $q$                                & $1$                    \\ \hline
       Initial disk mass                    &   $M_{\rm disk}$ [${\rm M_{\odot}}$] & $0.02 $                \\ \hline
       Semi-major axis                      &   $a$ [AU]                           & $10, 20, 30$           \\ \hline
       Inner boundary radius                &   $r_{\rm in}$ [AU]                  & $0.6 \cdot a$          \\ \hline
       Outer boundary radius                &   $r_{\rm out}$ [AU]                 & $1000$                 \\ \hline
       Eccentricity                         &   $\varepsilon$                      & $0$                    \\ \hline
       Simulation time                      &   $t_{\rm sim}$ [yr]                 & $5 \cdot 10^{5}$       \\ \hline \hline
 \end{tabular}
 
  \label{fig:hyd_param}
  \end{table*}

The result of the hydrodynamic simulations are the circumbinary disk surface density $\Sigma$, scale height $H$, and 
the positions of the binary components. Based on those, radiative transfer simulation are performed.

\subsection{Radiative transport simulation}%%%%%%%%%%%%%%%%%%%%%%%%%%%%%%%%%%%%%%%%%%%%%%%%%%%%%%%%%%%%%%%%%%%%%%%%%%%%%%%%%%%%%%%%%%%%%%%%%%%%%%%%%%%%%%%%%%%%%%%%%%%%%
\label{subsec:sim_mol}

In the following we describe the radiative transfer simulation. The density distribution from
the previous section is used to derive the corresponding temperature distribution 
and to calculate synthetic scattered light and thermal re-emission maps. 
These provide the basis for the assessment of whether the generated structures can be observed.

The radiative transfer simulations are performed with the 3D continuum and line radiative transfer code \texttt{Mol3D}
\citep{Ober_2015}. First, a temperature 
 is calculated solving the radiative transfer equation
 using a Monte Carlo algorithm and taking the optical properties of the dust into account. 
 The temperature is then used to simulate emission maps. In addition, scattered light maps are calculated 
 with the Monte Carlo algorithm as well.

\paragraph{Grid:}
A spherical grid is used with a logarithmic scaling of the 
$r$ coordinate to ensure that small structures near the binaries, in  particular the temperature 
gradient at the inner disk rim and  accretion arms, are sufficiently resolved. 

While the inner radius $r_{\rm in}$ is identical to that of the hydrodynamic simulations, the outer radius $r_{\rm out}$ is set to 200 AU because
most of the gas and all significant density structures are located inside this radius. All simulations are performed 
for a face-on inclination of $i=0^{\circ}$. 

\paragraph{Gas and dust distribution:}
\texttt{Mol3D} employs a perfectly mixed gas and dust 
with a gas-to-dust mass ratio of 100:1. 
Based on the isothermal hydrodynamically  calculated 2D surface density distribution, the 3D
density distribution $\varrho$ is constructed from

\begin{equation}
 \varrho = \frac{\Sigma}{H} \, \exp\Big[-\frac{1}{2}\left(\frac{z}{H}\right)^2\Big]  \, ,
 \end{equation}

 \noindent
 where $z$ is the $z$ coordinate \citep{LyndenBell_1969}. 

The normalised density distribution calculated with \texttt{Fosite} is scaled accordingly, allowing us to consider 
three different disk masses  $M_{\rm disk} = 10^{-1} {\rm M_{\odot}}, \,\, 10^{-2} {\rm M_{\odot}},     \,\, 10^{-3} {\rm M_{\odot}}$.
This provides the basis for studying the effect of optical depth on the emission maps.

\paragraph{Dust:}
The dust is considered to be homogeneous spheres of radius $a$ and with the following size distribution \citep{Dohnanyi_1969}:
\begin{equation}
 dn(a) \sim a^{-3.5} da \, .
 \end{equation} 
 Here, a minimum radius of $a_{\rm min}=5 \,\, {\rm nm}$ and maximum radius $a_{\rm max}= 250 \,\, {\rm nm}$ \citep{Mathis_1977} are assumed.
A composition of $62.5\%$ astronomical silicate and $37.5\%$ graphite is used with optical data from \cite{Weingartner_2001}.
Applying Mie theory, the optical properties are calculated at 100 logarithmically distributed wavelengths $\lambda_{\rm sim}$ in the interval between 
 $0.05$ and $2000$ $\rm \mu m$ \citep{Wolf_2004}.

 \paragraph{Binary luminosity:}
 To conduct the radiative transfer simulations of the dust density distribution constructed beforehand, the luminosities 
 of the binary components are required.
 The stars are considered to be main 
 sequence stars with a same chemical composition as the sun. The stellar luminosities corresponding to the masses 
 are derived from the main sequence tracks from \cite{Siess_2000}.
 
 The full parameter set for the radiative transfer 
 simulation is shown in Table~\ref{fig:rad_param}.

 \begin{table}
   \caption{Radiative transfer simulation parameters and ranges.}
\begin{tabular}{lc|c}
        \multicolumn{2}{c}{}                                                                 \\ \hline \hline
         Parameter                &                                        & Parameter value/range \\ \hline
         Binary temperature       & $T_{ B}$ [K]                           & $4500, 5600, 6600 $   \\ \hline
         Binary radius            & $R_{ B}$ [$R_{\odot}$]                 & $0.42, 0.98, 1.4$     \\ \hline
         Binary luminosity        & $L_{ B}$ [$M_{\odot}$]                 & $0.45, 0.87, 8.2$     \\ \hline
         Inner radius             & $r_{\rm in}$ [AU]                      & $0.6 \cdot a$         \\ \hline
         Outer radius             & $r_{\rm out}$ [AU]                     & $200$                 \\ \hline
         Simulation wavelengths   & $\lambda_{\rm sim}$ [$\rm \mu m$]      & $0.05 \dots 2000$     \\ \hline \hline
 \end{tabular}

  \label{fig:rad_param}
  \end{table}

\section{Results }%%%%%%%%%%%%%%%%%%%%%%%%%%%%%%%%%%%%%%%%%%%%%%%%%%%%%%%%%%%%%%%%%%%%%%%%%%%%%%%%%%%%%%%%%%%%%%%%%%%%%%%%%%%%%%%%%%%%%%%%%%%%%%%%%%%%%%%%%%%%%%
\label{sec:res}

In this chapter, we present the results of the simulations described in Sect.~\ref{sec:sim}. We discuss the most notable 
features of binary systems and their influence on the radiation emitted by the surrounding disk, 
i.e. observable quantities.

\subsection{Characteristic structures in the density distribution of circumbinary disks}%%%%%%%%%%%%%%%%%%%%%%%%%%%%%%%%%%%%%%%%%%%%%%%%%%%%%%%%%%%%%%%%%%%%%%%%%%%%%%%%%%%%%%%%%%%%%%%%%%%%%%%%%%%%%%%%%%%%%%%%%%%%%
  \label{subsec:res_fos}

  First we present the results of the hydrodynamic simulations and discuss the structures
  caused by the disk-binary interaction and the mechanism creating them. 
  
\begin{figure}
         \resizebox{\hsize}{!}{\includegraphics{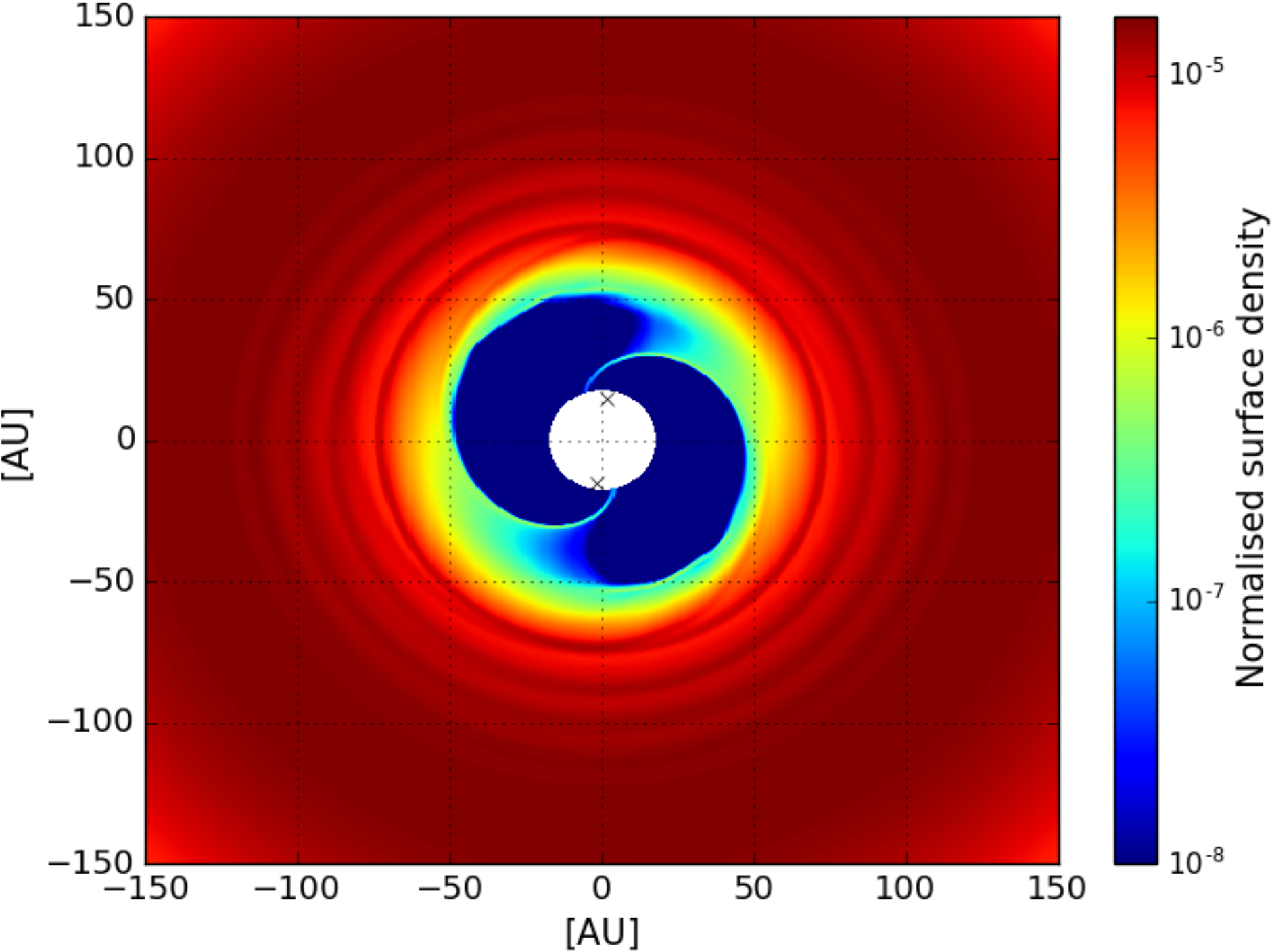}}
         \caption{Normalised surface density for $a=30$ AU, $M_{\rm B}= 1{\rm M_{\odot}}$. The crosses indicate the positions of the binary components.}
         \label{fig:surfacedensity_30}
\end{figure}

\begin{figure}
         \resizebox{\hsize}{!}{\includegraphics{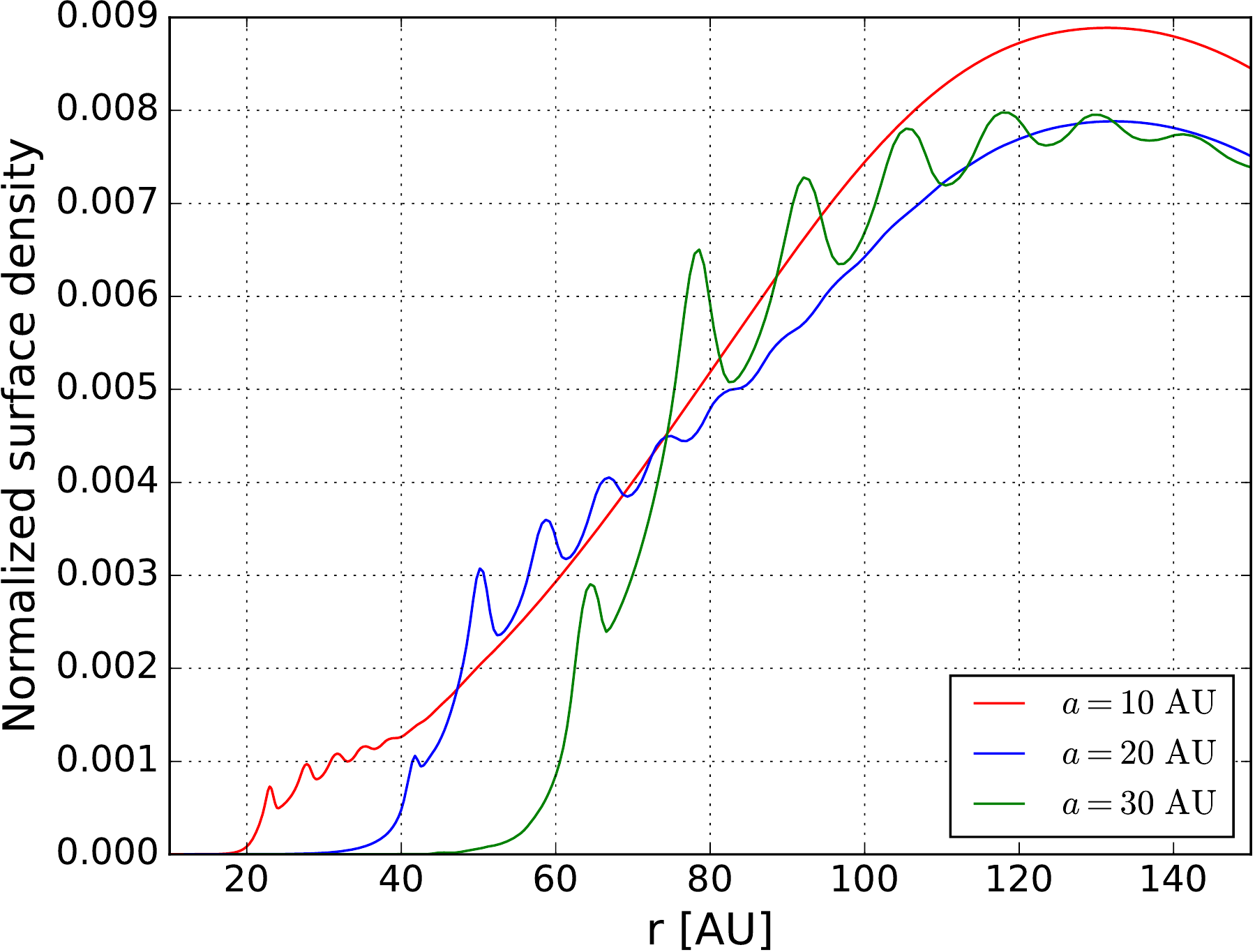}}
         \caption{Normalised by the total mass and azimuthally averaged surface density for semi-major axes $a=10$ AU, $20$ AU, and $30$ AU; $M_{\rm B}= 1\,\, {\rm M_{\odot}}$.}
         \label{fig:radial_density}
\end{figure}

  Fig.~\ref{fig:surfacedensity_30} shows an image of the  surface density distribution for a semi-major axis $a = 30$ AU.
  The most notable feature of the distribution is the cavity in the inner region of the disk. The cavity appears 
  if the binary can exert enough torque on the surrounding disk to counter the torque generated by the viscous stresses. As can be gathered
  from Fig.~\ref{fig:radial_density}, the radius of the cavity is approximately $2-2.5$ times  the binary separation. 
  This is consistent with the theoretical predictions   
  \citep{Artymowicz_Lubow_1994}. 
   As mentioned in Sect.~\ref{subsec:sim_fos} the choice 
  of equally massive stars and an eccentricity of $\varepsilon = 0$ leads to an almost circular cavity centred 
  on the barycentre of the binary. Since the accretion arms have to stretch from the inner edge of the disk to 
  the binary orbit, larger cavities result in longer accretion arms.
  The accretion arms also show a distinct point symmetry that remains undisturbed for at least 100 binary orbits. 
  This finding is in agreement with the results of \cite{Gunther_2002}.
  
  The hydrostatic scale height $H$ only depends on the temperature and
  gravitational potential in our simulations. These are either provided as a fixed initial condition or only changes slightly for radii $r>>a$ during the simulation.
  Consequently, the hydrostatic scale height profile is smooth and monotonically is increasing. This means that in the current model only the shadowing 
  effects that result from the increase of the surface density, in contrast to variation of $H$, can be studied.%(Fig.~\ref{fig:scaleheight}) 

  Another striking feature of the investigated binary systems are the 
  density waves that extend radially from the inner edge of the disk.
  In Fig.~\ref{fig:radial_density} the surface density $\Sigma$, azimuthally averaged and normalised by the total mass is shown for three different values of 
  the binary semi-major axis $a$. 
  We identify two important quantities: the wavelength of the surface density oscillation $\lambda_{a}$, measured as the distance between two maxima, and 
  its magnitude $\mathcal{M}_{a}$, measured as the difference of the maximum and its subsequent minimum. One can clearly see that both quantities 
  depend on the semi-major axis $a$.
  To quantify this dependence, the wavelength $\lambda_{a}$ was calculated  
  by first averaging over the distance of the maxima in each time step of the simulation and averaging the result over 100 time steps in a time 
  interval of $10^4$ years. In Table~\ref{fig:lam_mag}
  the resulting wavelength $\lambda_{a}$ and magnitude $\mathcal{M}_{a}$ are shown. One can see a trend of increasing wavelength 
  $\lambda_{a}$ with increasing $a$. The same applies to the magnitude $\mathcal{M}_{a}$ of the density oscillation, i.e. an increase of the semi-major 
  axis leads to an increase of the magnitude of the density oscillation.

        \begin{figure}
         \resizebox{\hsize}{!}{\includegraphics{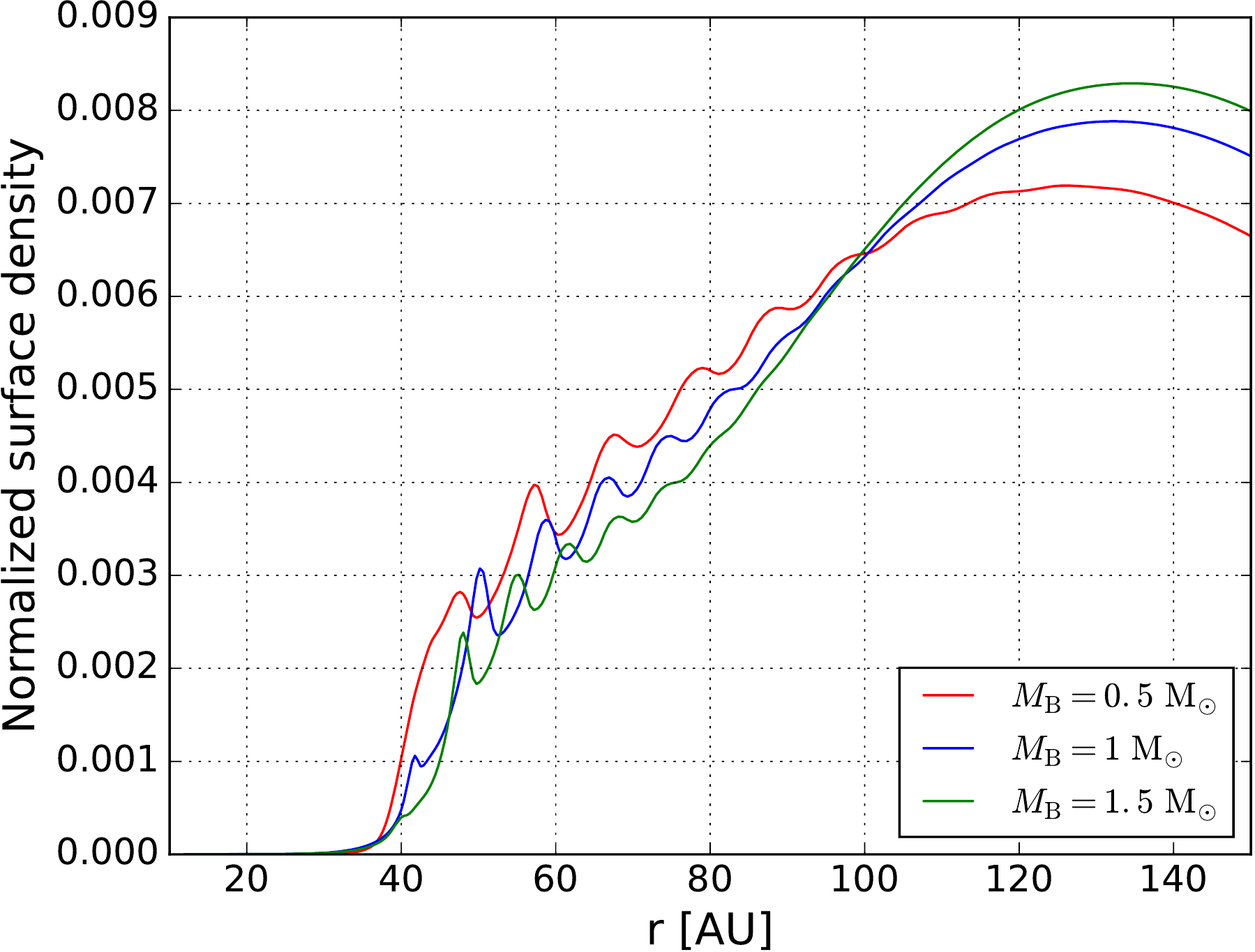}}
         \caption{Azimuthally averaged normalised surface density for $a=20$ AU, and $M_{\rm B} = 0.5 \,\, {\rm M_{\odot}}, 1\,\,  {\rm M_{\odot}},\,\,\,\, {\rm and}\,\,\,\, 1.5 \,\, {\rm M_{\odot}}$.}
         \label{fig:radial_density_mass}
\end{figure}

  \begin{table}
    \caption{Wavelength and magnitudes for different semi-major axes $a$ and binary masses $M_{\rm B}$.}
\begin{tabular}{c|c|c}
        \multicolumn{3}{c}{}                                                                                                           \\ \hline \hline
                    Parameter                                     & $\lambda_{a}$ [AU]           & $\mathcal{M}_{a}$                   \\ \hline
         $M_{\rm B} = 1.0 \,{\rm M_{\odot}} \, ,\,\,\,\, a = 10$ AU & $5.69$                       & $1.13  \cdot 10^{-4} $              \\ \hline
         $M_{\rm B} = 1.0\, {\rm M_{\odot}} \, ,\,\,\,\, a = 20$ AU & $8.36$                       & $4.61  \cdot 10^{-4} $              \\ \hline
         $M_{\rm B} = 1.0\, {\rm M_{\odot}} \, ,\,\,\,\, a = 30$ AU & $11.0$                       & $6.52  \cdot 10^{-4} $              \\ \hline
                                                                  &$\lambda_{M_{\rm B}}$ [AU]    & $\mathcal{M}_{M_{\rm B}}$           \\ \hline
         $M_{\rm B} = 0.5\, {\rm M_{\odot}} \, ,\,\,\,\, a = 20$ AU & $10.4$                       & $2.50  \cdot 10^{-4} $              \\ \hline
         $M_{\rm B} = 1.0\, {\rm M_{\odot}} \, ,\,\,\,\, a = 20$ AU & $8.36$                       & $4.61  \cdot 10^{-4} $              \\ \hline
         $M_{\rm B} = 1.5\, {\rm M_{\odot}} \, ,\,\,\,\, a = 20$ AU & $6.96$                       & $3.44  \cdot 10^{-4} $              \\ \hline \hline
 \end{tabular}
  \label{fig:lam_mag}
  \end{table}

  In Fig.~\ref{fig:radial_density_mass} the impact 
  of binaries with different masses on the circumbinary disk is illustrated. We find that the 
  radius of the cavity does not depend on the binary mass. By applying the same procedure as for the calculation of wavelength $\lambda_{a}$ 
  and the magnitude $\mathcal{M}_{a}$, the values of $\lambda_{M_{\rm B}}$ and $\mathcal{M}_{M_{\rm B}}$ can be determined. Here, 
  $\lambda_{M_{\rm B}}$ is the oscillation wavelength of the surface density  and $\mathcal{M}_{M_{\rm B}}$ magnitude with regard to binary mass $M_{\rm B}$. 
  We find that with increasing binary mass $M_{\rm B}$ the wavelength of the oscillation $\lambda_{M_{\rm B}}$ decreases. 
  At the same time the trend for the mass-dependent magnitude $M_{\rm B}$ appears to be more complex, as the values for 
  $M_{\rm B} = 0.5 \,\, {\rm M_{\odot}}$ and $M_{\rm B} = 1.5 \,\, {\rm M_{\odot}}$ are smaller than for $M_{\rm B} = 1 \,\, {\rm M_{\odot}}$. 
  This suggests that there is a value of  $M_{\rm B}$ for which the magnitude reaches a maximum.
   Similar density waves were shown in Fig. 2 of the work from \cite{Gunther_2004}.

  At this point it appears necessary to discuss the theoretical basis of these density waves. 
  Starting with the 2D continuity and Navier-Stokes equations we calculate
  
    \begin{eqnarray} 
    \partial_t^2 \Sigma - \frac{1}{r} \partial_r [ r \partial_r (\Sigma c_s^2)  ] &=& 0 
  ,\end{eqnarray}
  
  \noindent
  which is a 1D wave equation in polar coordinates (full derivation can be found in App.~\ref{sec:app}). 
  The interpretation of this result is the following: while the gas is accreted inward with velocity $v$, it provides a moving frame for 
  density waves. These density waves are driven by the binary in the centre and propagate outward with sound speed $c_s$.
  This interpretation also explains the trend of wavelength  ($\lambda_{a}$, $\lambda_{M_{\rm B}}$) with regard to the semi-major axis $a$ and binary mass $M_{\rm B}$. 
  The system behaves as a driven harmonic oscillator that swings with the frequency of the instigator $f$. This frequency is coupled with the binary 
  period $P$. Neglecting the dispersion, one would get the formula for the wavelength $\lambda = c_s/f\approx c_s \cdot P$. By increasing the semi-major
  axis $a$ or decreasing the binary mass $M_{\rm B}$, we increase the orbital period and with that the wavelength ($\lambda_{a}$, $\lambda_{M_{\rm B}}$). 
  The magnitude of the oscillation ($\mathcal{M}_{a}$, $\mathcal{M}_{M_{\rm B}}$) should be proportional to the amount of torque that the binary can exert on the disk. 
  This seems to be the case with increasing values of the semi-major axis $a$. However, one would assume that a binary with higher mass $M_{\rm B}$ would be able to generate 
  a stronger torque. Instead, we find that the magnitude $\mathcal{M}_{M_{\rm B}}$ first increases with increasing binary mass, but then decreases for value of
  $M_{\rm B} = 1.5 \,\, {\rm M_{\odot}}$. The exact nature of this phenomenon should be investigated in a future study.

\subsection{Observability of characteristic structures: Analysis of ideal observations}%%%%%%%%%%%%%%%%%%%%%%%%%%%%%%%%%%%%%%%%%%%%%%%%%%%%%%%%%%%%%%%%%%%%%%%%%%%%%%%%%%%%%%%%%%%%%%%%%%%%%%%%%%%%%%%%%%%%%
\label{subsec:res_mol}

  The  goal of this study is to investigate whether the characteristic structures in a circumbinary disk can be observed 
  with currently operating and future instruments/observatories. For this purpose we simulate various observable quantities, such as scattered light and re-emission images, as 
  well as the spectral energy distribution (SED) in the wavelength range between $0.05 \,\, {\mu \rm m}$ and $2000 \,\, {\mu\rm m}$ of the circumbinary disks discussed in Sect.~\ref{subsec:res_fos}. 
  A distance of 140 parsec to the object is assumed for all simulations.

  In the hydrodynamic simulations it was shown that the most striking differences from an undisturbed 
  circumstellar disk are the large gas and dust depleted cavities in the centre, accretion arms in the 
  cavity, and density waves at the inner rim of the disk. Of course, a central cavity is also characteristic for circumstellar transitional disk. 
  However, in the case of a transitional 
  disk this cavity results from disk evolution. With the calculated surface density and scale height 
  it is now possible to derive a 3D density distribution $\varrho(r,\theta , \phi)$ 
  according to Eq. 2.2.1. This density is post-processed with \texttt{Mol3D} to derive observable quantities (see Sect.~\ref{subsec:sim_mol} for details).
  
\begin{figure*}
          \resizebox{\hsize}{!}{\includegraphics{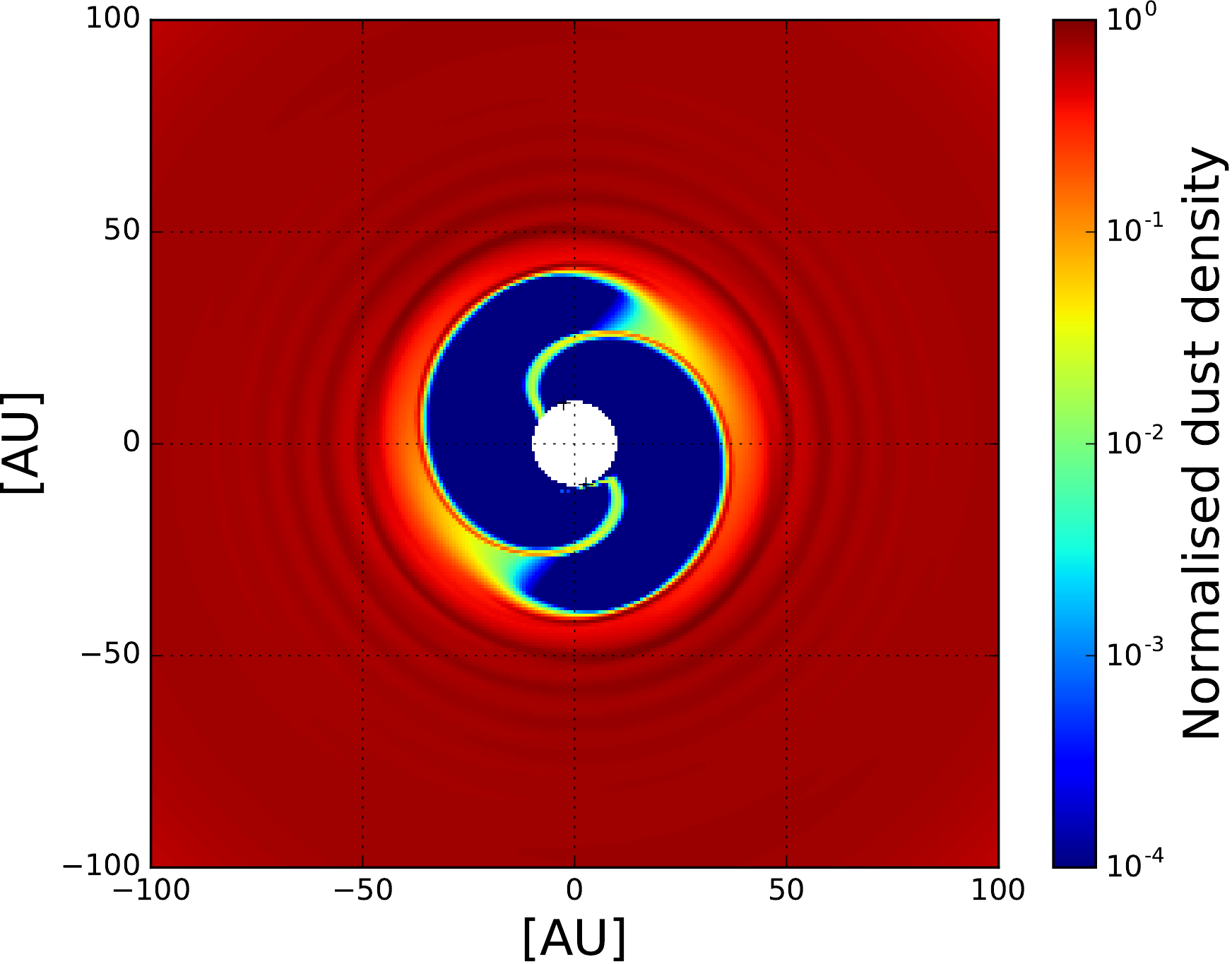} \quad
                                \includegraphics{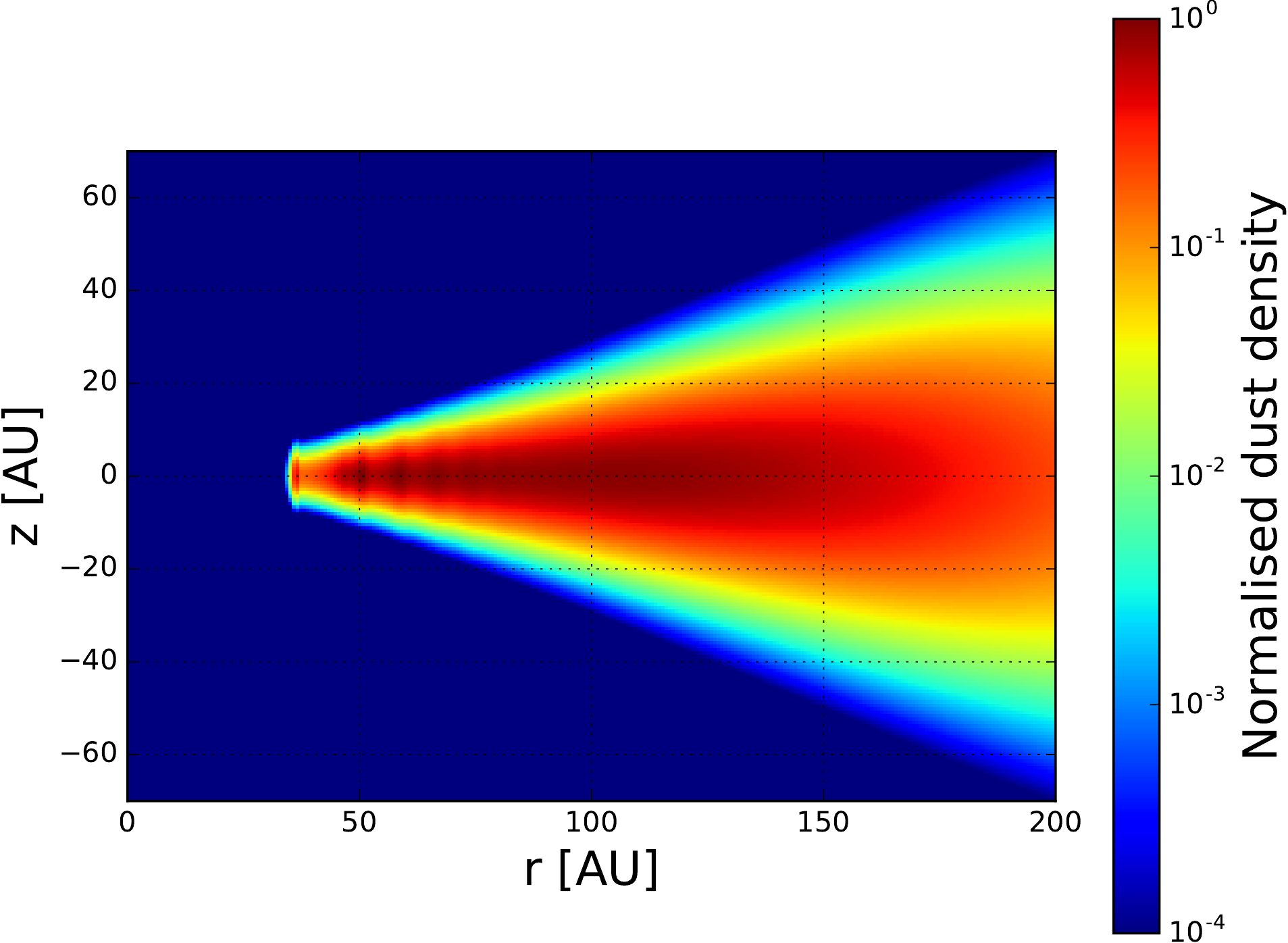}}
          \caption{Normalised dust density distribution  for $a=20$ AU, $M_{\rm B}= 1 \,\, {\rm M_{\odot}}$. The crosses indicate the positions of the binary components.}
          \label{fig:dust_density}
   \end{figure*}

\begin{figure*}
         \resizebox{\hsize}{!}{\includegraphics{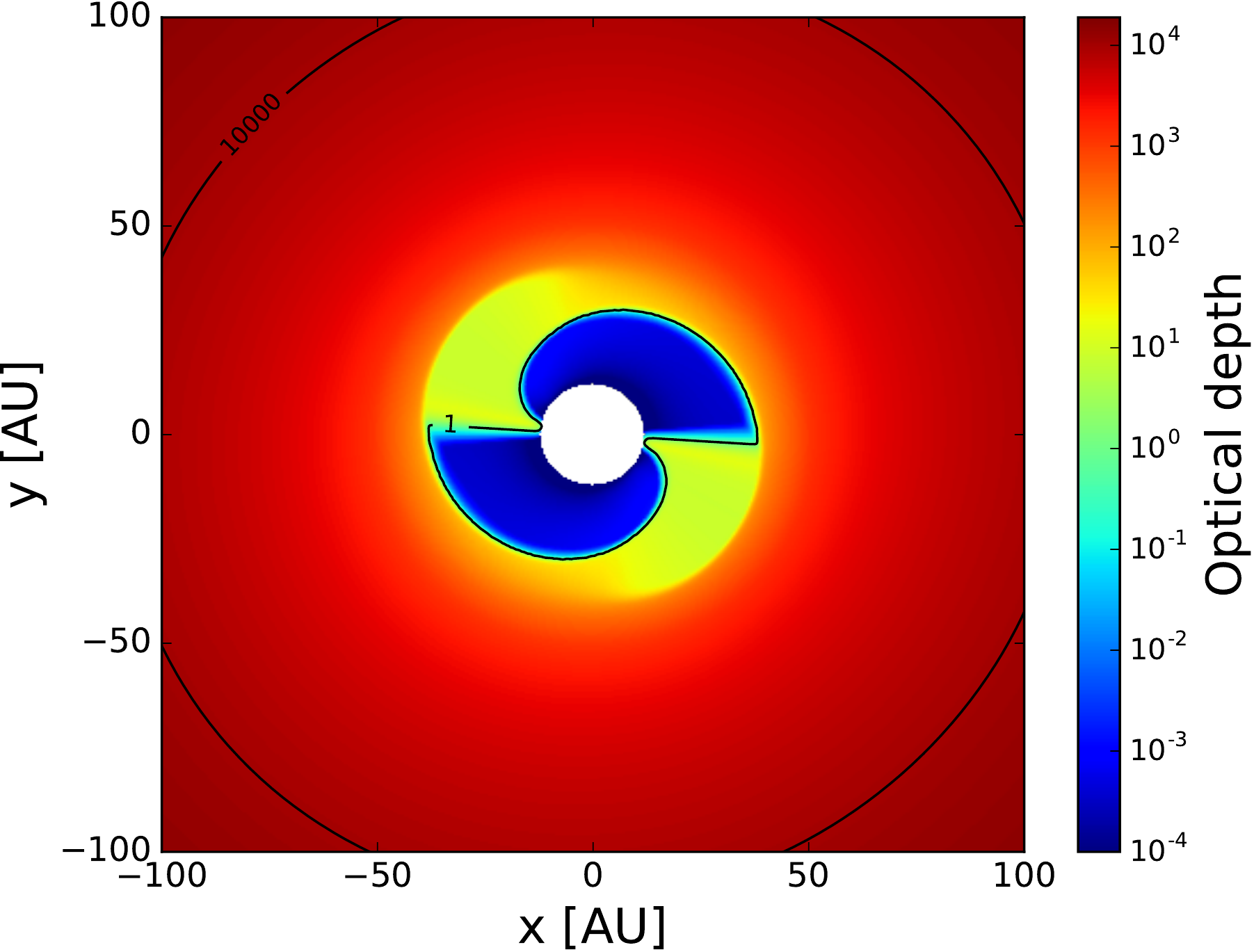} \quad
                               \includegraphics{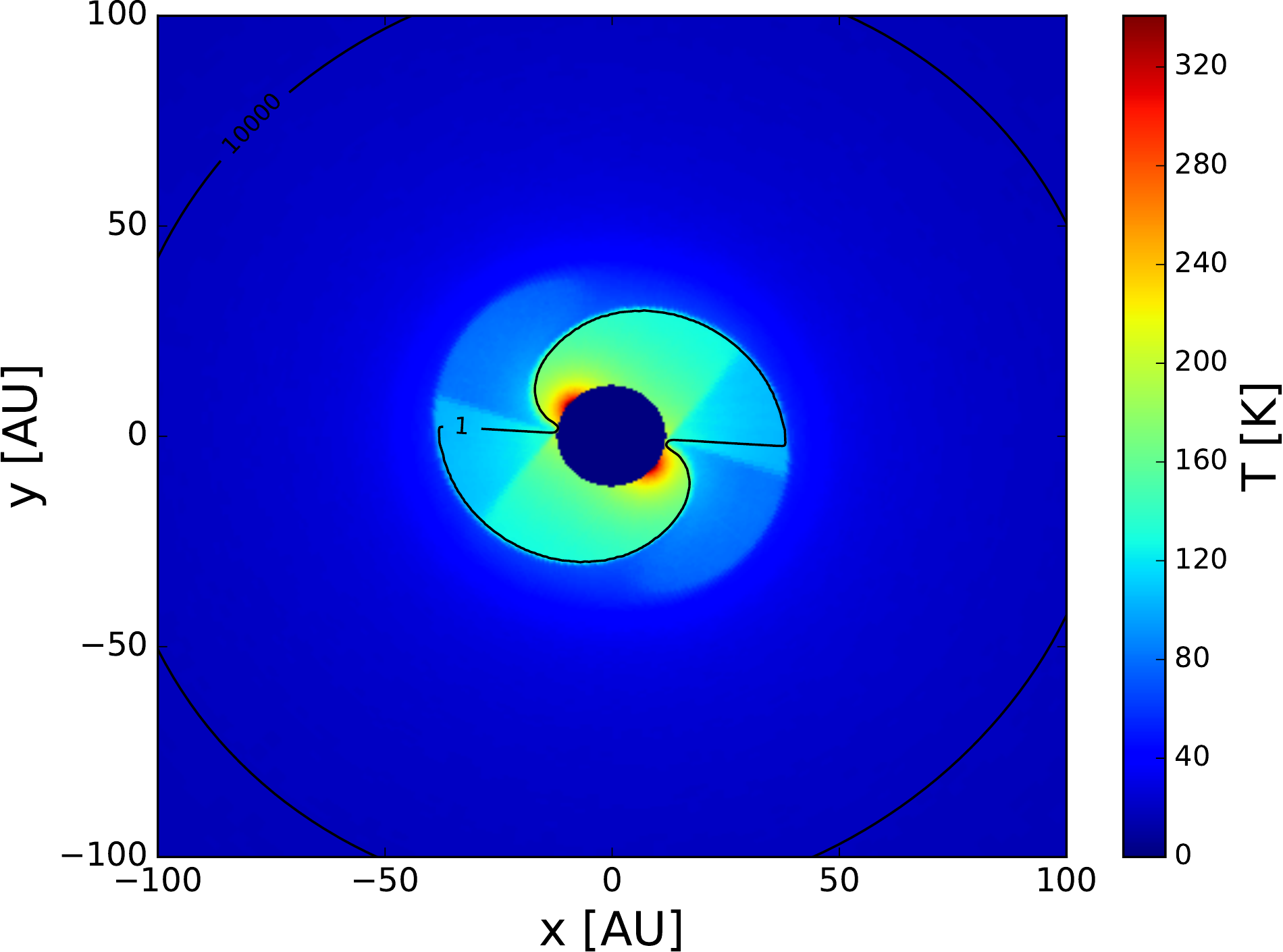}}
         \caption{\textit{Left}: Optical depth $\tau$ at $\lambda = 526$ nm calculated in the midplane. \textit{Right}: resulting midplane temperature of the same system.  Black line denotes where the 
                  optical depth reaches values of $\tau = 1$ and $\tau = 10^4$, respectively.}
         \label{fig:tau_1}
  \end{figure*}

  Fig.~\ref{fig:dust_density} shows an exemplary density distribution for $a=20$ AU and 
  $M_B=1 \,\, {\rm M_{\odot}}$.   Although the hydrostatic scale height $H$ is 
  fairly smooth as a function of the radial distance to the central star, a variation of the disk profile in the plane perpendicular 
  to the midplane can be observed. From Eq. 2.2.1 it is obvious that the variation of the surface density causes a 
   shadow that screens parts of the inner disk edge 
  (Fig.~\ref{fig:tau_1}).
   Accretion arms caused by the presence of a binary, 
  although not very dense, have enough mass to reach optical depths $\tau \gtrsim 1$ (see Fig.~\ref{fig:tau_1}).
  
    In the following we discuss the influence of the binary on the observational properties of the disk.

 \subsubsection{Spectral energy distribution}%%%%%%%%%%%%%%%%%%%%%%%%%%%%%%%%%%%%%%%%%%%%%%%%%%%%%%%%%%%%%%%%%%%%%%%%%%%%%%%%%%%%%%%%%%%%%%%%%%%%%%%%%%%%%%%%%%%%%
\label{subsubsec:sed}

\begin{figure}
           \resizebox{\hsize}{!}{\includegraphics{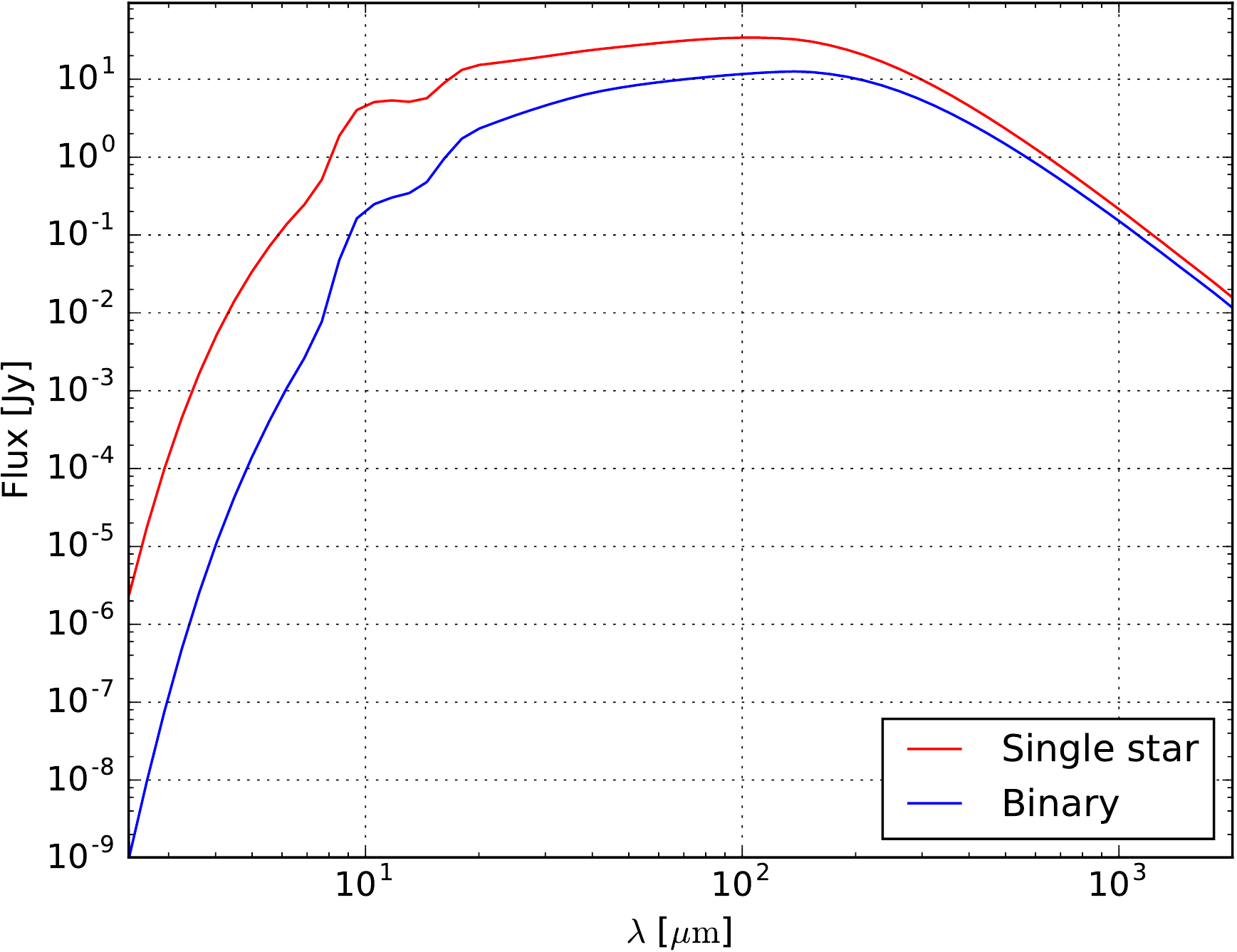}}
            \caption{Spectral energy distribution for two simulated disks with $M_{\rm disk} = 10^{-1} \,\, {\rm M_{\odot}}$, stellar mass of $2 \,\, {\rm M_{\odot}}$.
            Blue: single star; red: binary with two components of  $M_{\rm B}=1 \,\, {\rm M_{\odot}}$. Inner cavity radius for the binary $ \approx 40 $ AU.}
           \label{fig:sed_2}
  \end{figure}

\begin{figure}
          \resizebox{\hsize}{!}{\includegraphics{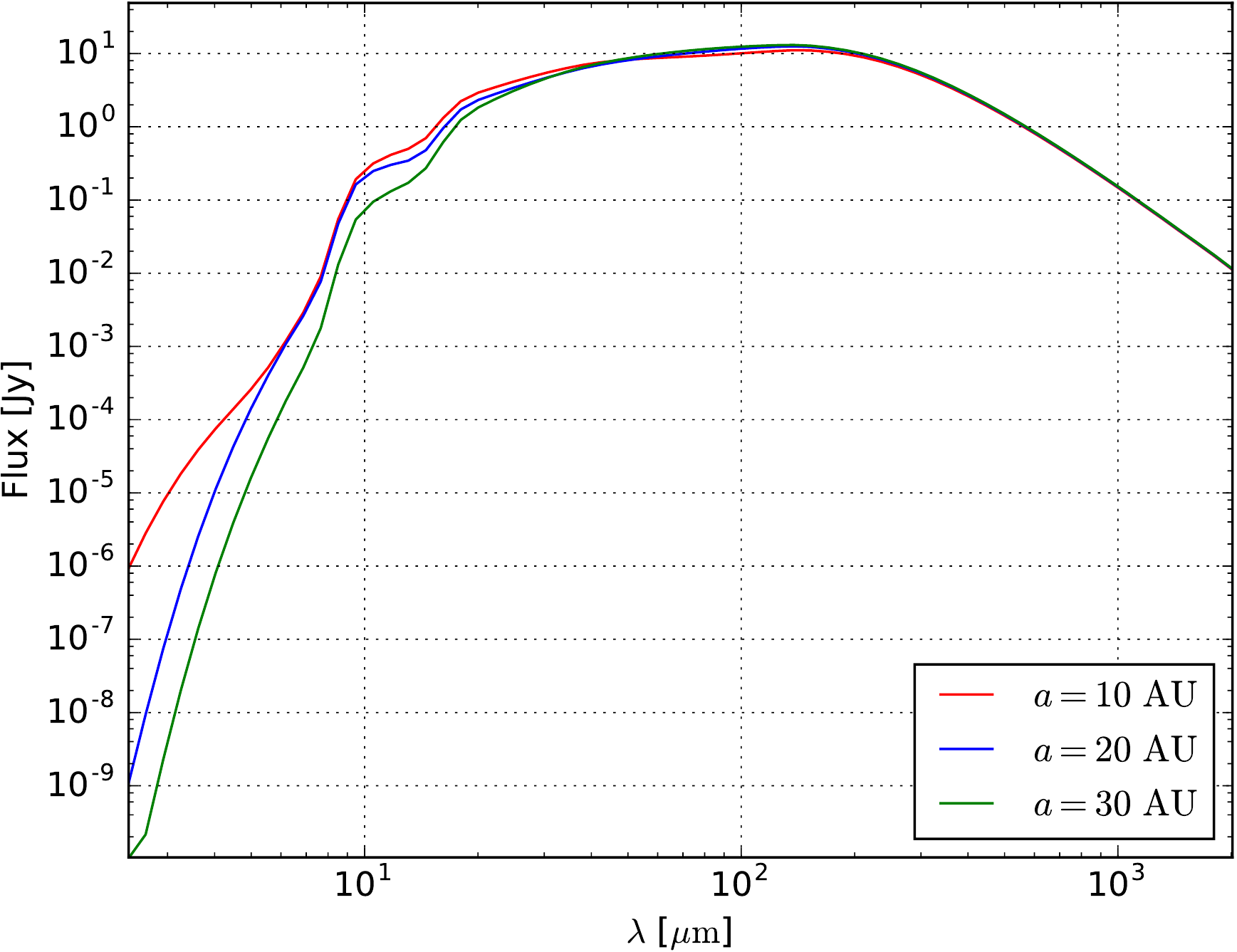}}
          \caption{Dependence of the SED on the semi-major axes  $a = 10, 20, 30 $ AU ($M_{\rm B}=1 \,\, {\rm M_{\odot}}$). }
          \label{fig:sed_3}
 \end{figure}

   The cavity in the centre of the disk is clearly visible in all calculated density distributions and emission maps and its 
   observability is the topic of Sect.~\ref{subsec:obs}. Because of the cavity, circumbinary disks tend to have less dust, which can be heated 
    to a high temperature. This results in much lower flux at near- and mid-infrared spectrum in comparison to 
    circumstellar disks (Fig.~\ref{fig:sed_2}). Because the cavity radius scales with the  semi-major axis $a$, the flux at short wavelengths is in indirect proportion to  $a$
    (Fig.~\ref{fig:sed_3}). The flux of the corresponding circumstellar disk (i.e. same disk mass and net stellar mass) is higher for all wavelengths (Fig.~\ref{fig:sed_2}).
    It was simulated using the standard set-up presented in Sect.~\ref{subsec:sim_fos} and a star with the mass of $M_{\rm star}=2 \,\,{\rm M_{\odot}}$.
    The luminosity of this star is higher than the combined 
    luminosity of two stars with individual masses $M_{\rm star}=1 \,\, {\rm M_{\odot}}$. This leads to an overall higher temperature and 
    thus to higher fluxes. This opens a potential way for distinguishing between disks around binaries and single stars: 
    The outer edges of the circumbinary disk behave nearly Keplerian. Through the line broadening one can determine the 
    mass of the central star. For the same mass of the central star (or binary), the decreased flux due to the cavity {\em and} the lower
    disk temperature (due to the lower net stellar luminosity) result in a lower infrared emission in the case of the circumbinary disk.

 \begin{figure}
          \resizebox{\hsize}{!}{\includegraphics{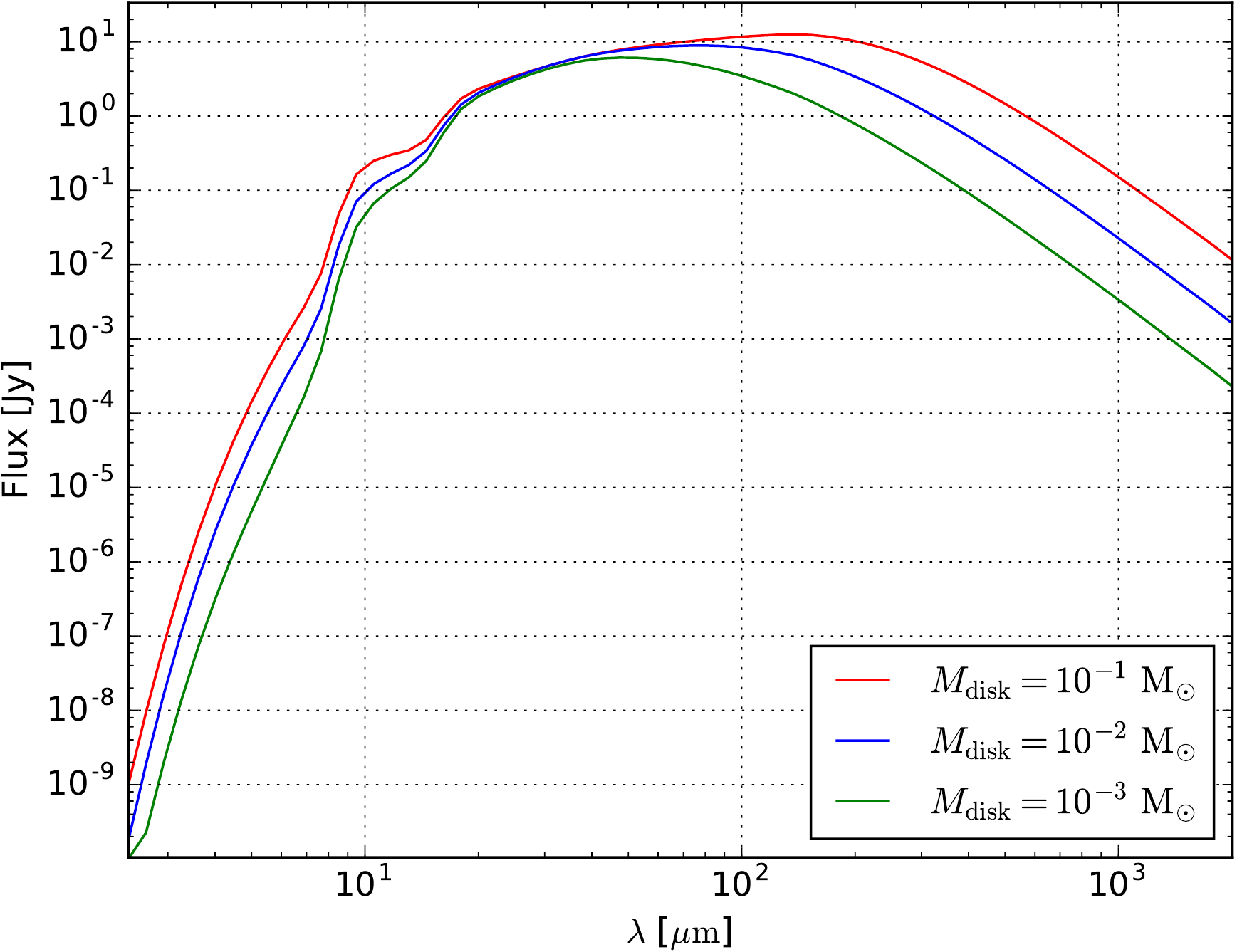}}
          \caption{Dependence of the SED on the binary disk mass.  $a = 20 $ AU, $M_{\rm B}=1 \,\, {\rm M_{\odot}}$, and $M_{\rm disk} = 10^{-1} - 10^{-3} \,\,{\rm M_{\odot}}$ }
          \label{fig:sed_1}
 \end{figure}

  At longer far-infrared to millimetre wavelengths the differences of the fluxes arising from circumbinary disks with different cavity radii 
  (i.e. different semi-major axes of the central binary) become negligible (Fig.~\ref{fig:sed_3}).
  This is expected since, for large distances from the central binary, the disks structures are similar.
  
  The dependence on the disk mass can be seen in Fig.~\ref{fig:sed_1}. Both the short wavelength and long wavelength 
  radiation increases with disk mass, although the increase is caused by different mechanisms. The presence of more hot dust in 
  the vicinity of the binary leads to higher  levels of short wavelength radiation. For long wavelength radiation, for which the disk is optically 
  thin in the vertical direction, higher dust mass leads to a higher column density of radiating particles.

   \subsubsection{Surface brightness distribution}
   \label{subsubsec:spat_res}
   
   So far, we only considered the synthetic SEDs of the simulated disks. However, as both the contributions of
   the scattered and re-emitted radiation depend solely on the radial density distribution, an interpretation of the SED to reveal 
   the characteristic disk structures discussed in Sect.~\ref{subsec:res_fos} is rather limited. We now 
   investigate the impact of these characteristic structures on spatially resolved observations in various wavelength regimes.

\begin{figure*}
          \resizebox{\hsize}{!}{\includegraphics{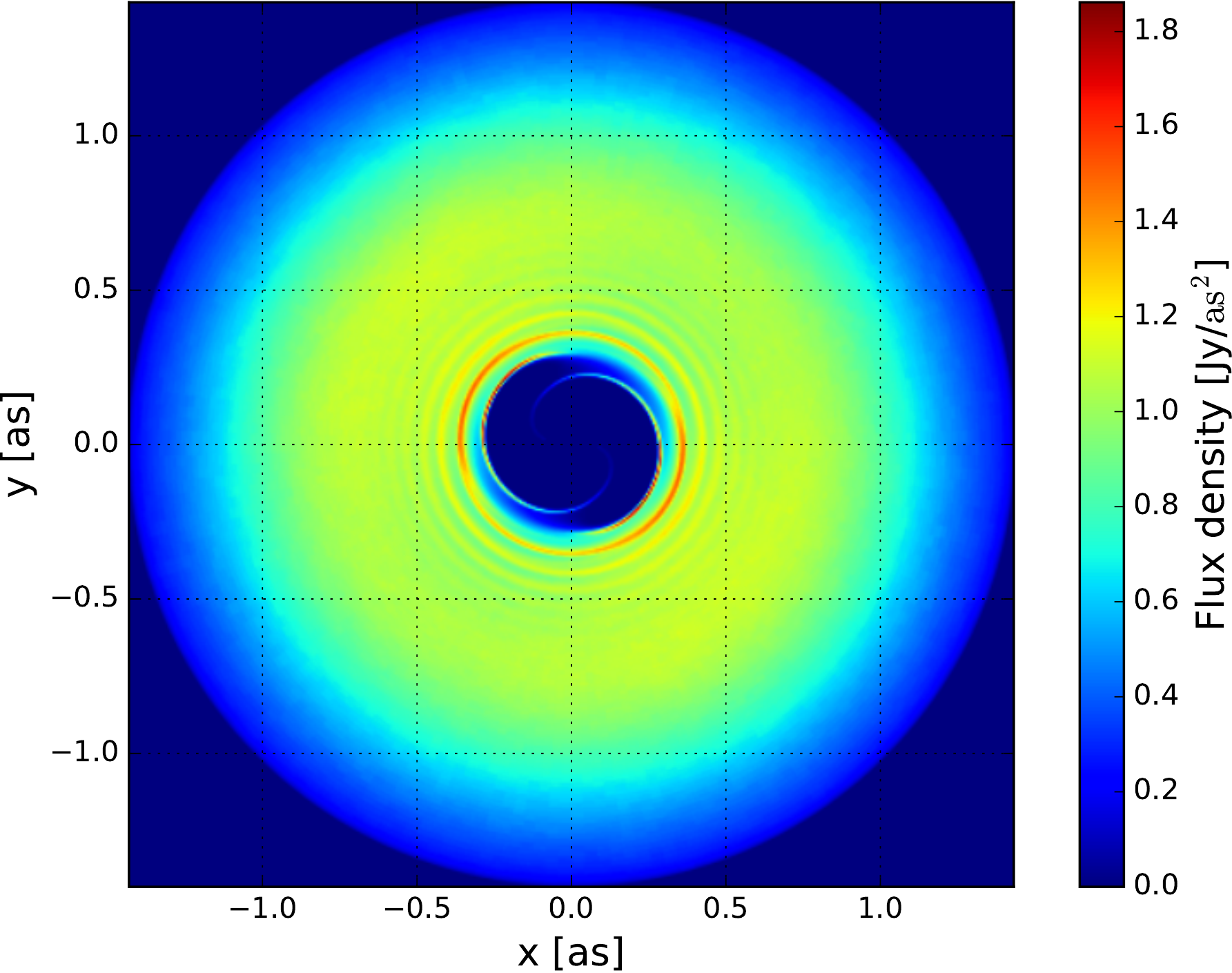} \quad
                                \includegraphics{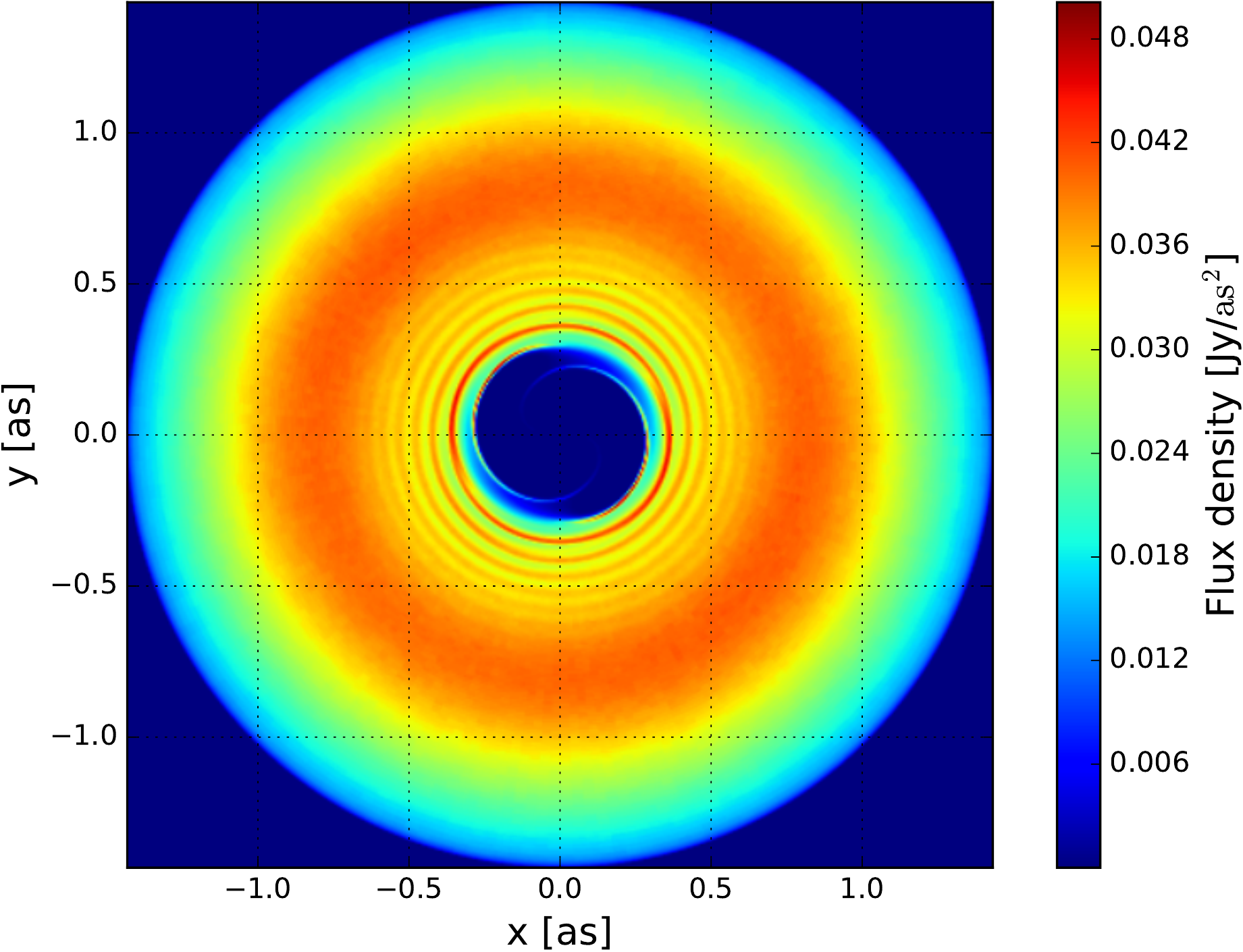}}
          \caption{Surface brightness distribution at $\lambda = 324$ $\mu m$ for disk masses $M_{\rm disk}=10^{-1}{\rm M_{\odot}}$ 
                   (\textsl{left}) and $M_{\rm disk}=10^{-3}{\rm M_{\odot}}$ (\textsl{right}).}
          \label{fig:flux_mass_1}
    \end{figure*}

   \paragraph{Disk mass:}
   The effect the disk mass has on observable features is best seen in the (sub)millimetre regime, since the observed column
   density is much higher. Fig.~\ref{fig:flux_mass_1} shows the two surface brightness distribution for two disk masses at  $\lambda = 324$ $\mu m$.
   The cavity in the centre of the disk is not affected by the mass increase. As the disks are optically thin in this wavelength range, 
   the re-emission flux scales approximately linearly with the disk mass.

\begin{figure*}
          \resizebox{\hsize}{!}{\includegraphics{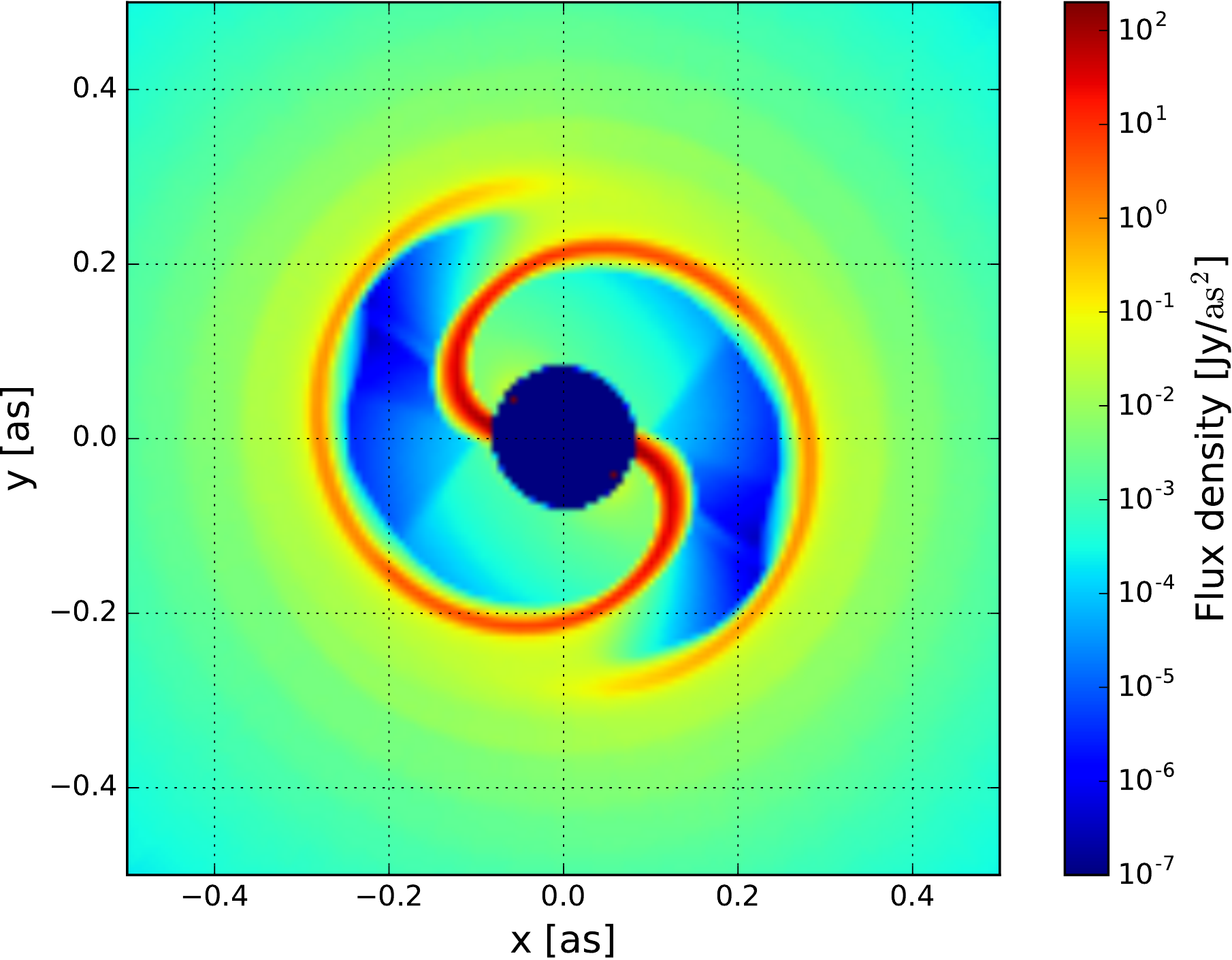} \quad
                                \includegraphics{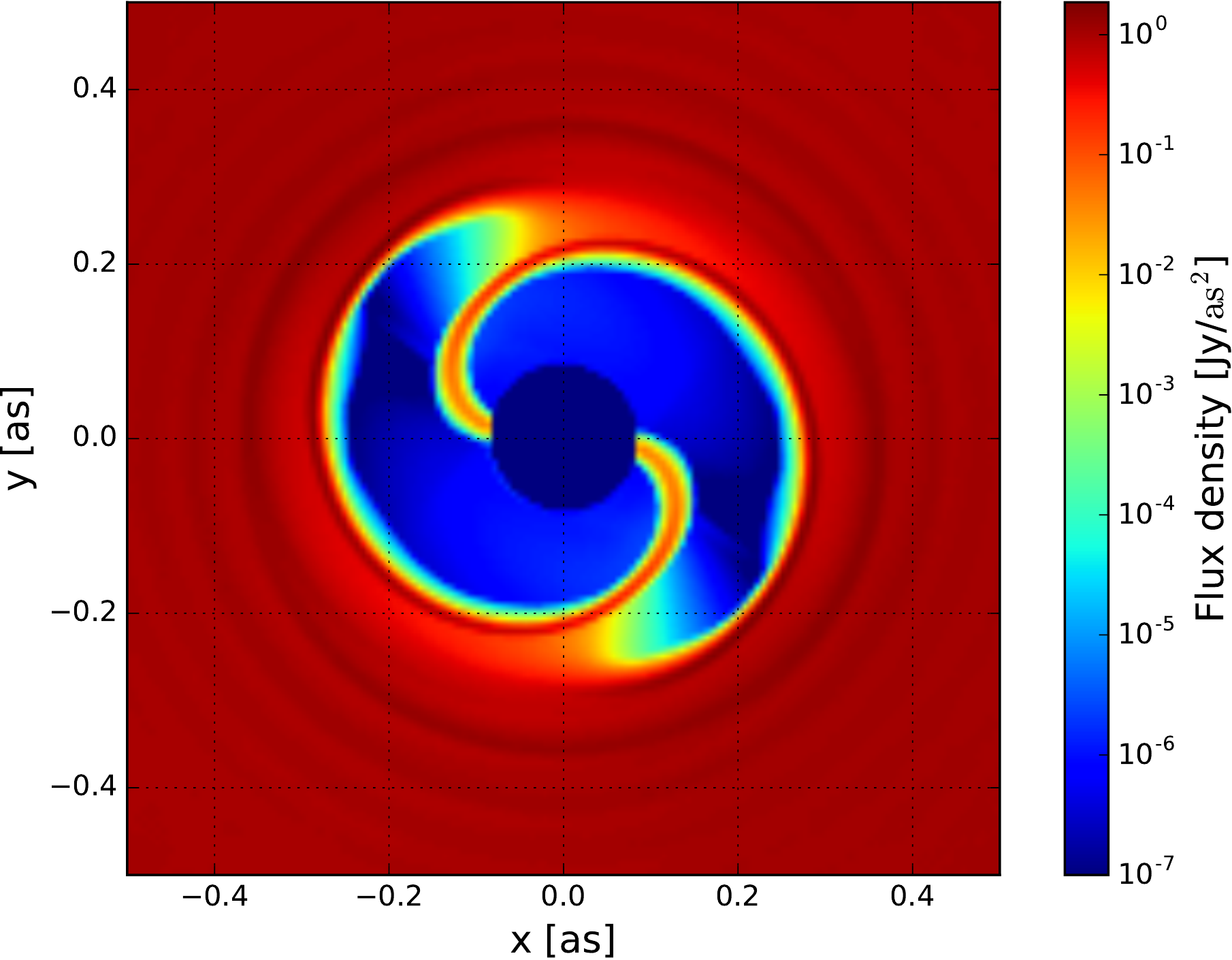}}
          \caption{Surface brightness distribution at $\lambda = 10.5$ $\rm \mu m$ (\textsl{left}) and $324$ $\rm \mu m$ (\textsl{right}) ($a = 20$ AU, $M_{ \rm B}=1 \,\, {\rm M_{\odot}}$).}
          \label{fig:flux_1}
 \end{figure*}

   \paragraph{Flux maximum:}
   Fig.~\ref{fig:flux_1} shows the spatial flux density distribution at different wavelengths. The near- and mid-infrared radiation reaches its maximum 
   along the accretion arms, which are the innermost structures, directly exposed to the stellar radiation.
   The thermal re-emission from the inner disk edge, where the density waves are strongest reaches its maximum at submillimeter wavelengths.

  \paragraph{Density waves:}
   In Sect.~\ref{subsec:res_fos} we presented a quantitative description of the dependence of the density waves 
    generated by the binary on the semi-major axis $a$ and the binary mass $M_{\rm B}$. 
% From Fig.~\ref{fig:flux_mass_1}~-~\ref{fig:semi_1} we can summarise that these waves 
%   influence the spatial flux distribution in the disk and hence can potentially be observed.
  We now investigate the wave
  structure in the resulting surface brightness distributions employing a similar algorithm as in Sect.~\ref{subsec:res_fos}.
  At first, we average the brightness distribution in azimuthal direction. Subsequently,
  we determine the locations of maxima  and minima and calculate the flux difference between the maxima and their surrounding 
  minima. From both differences, the smaller one is chosen, which ensures that in case of sufficient sensitivity at least one flux maximum is detected. 
  After performing this procedure for all maxima, the largest difference for each parameter configuration, defined by the binary mass $M_{\rm B}$, 
  disk mass $M_{\rm disk}$, semi-major axis $a$, and wavelength $\lambda$, is chosen. This allows us to derive the lower boundary for the sensitivity  
  required to detect at least one flux maximum.
  
\begin{figure*}
%       \begin{subfigure}
          \resizebox{\hsize}{!}{\includegraphics{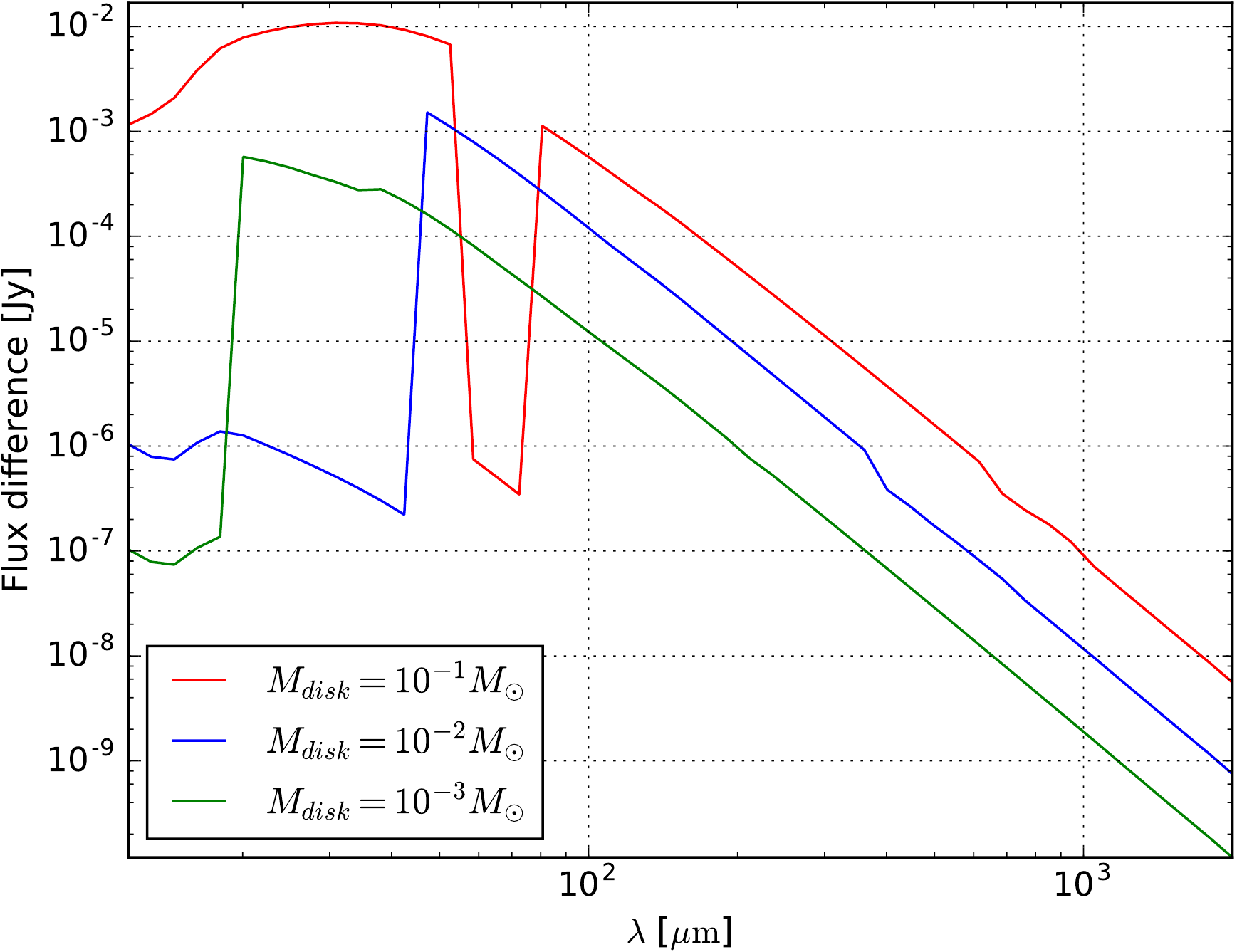} \quad
%           \subcaption{(a)}
%       \end{subfigure}
          \includegraphics{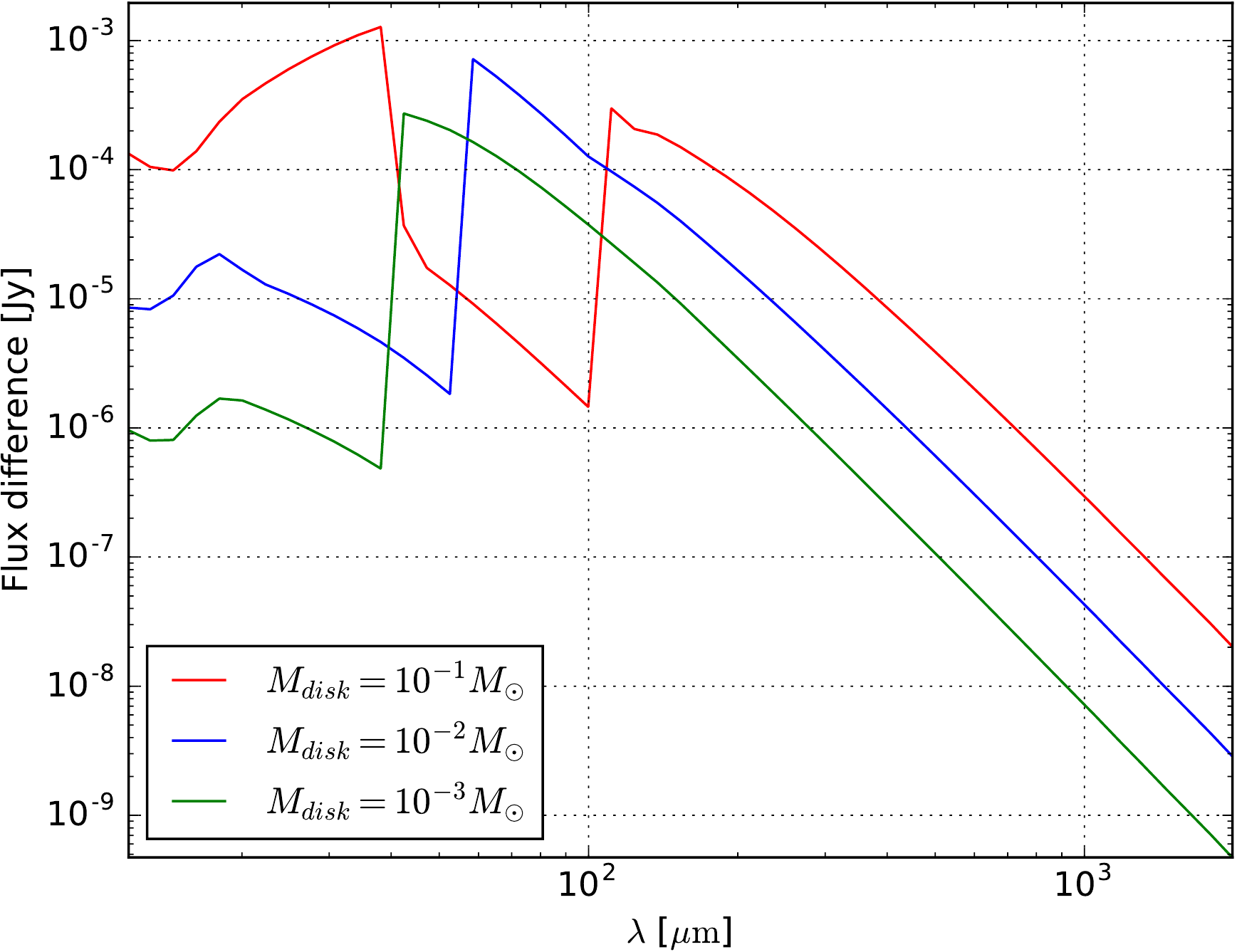} \quad
          \includegraphics{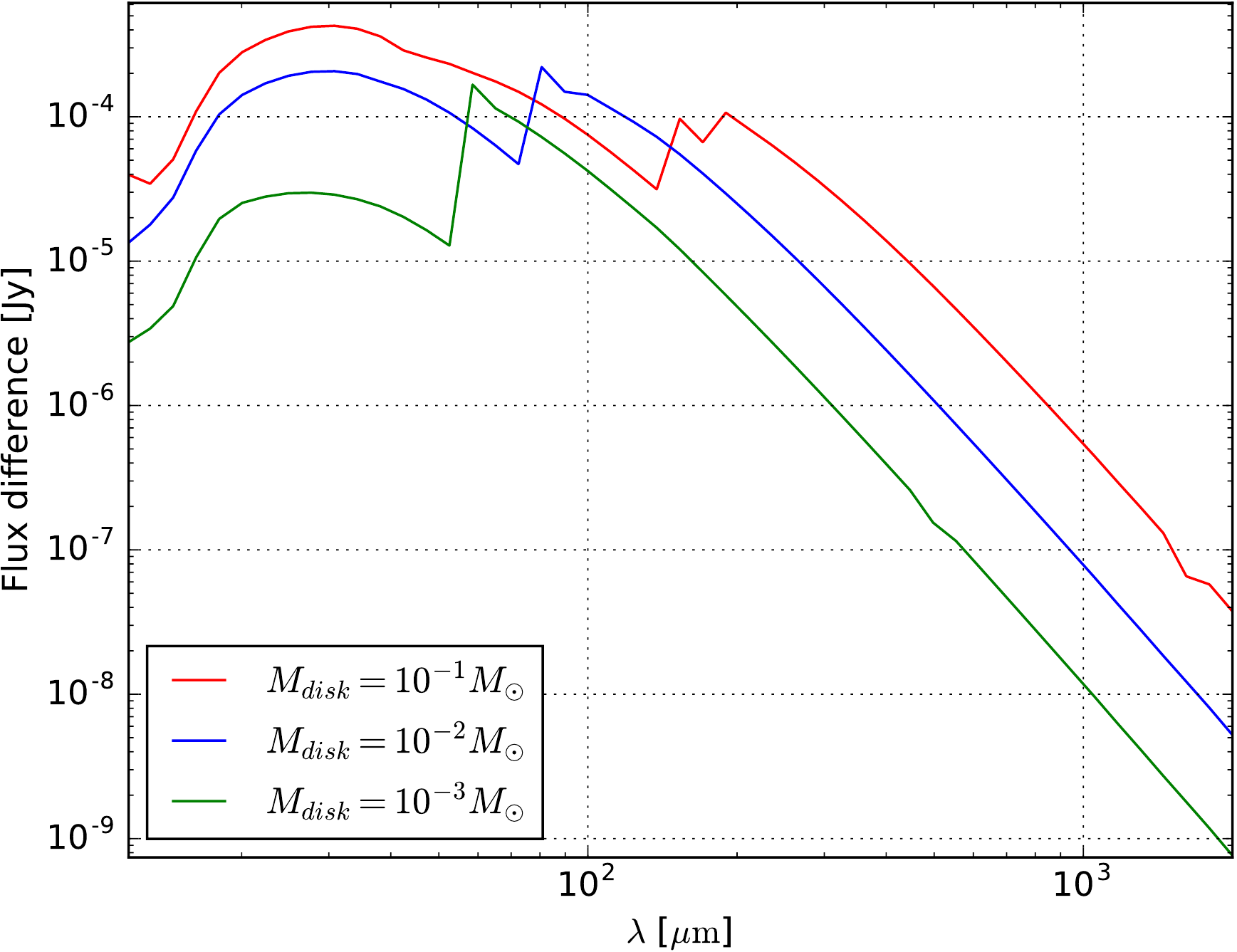}}
          \caption{Flux difference for  $M_{\rm B}=1 \,\,{\rm M_{\odot}}$ and  $a = 10,\,\,  20,\,\, 30$ AU  (see Sect.~\ref{subsec:res_fos} for details).}
          \label{fig:flux_diff_a}
 \end{figure*}

\begin{figure*}
          \resizebox{\hsize}{!}{\includegraphics[width=0.96\textwidth]{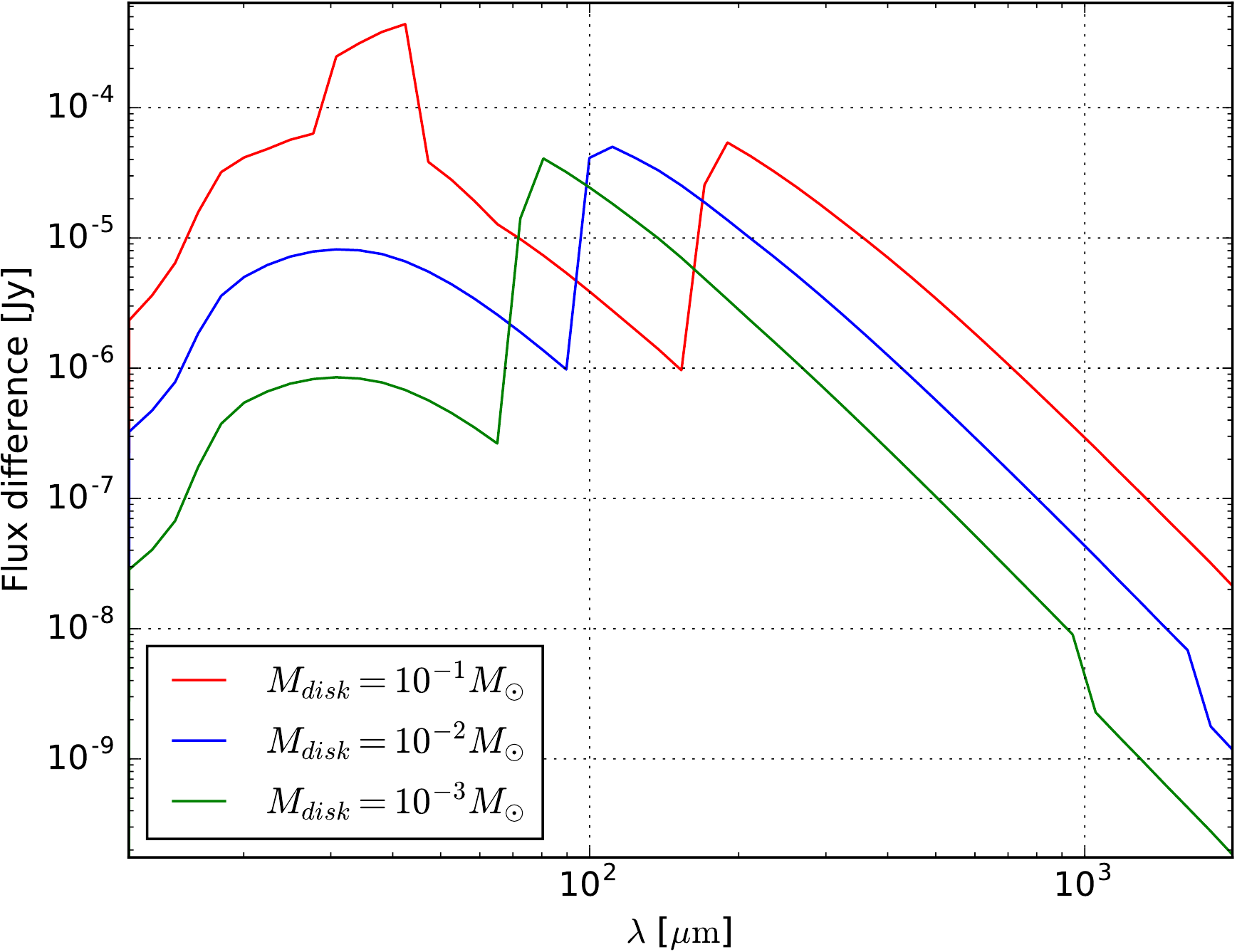} \qquad
                                \includegraphics[width=\textwidth]{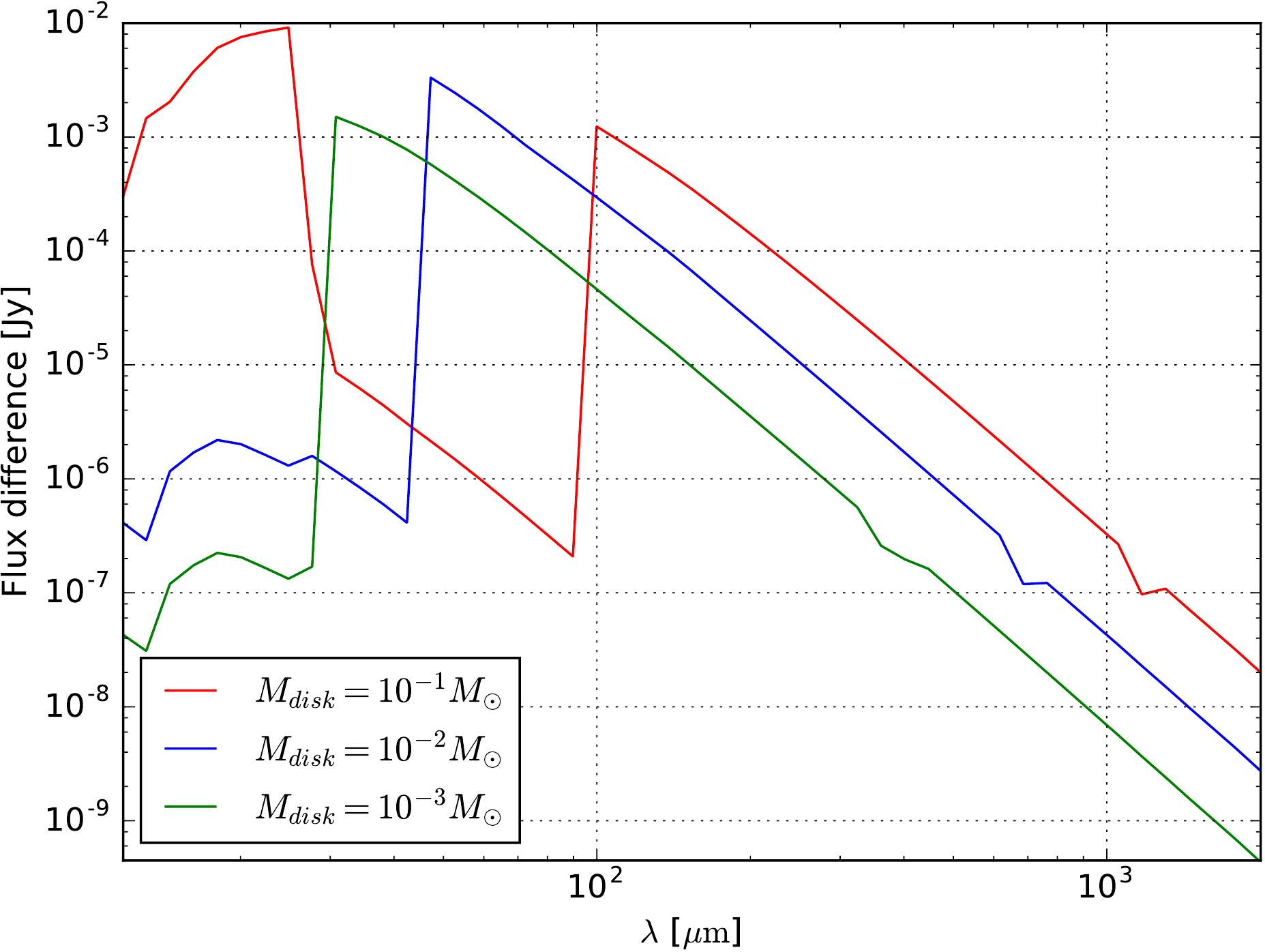}}
          \caption{Flux difference for  $M_{\rm B}=0.5,  \,\, 1.5 \,\, {\rm M_{\odot}}$ and  $a = 20$ AU (see Sect.~\ref{subsec:res_fos} for details).}
          \label{fig:flux_diff_M}
 \end{figure*}

  Fig.~\ref{fig:flux_diff_a} shows the flux differences as a function of the wavelength of the emitted radiation. Each figure represents a different semi-major axis $a$ and the colours 
  denote different disk masses for each binary configuration. Fig.~\ref{fig:flux_diff_M} depicts the same quantities for different binary masses $M_{\rm B}$. 
  
  We start at the long wavelength end of these graphs to find an explanation for the trend observed here. At wavelengths $\lambda > 1$ mm the disks are 
  optically thin ($\tau \approx 1$ in the most dense regions for $M_{\rm disk}=10^{-1} \,{\rm M_{\odot}}$). 
  The linear increase of the flux towards shorter wavelengths is due to the higher fluxes at these wavelengths (see e.g. Fig.~\ref{fig:sed_1}).
  Additionally, the amplitude of the density waves is the highest at the inner disk rim (Fig.~\ref{fig:radial_density} and~\ref{fig:radial_density_mass}).
  As the flux maximum moves inward for shorter wavelengths, the flux at the location of the largest magnitudes increases, resulting in the further increase of the flux differences.
  At about 100 $\mu m$ the flux difference decreases abruptly.
  This is the wavelength at which the disk becomes optically thick. At even shorter wavelengths, one no longer traces the full column density, but only the wavelengths-dependent photosphere of the disk. 
  This interpretation also explains why this decrease happens at shorter wavelengths for disks with less mass.
  
  The remaining feature to be discussed is the bump at wavelengths 
  between the increase of the optical depth and about 10 $\rm \mu m$. In this wavelength range the flux difference shows a more complex dependence 
  on the parameters than in the cases discussed previously. It increases with disk mass, but not linearly (in a log-log diagram) as before. 
  Here, the temperature at the inner disk rim influences the maximum differences. These  are larger for smaller semi-major 
  axis $a$ values and higher binary masses $M_{\rm B}$. In both cases the temperature increases owing to the proximity of the binary orbit to the disk inner rim or owing to higher stellar luminosity.
  At these short wavelengths the regions of the disk contributing most to the flux no longer have azimuthal symmetry 
  that is necessary for the applied algorithm. For this reason these parts of the graphs are not discussed further.

   \paragraph{Accretion arms:}
   As discussed earlier, the appearance of the disk at infrared wavelengths is dominated by the thermal re-emission and scattered light of the accretion arms (see Fig.~\ref{fig:flux_1}). 
   In this part of the disk, 
   the gravitational potential differs significantly from a standard Keplerian potential. This leads to a high temporal variability and to the lack of the 
   azimuthal symmetry. For those reasons, we abstain from attempting to quantify our results as  we did with density waves. Instead, we  only outline general trends with regard to the various parameter spaces.

\begin{figure}
          \resizebox{\hsize}{!}{\includegraphics{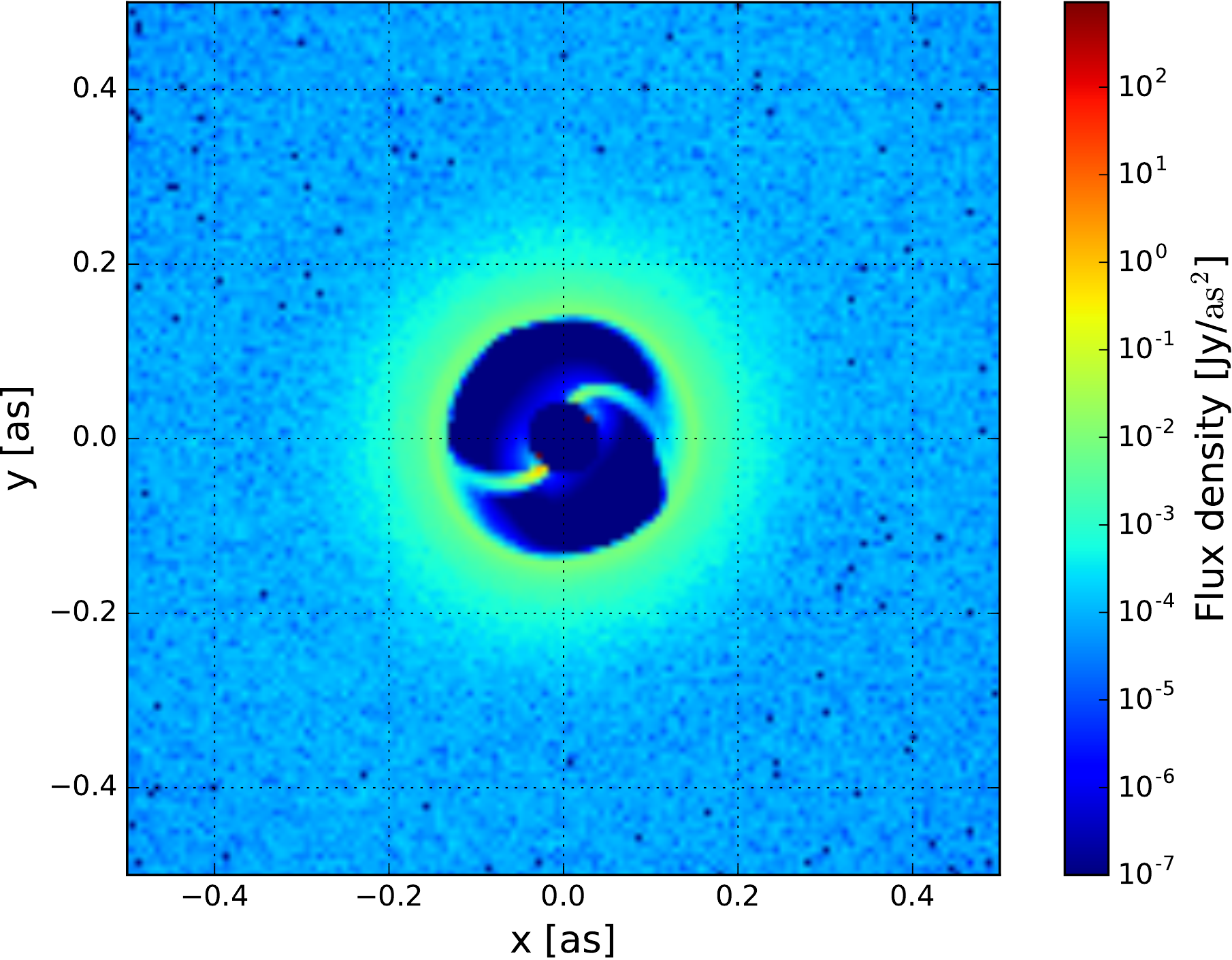}} \\
          \resizebox{\hsize}{!}{\includegraphics{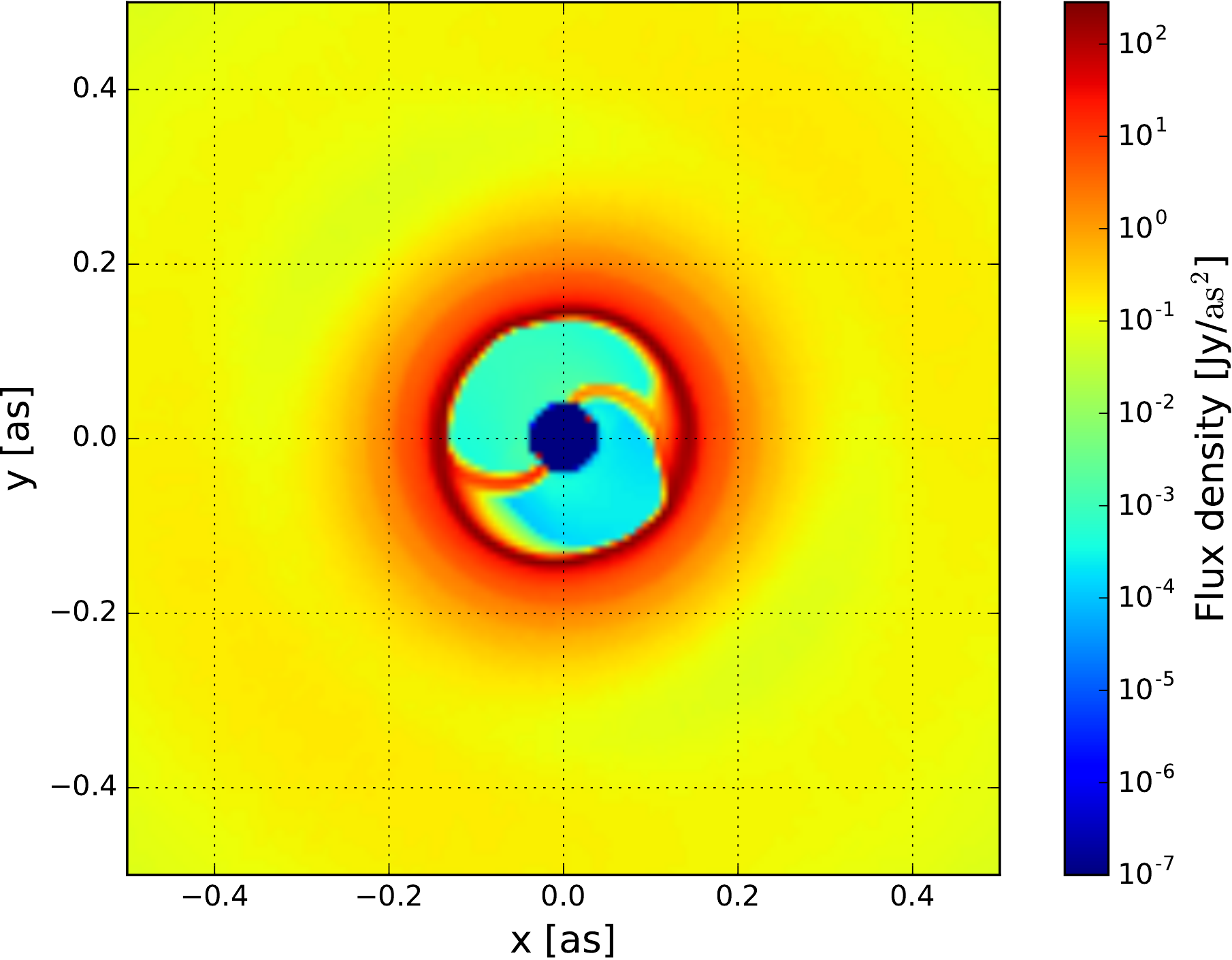}}
          \caption{Surface brightness distribution at  $\lambda =4.5$ $\rm \mu m$ (\textsl{left}) and $\lambda =20$ $\rm \mu m$ (\textsl{right})  for $M_{\rm B}=1\,\, {\rm M_{\odot}}$ and  $a = 10$ AU. }
          \label{fig:IR_10}
 \end{figure}

\begin{figure}
          \resizebox{\hsize}{!}{\includegraphics{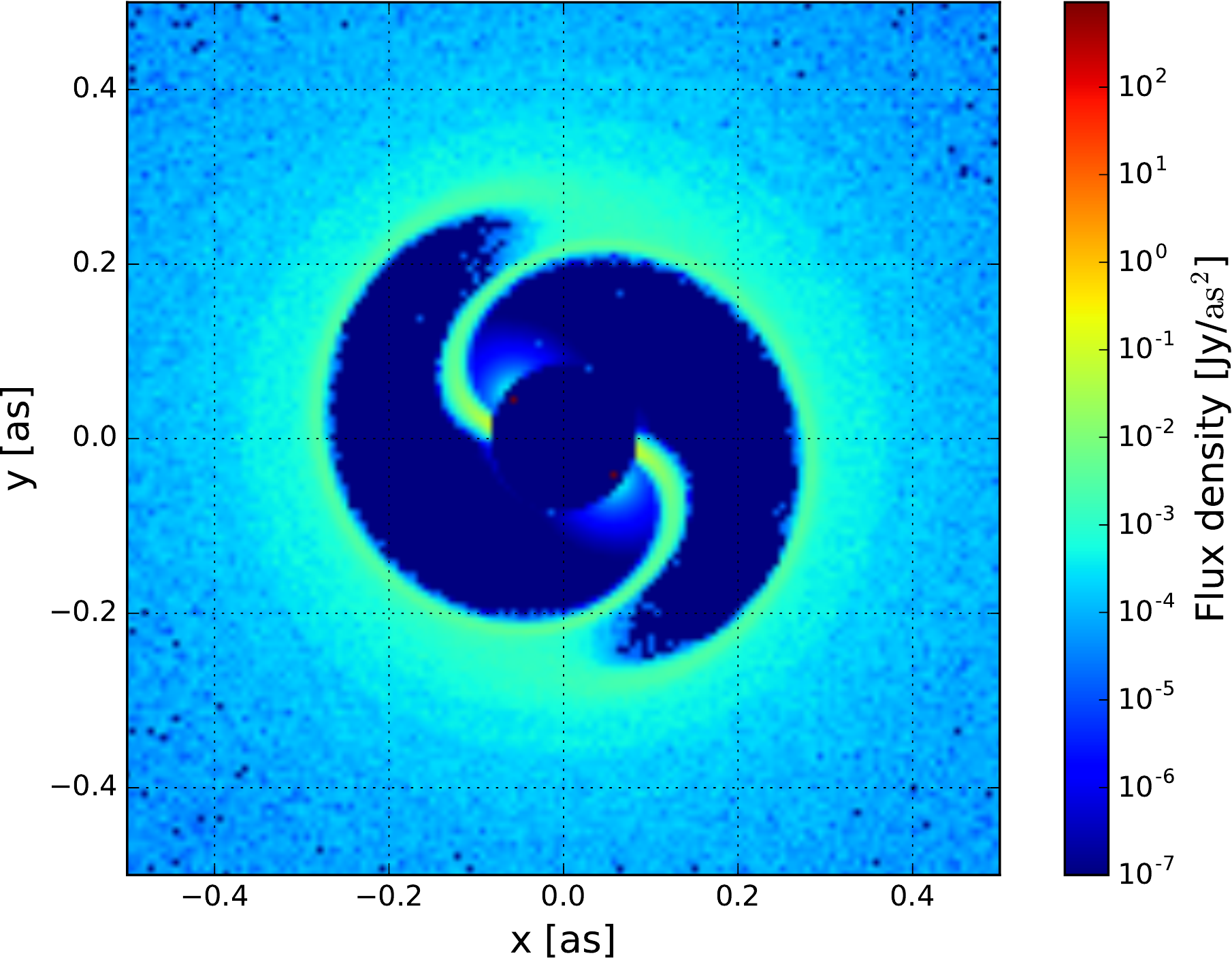}} \\
          \resizebox{\hsize}{!}{\includegraphics{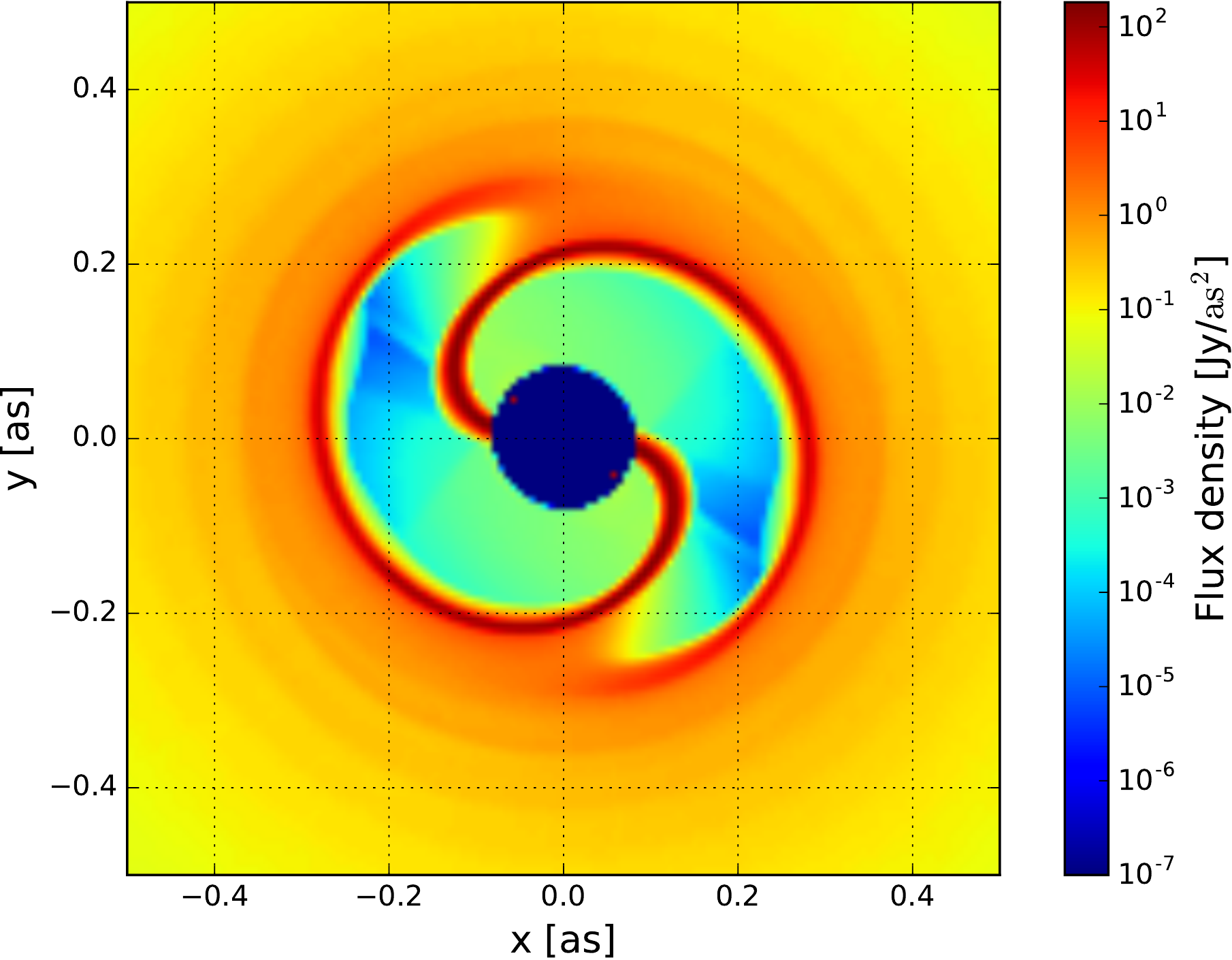}}
          \caption{Surface brightness distribution at  $\lambda =4.5$ $\rm \mu m$ (\textsl{left}) and $\lambda =20$ $\rm \mu m$ (\textsl{right})  for $M_{\rm B}=1\,\, {\rm M_{\odot}}$ and  $a = 20$ AU. }
          \label{fig:IR_20}
 \end{figure}

\begin{figure}
          \resizebox{\hsize}{!}{\includegraphics{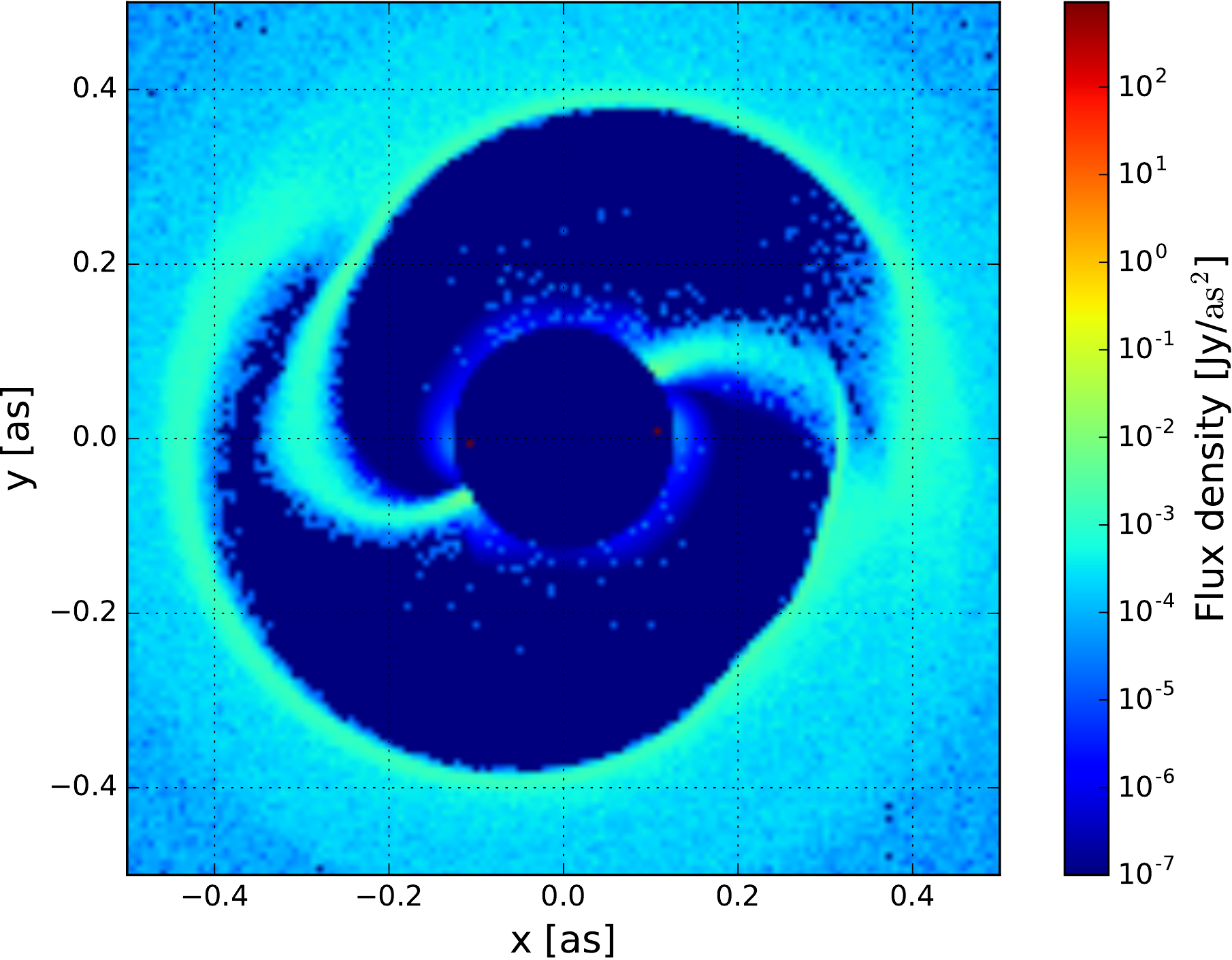}} \\
          \resizebox{\hsize}{!}{\includegraphics{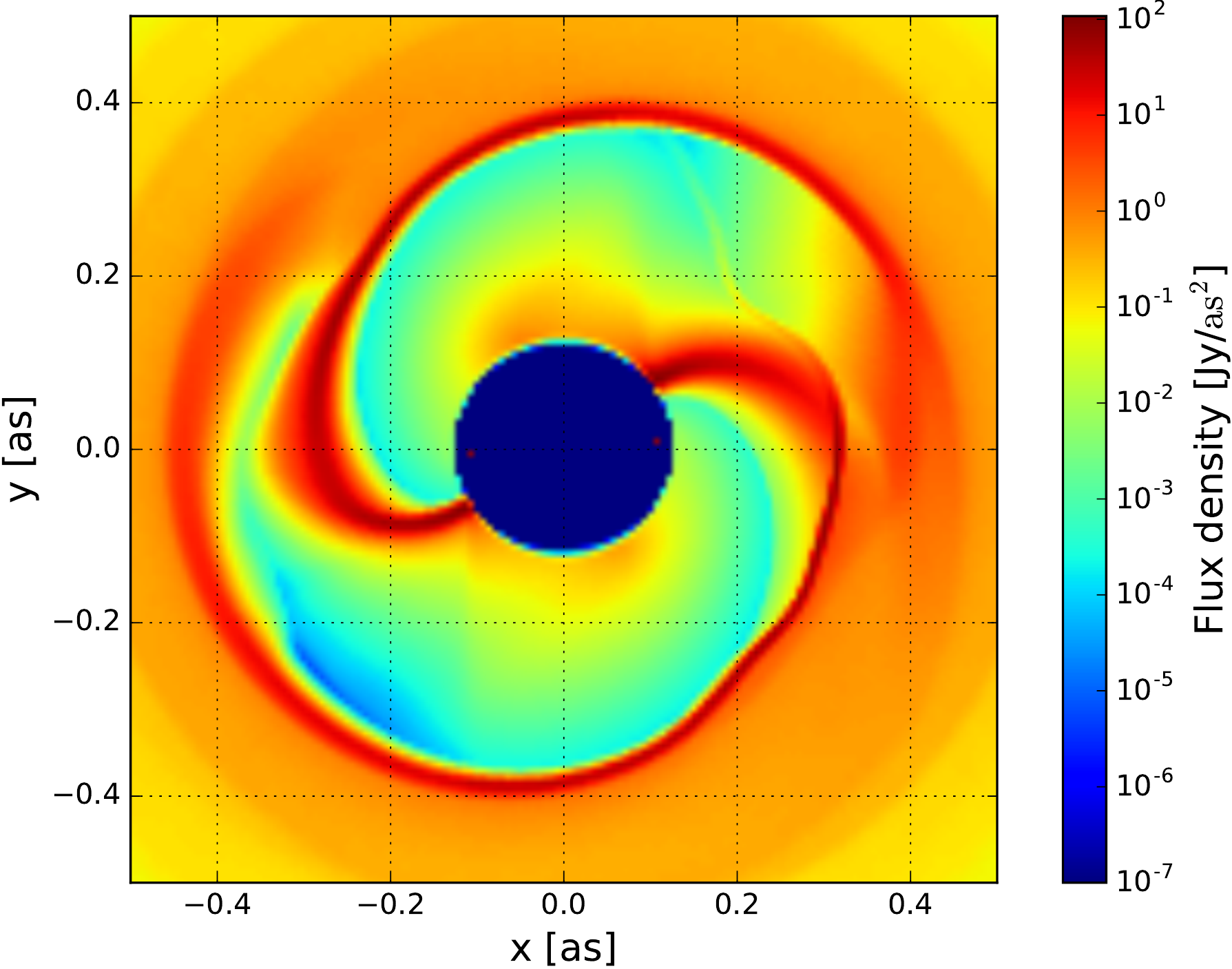}}
          \caption{Surface brightness distribution at  $\lambda =4.5$ $\rm \mu m$ (\textsl{left}) and $\lambda =20$ $\rm \mu m$ (\textsl{right})  for $M_{\rm B}=1 \,\,{\rm M_{\odot}}$ and  $a = 30$ AU. }
          \label{fig:IR_30}
 \end{figure}

\begin{figure}
          \resizebox{\hsize}{!}{\includegraphics{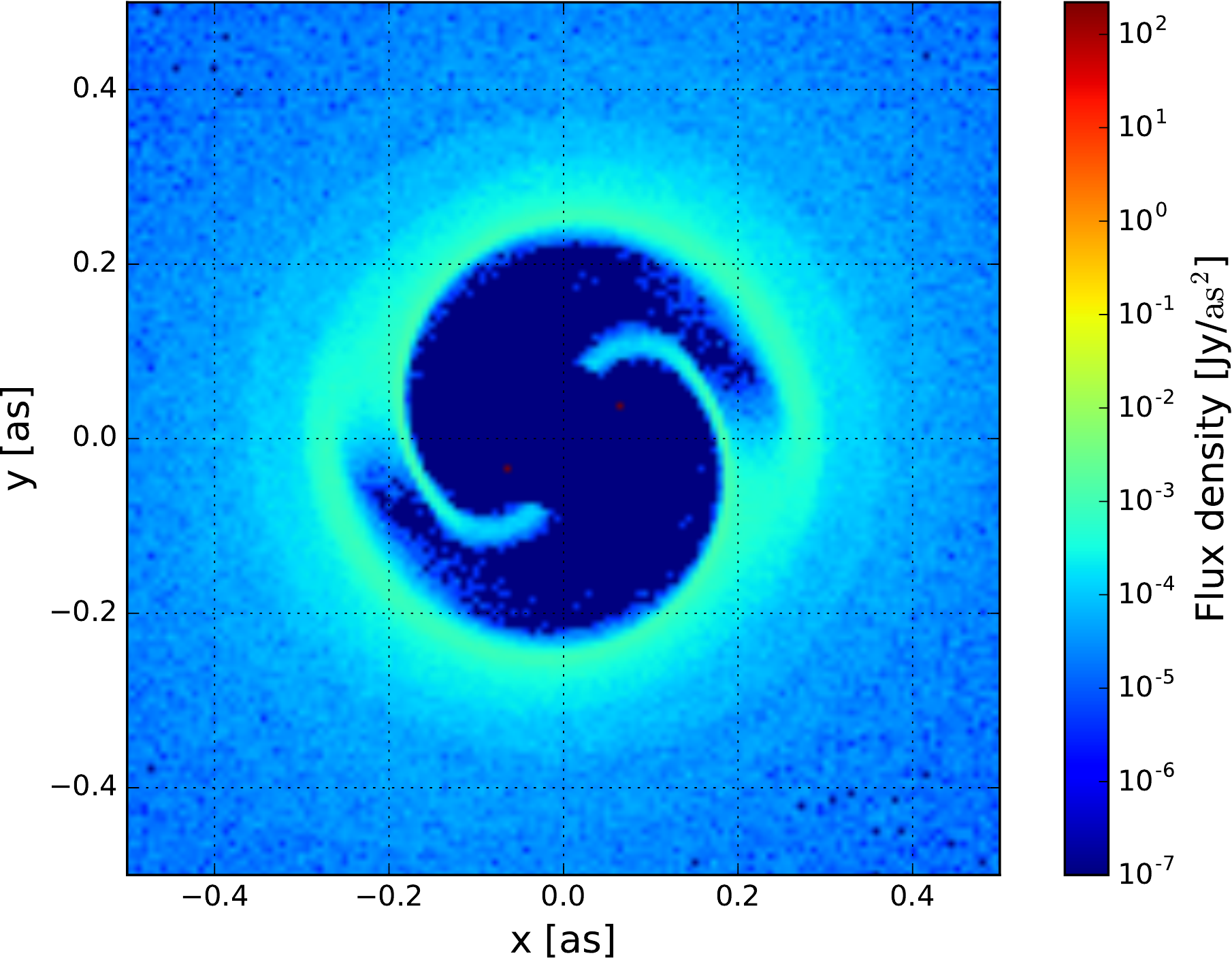}} \\
          \resizebox{\hsize}{!}{\includegraphics{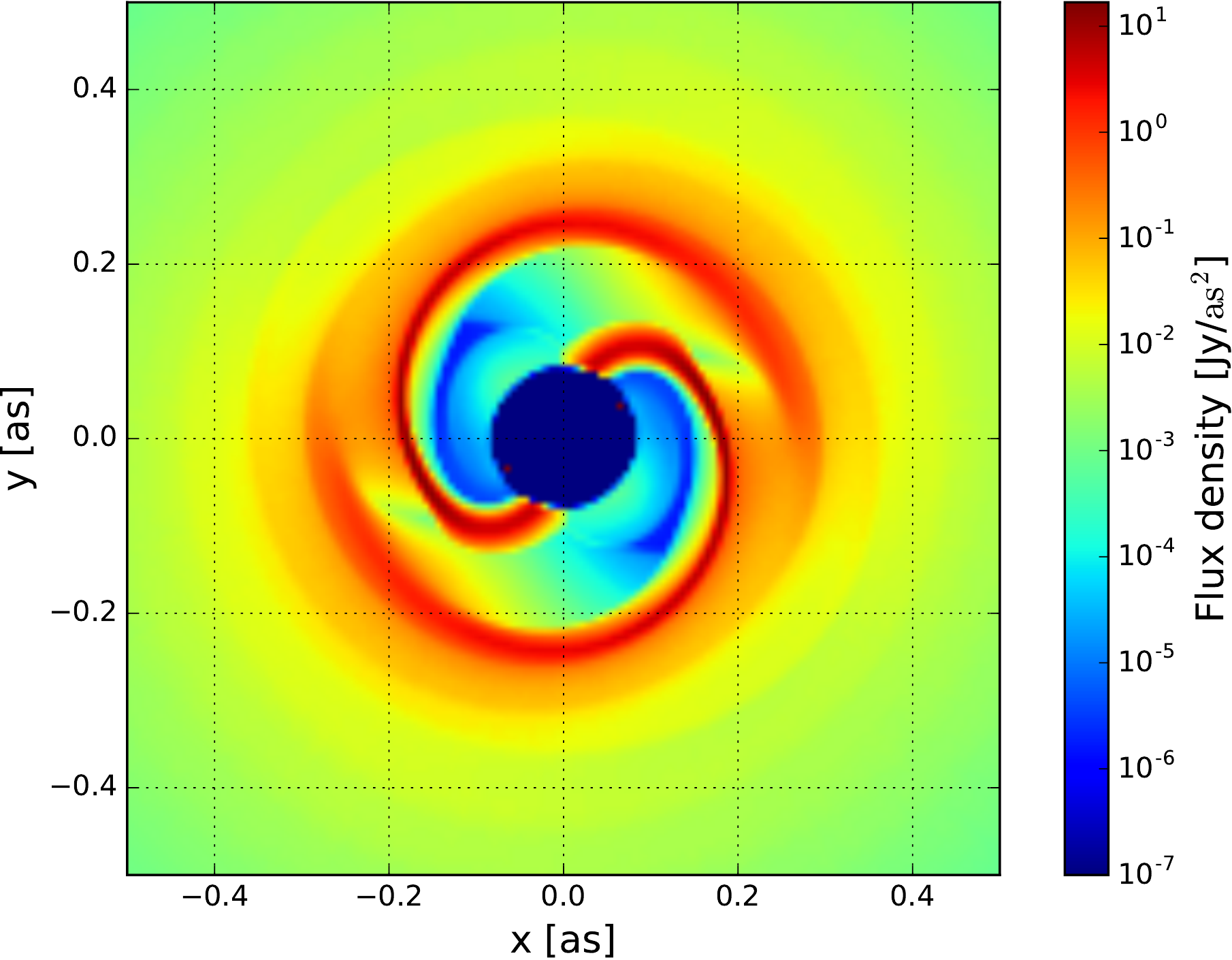}}
          \caption{Surface brightness distribution at  $\lambda =4.5$ $\rm \mu m$ (\textsl{left}) and $\lambda =20$ $\rm \mu m$ (\textsl{right})  for $M_{\rm B}=0.5 \,\,{\rm M_{\odot}}$ and  $a = 20$ AU. }
          \label{fig:IR_20_05}
   \end{figure}

\begin{figure}
          \resizebox{\hsize}{!}{\includegraphics{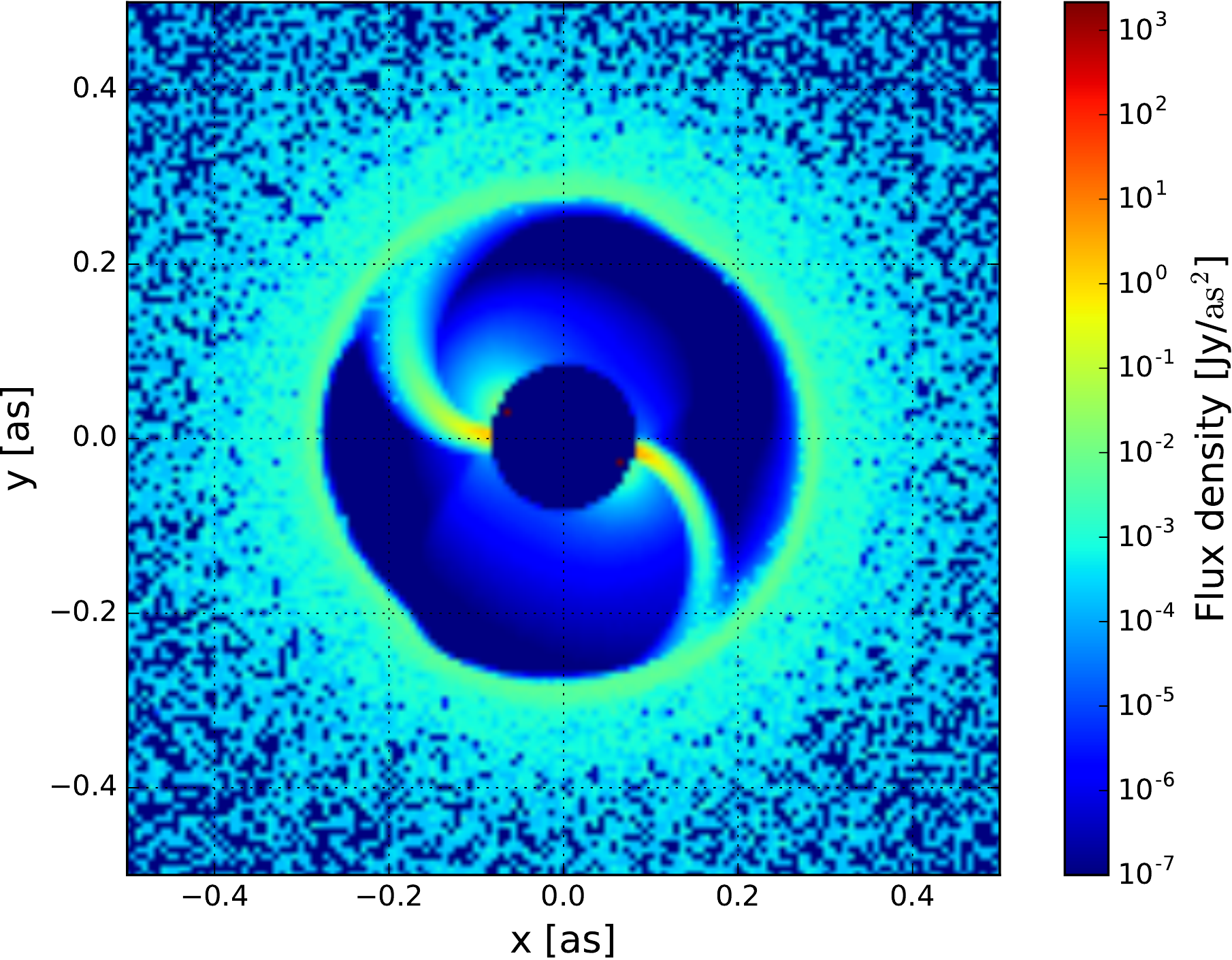}} \\
          \resizebox{\hsize}{!}{\includegraphics{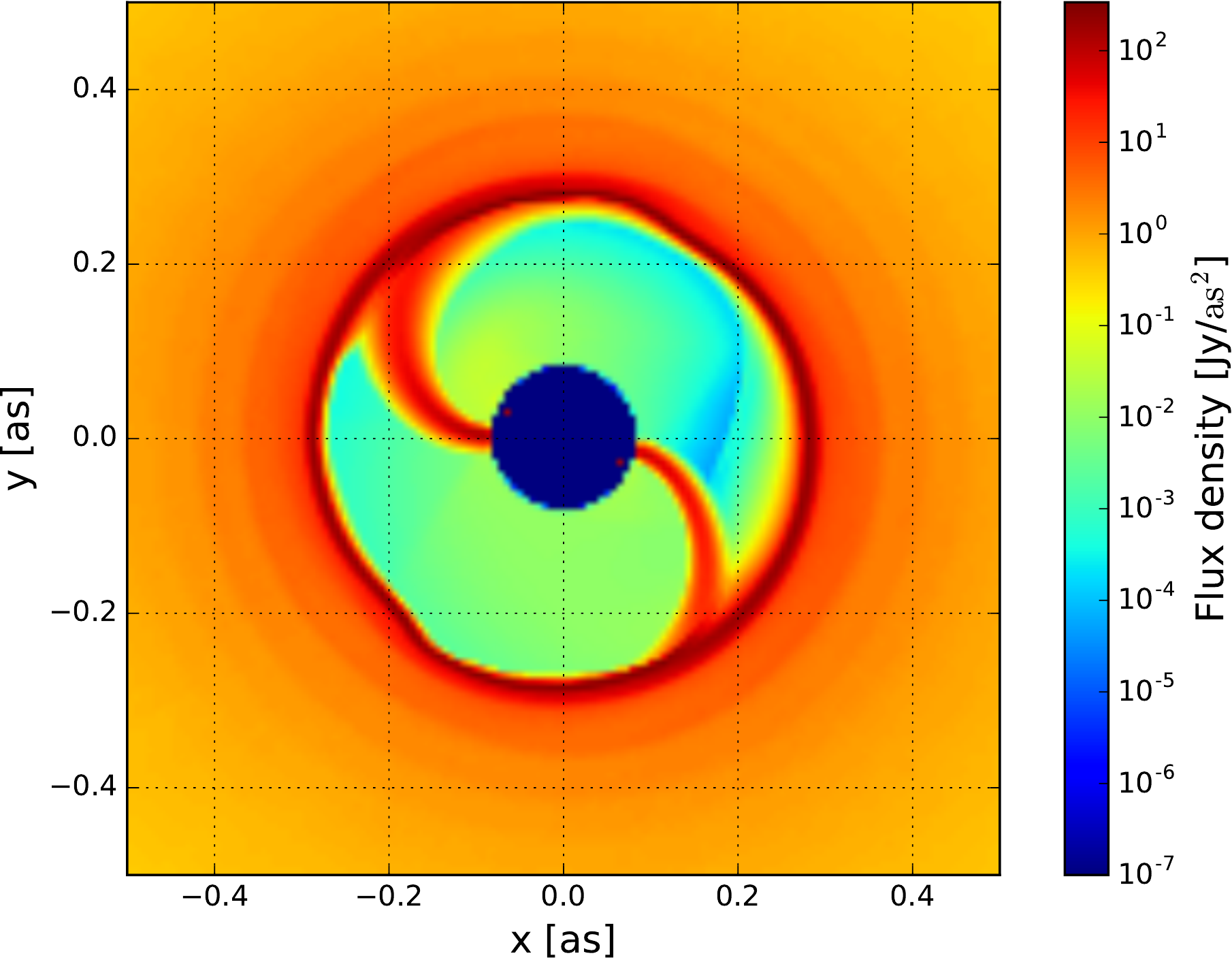}}
          \caption{Surface brightness distribution at  $\lambda =4.5$ $\rm \mu m$ (\textsl{left}) and $\lambda =20$ $\rm \mu m$ (\textsl{right})  for $M_{\rm B}=1.5 \,\,{\rm M_{\odot}}$ and  $a = 20$ AU. }
          \label{fig:IR_20_15}
 \end{figure}

   Figs.~\ref{fig:IR_10} -~\ref{fig:IR_20_15} show the surface brightness distribution for semi-major axes $a = 10,\,\,  20,\,\, 30$ AU, $M_{\rm B}=1 \,\,{\rm M_{\odot}}$, 
   and masses $M_{\rm B}=0.5,\,\,1.5 \,\, {\rm M_{\odot}}$, $a = 20$ AU
   at two different wavelength, $\lambda =4.5$ $\rm \mu m$ and $\lambda =20$ $\rm \mu m$.

   Here we can note a major trend that was partially mentioned before, namely that greater 
   values of the semi-major axis $a$ lead to greater distances of disk edge to the binary orbit and therefore lower fluxes. Similarly, a higher binary mass leads to a higher stellar luminosity 
    and consequently to a higher thermal re-emission flux. A further trend is the increase of flux with wavelength 
    (see SEDs in Figs.~\ref{fig:sed_2} -~\ref{fig:sed_1}). In Table~\ref{fig:eelt_pre_falt} the fluxes originating from 
    inside a circle with radius of $2 \times a$ are compiled. From these we can deduce that the binary mass $M_{\rm B}$ and thus its luminosity has the biggest impact on the fluxes.

   \begin{table}
     \caption{Integrated flux of the scattered and re-emitted radiation in units of mJy originating from inside a circle with the radius of $2 \times a$.
              The direct stellar radiation is not considered.}
\begin{tabular}{c|c|c}
        \multicolumn{3}{c}{}                                                                                                        \\ \hline \hline
                    Parameter                                     & $\lambda = 4.5 {\rm \mu m}$  & $\lambda = 20 {\rm \mu m}$       \\ \hline
         $M_{\rm B} = 1.0 {\rm M_{\odot}} \, ,\,\,\,\, a = 10$ AU & $0.43$                       & $ 2819.76 $                      \\ \hline
         $M_{\rm B} = 1.0 {\rm M_{\odot}} \, ,\,\,\,\, a = 20$ AU & $0.27$                       & $ 2137.45 $                      \\ \hline
         $M_{\rm B} = 1.0 {\rm M_{\odot}} \, ,\,\,\,\, a = 30$ AU & $0.1 $                       & $ 1363.06 $                      \\ \hline
         $M_{\rm B} = 0.5 {\rm M_{\odot}} \, ,\,\,\,\, a = 20$ AU & $0.079$                      & $ 152.56  $                      \\ \hline
         $M_{\rm B} = 1.5 {\rm M_{\odot}} \, ,\,\,\,\, a = 20$ AU & $1.71$                       & $ 7129.55 $                      \\ \hline \hline
 \end{tabular}    
  \label{fig:eelt_pre_falt}
  \end{table}

   Taking the trends discussed above  into account, one can derive predictions for the observability of circumbinary disk features
   (at this stage we disregard the influence of a realistic observing instruments; this is the topic of the 
   following sections). 
   
   \begin{itemize}
    \item The accretion arms can easiest be observed in the infrared, where we find the highest fluxes 
    for them. Systems with smaller semi-major axis values $a$ and higher binary masses $M_{\rm B}$ are preferable for such observations.
   \item The flux of a disk with a higher dust mass is significantly larger, which leads to the conclusion that more 
    massive disks are more suitable for observations in infrared and in millimetre radiation. 
    \item For observations at submillimetre/millimetre wavelengths we 
    find that higher values of the semi-major axis $a$ result in larger flux differences for the density waves. 
    The binary mass has no significant impact on the flux difference, but one should expect the overall flux to be higher for systems with higher 
    binary masses.
   \end{itemize}

    %%%%%%%%%%%%%%%%%%%%%%%%%%%%%%%%%%%%%%%%%%%%%%%%%%%%%%%%%%%%%%%%%%%%%%%%%%%%%%%%%%%%%%%%%%%%%%%%%%%%%%%%%%%%%%%%%%%%%%%%%%%%%%%%%%%%%%%%%%%%%%%%%%%%
   \subsection{Simulated observations: Specific instrument studies}%%%%%%%%%%%%%%%%%%%%%%%%%%%%%%%%%%%%%%%%%%%%%%%%%%%%%%%%%%%%%%%%%%%%%%%%%%%%%%%%%%%%%%%%%%%%%%%%%%%%%%%%%%%%%%%%%%%%%%%%%%%%%%%%%%%%%%%%%%%%%%%%%%%%
   \label{subsec:obs}
   
   We now study synthetic observations based on the analysis of ideal surface brightness distributions.
   First, we consider simulated observations at submillimeter/millimetre wavelengths. Subsequently, 
   simulated observations with the future E-ELT are presented, 
   which will operate at optical to infrared wavelengths.

  \subsubsection{ALMA}%%%%%%%%%%%%%%%%%%%%%%%%%%%%%%%%%%%%%%%%%%%%%%%%%%%%%%%%%%%%%%%%%%%%%%%%%%%%%%%%%%%%%%%%%%%%%%%%%%%%%%%%%%%%%%%%%%%%%%%%%%%%%%%%%%%%%%%%%%%%%%%%%%%%
    \label{subsubsec:obs_ALMA}
    
    In this subsection we evaluate the observability of the characteristic structures discussed in Sect.~\ref{subsec:res_mol}
    with the (sub)-millimetre interferometer ALMA. In particular these are the inner cavity and  density waves. At first, we 
    specify basic technical data assumed in this observational feasibility study.

     \paragraph{ALMA set-up:}
     We perform simulations within the frame of capabilities in Observational Cycle 4, which employs 
     40 of the 12 m antennas in nine different configurations with maximum baselines between $155$ m and $12\,644$ m
     (seven wavelength bands between $0.32$ mm and $3.6$ mm).
     Despite the fixed number of antennas and configurations, the results presented here are of general nature because they only depend on the simulated scattered light and re-emission maps and the chosen angular resolution and sensitivity. 
     The synthetic observations are performed with the ALMA simulation tool kit 
     \texttt{CASA} \citep{McMullin_2007}. The on-source time for all simulated observations is set to 6 h 
     with the thermal noise option enabled and recommended water vapours values applied. 
     A declination representative for the nearby star-forming region in Taurus is used ($\delta \approx 22^\circ$).

\begin{figure}
          \resizebox{\hsize}{!}{\includegraphics{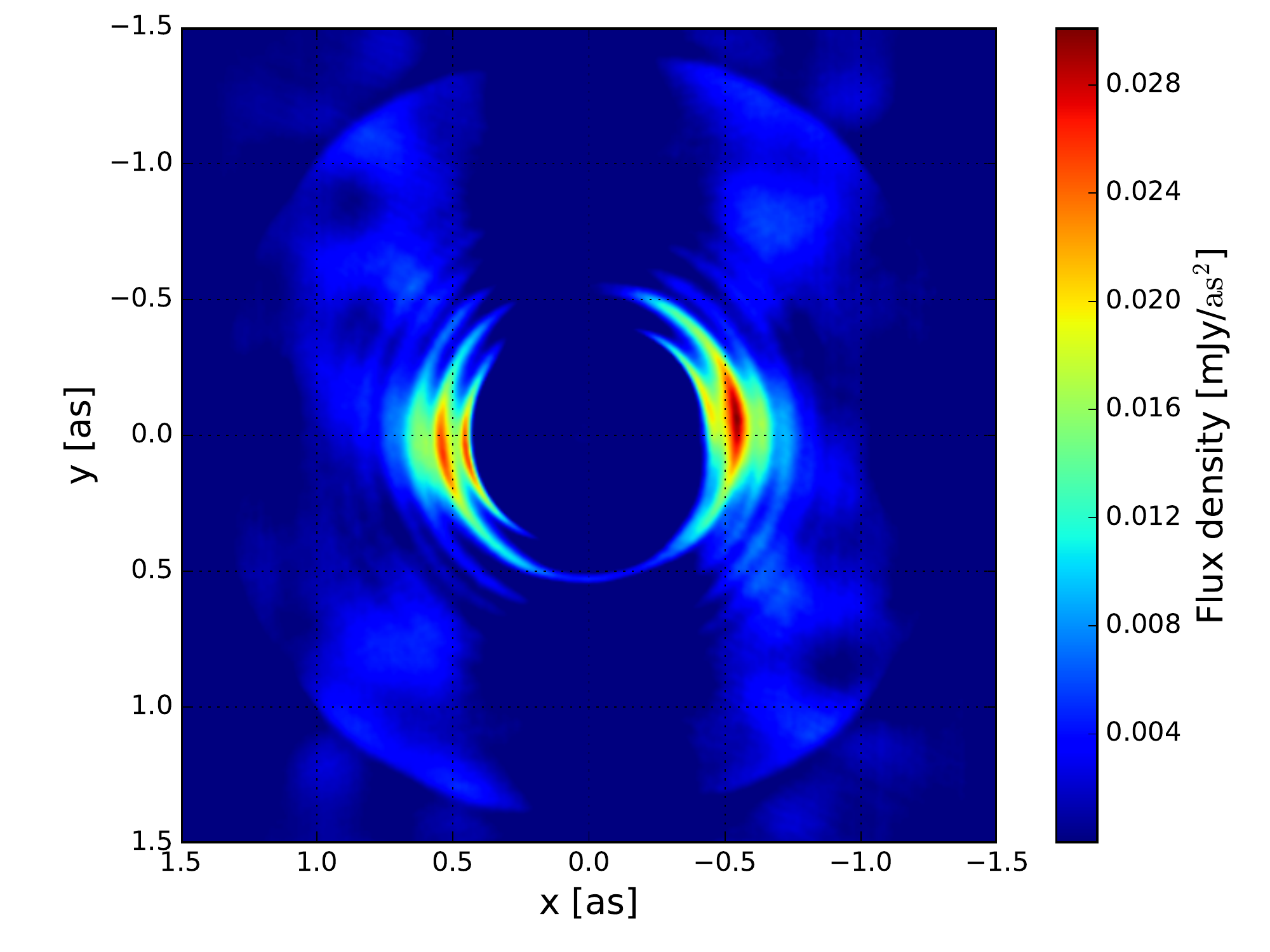}}
          \caption{Synthetic ALMA observation at $\lambda =320$ $\rm \mu m$ with configuration \# 7; $M_{\rm B}=1\,\, {\rm M_{\odot}}$ and  $a = 30$ AU.}
          \label{fig:alma_a}
 \end{figure}

\begin{figure}
           \resizebox{\hsize}{!}{\includegraphics{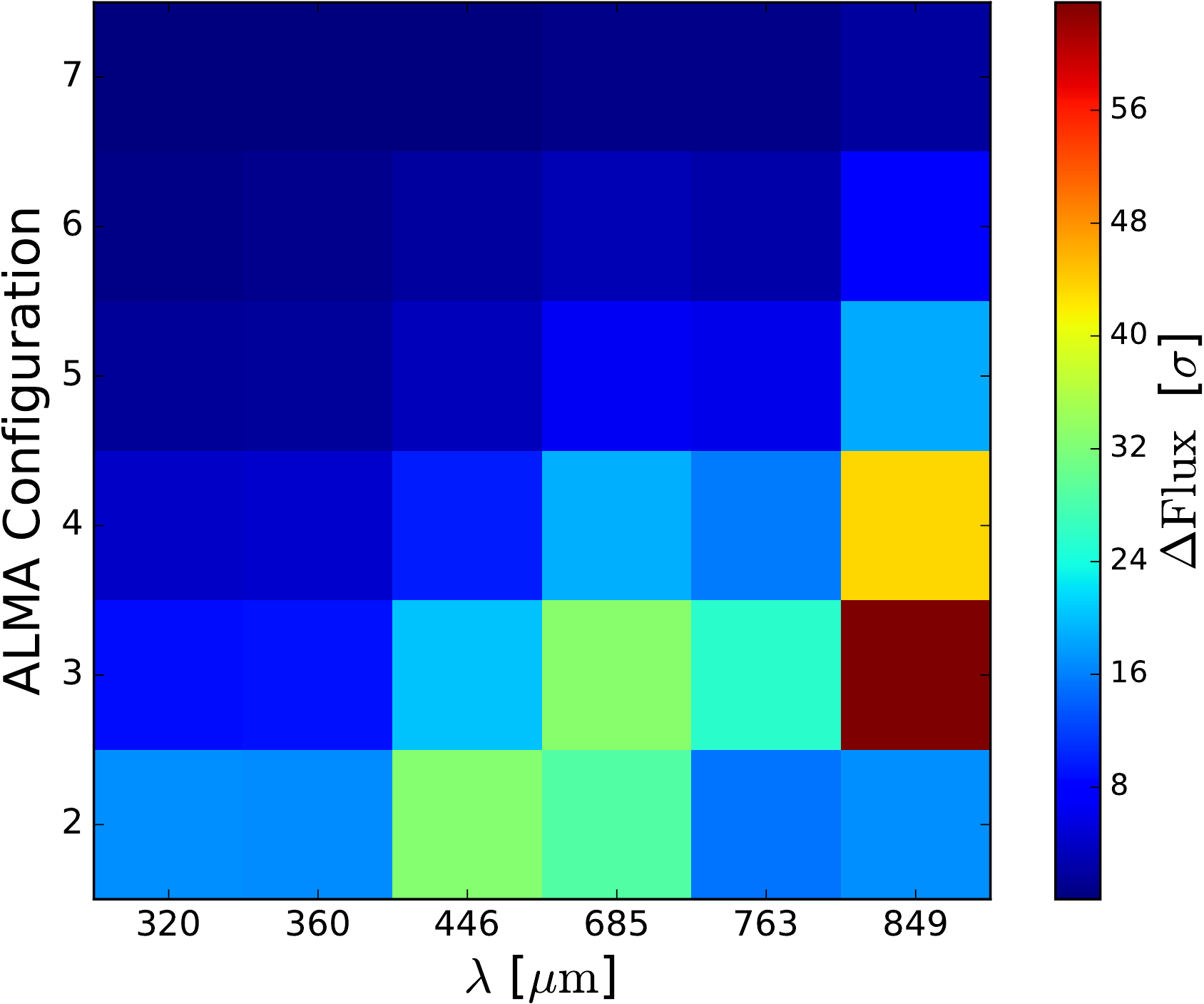}}
           \caption{Flux difference between density wave maximum and minimum in units of sensitivity for $a= 10$ AU,  $M_{\rm B} = 1\,\, {\rm M_{\odot}}$, and $M_{\rm disk} = 10^{-1}\,\, {\rm M_{\odot}}$.}
          \label{fig:sigma_10_1}
    \end{figure}

\begin{figure}
           \resizebox{\hsize}{!}{\includegraphics{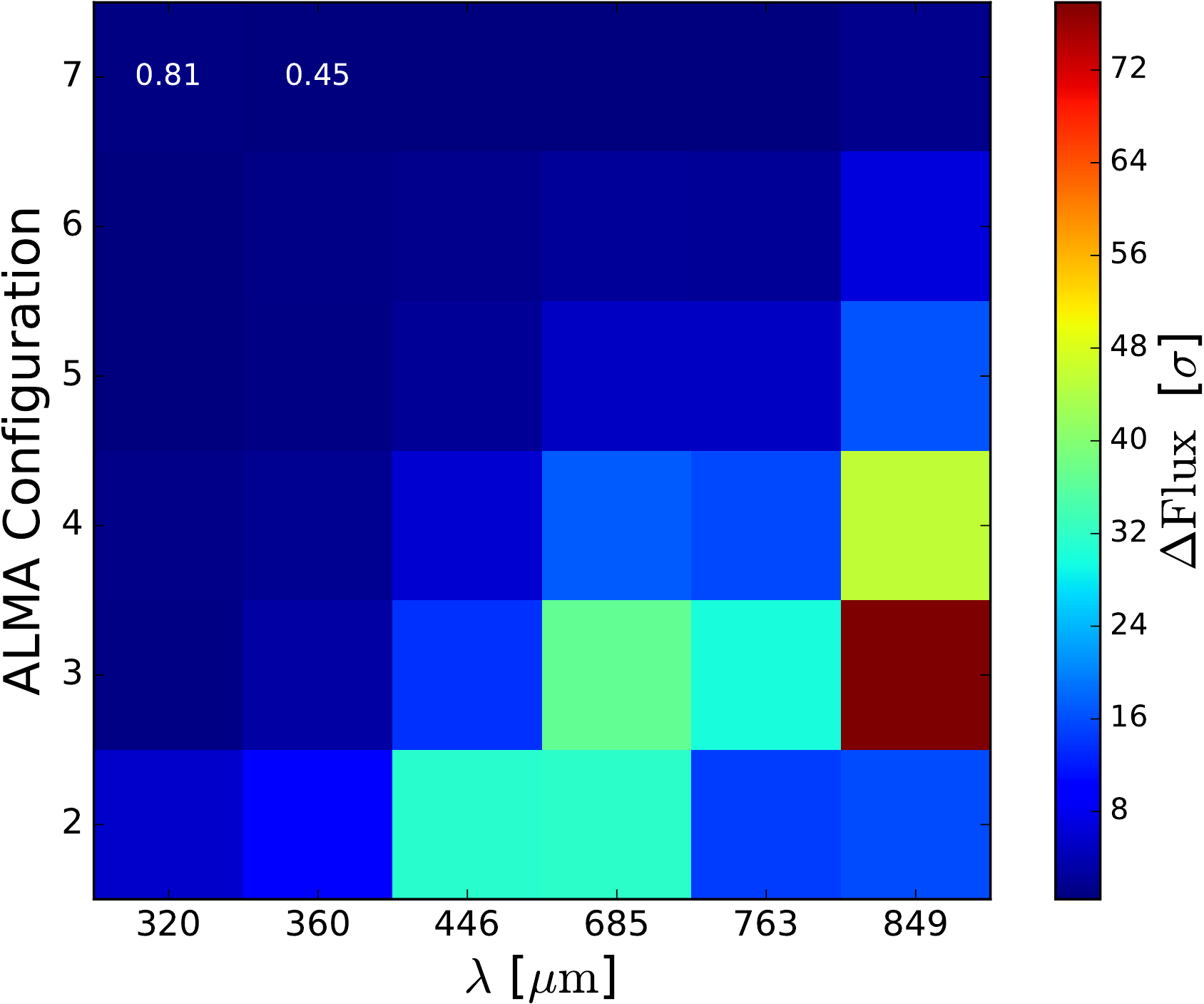}}
           \caption{Flux difference between density wave maximum and minimum in units of sensitivity for $a= 20$ AU,  $M_{\rm B} = 1\,\, {\rm M_{\odot}}$, and $M_{\rm disk} = 10^{-1}\,\, {\rm M_{\odot}}$.}
          \label{fig:sigma_20_1}
    \end{figure}

\begin{figure}
           \resizebox{\hsize}{!}{\includegraphics{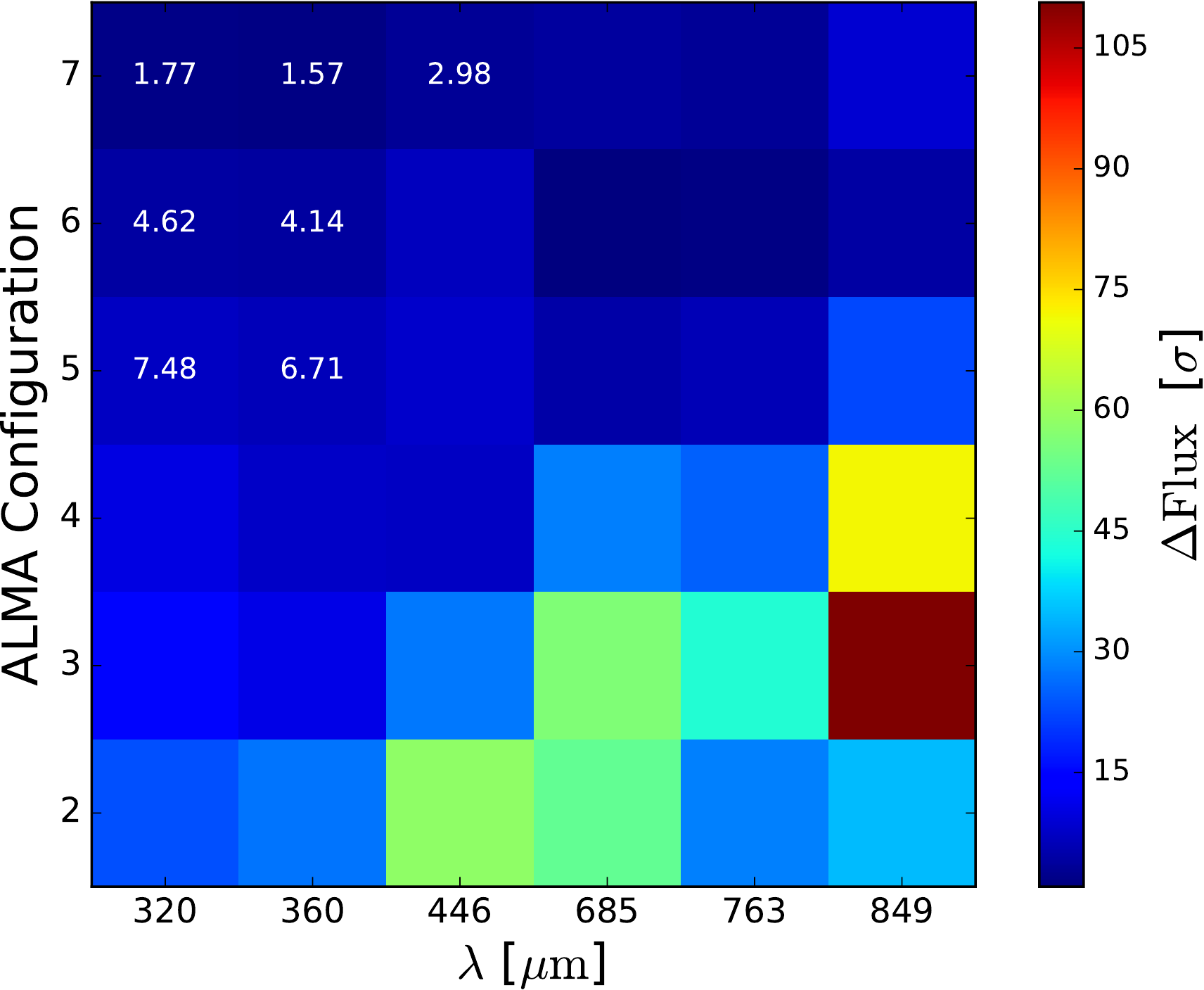}}
           \caption{Flux difference between density wave maximum and minimum in units of sensitivity for $a= 30$ AU,  $M_{\rm B} = 1\,\, {\rm M_{\odot}}$, and $M_{\rm disk} = 10^{-1}\,\, {\rm M_{\odot}}$.}
          \label{fig:sigma_30_1}
    \end{figure}

\begin{figure}
           \resizebox{\hsize}{!}{\includegraphics{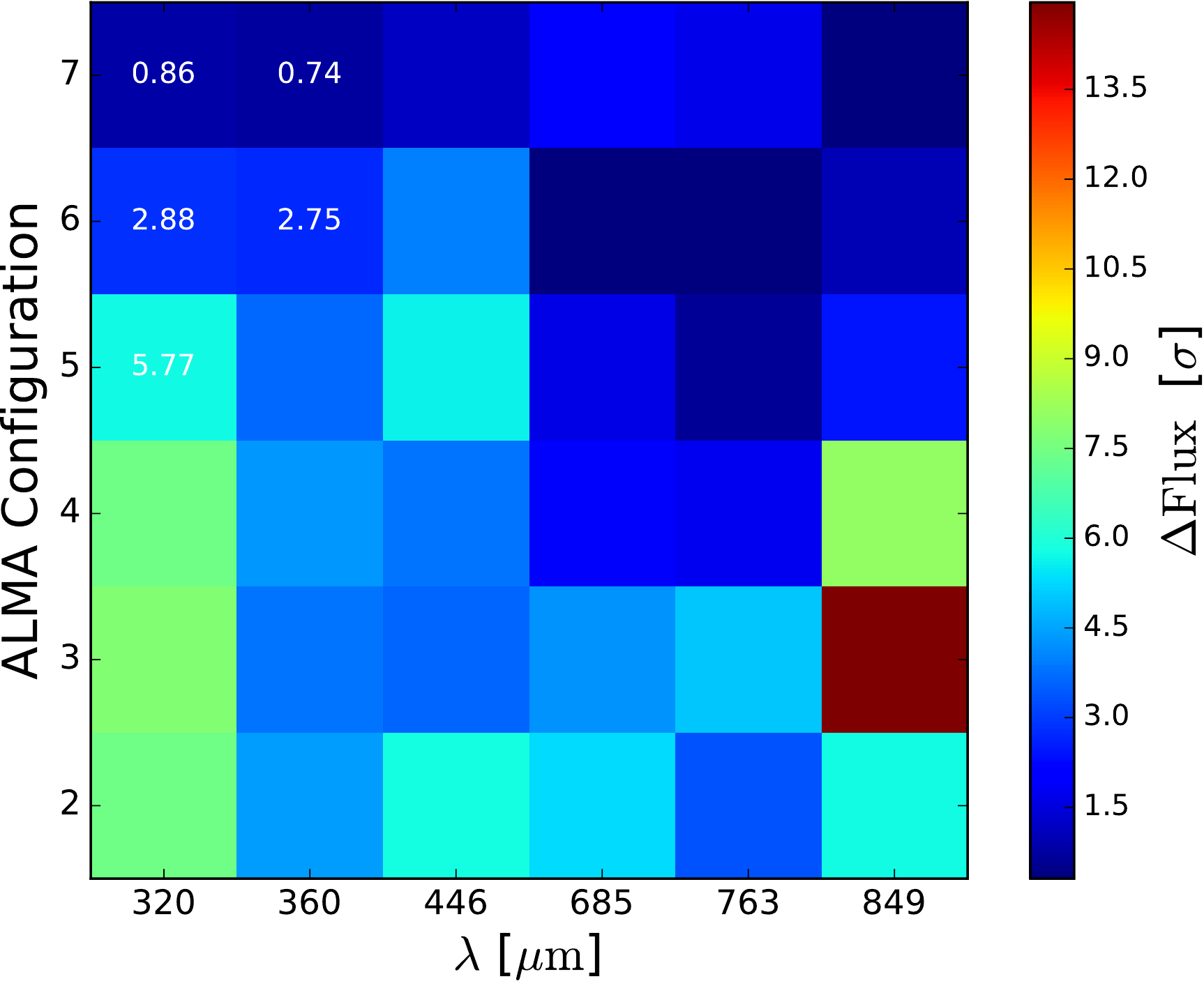}}
           \caption{Flux difference between density wave maximum and minimum in units of sensitivity for $a= 20$ AU,  $M_{\rm B} = 0.5\,\, {\rm M_{\odot}}$, and $M_{\rm disk} = 10^{-1}\,\, {\rm M_{\odot}}$.}
          \label{fig:sigma_20_05}
    \end{figure}

\begin{figure}
           \resizebox{\hsize}{!}{\includegraphics{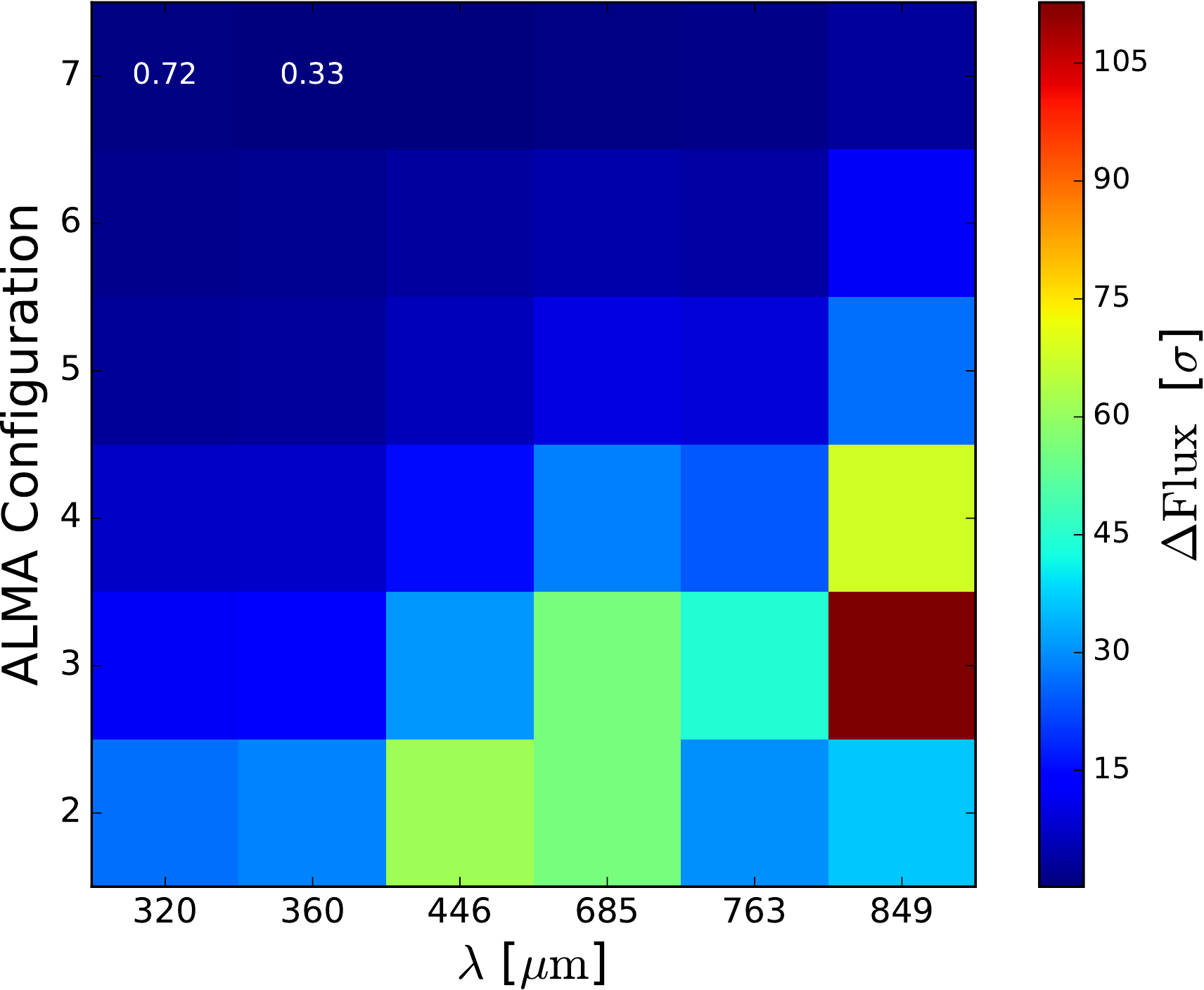}}
           \caption{Flux difference between density wave maximum and minimum in units of sensitivity for $a= 20$ AU,  $M_{\rm B} = 1.5\,\, {\rm M_{\odot}}$, and $M_{\rm disk} = 10^{-1} \,\, {\rm M_{\odot}}$.}
          \label{fig:sigma_20_15}
    \end{figure}

     \paragraph{Density waves:}
     The goal of this study is to determine whether the density waves can be observed with ALMA and, if so, which configuration and band 
     are the most suitable to perform these observations with regard to the considered binary system. We begin with an example of a synthetic ALMA 
     observation (see Figs.~\ref{fig:alma_a}). For this, we choose the shortest possible wavelength 
     ALMA is capable of observing ($\lambda = 320$ $\rm \mu m$, band 10)  and the configuration with the maximum baseline in that band ($D = 3697$ m), which results
     in an angular resolution of $\theta_{\rm res} = 0.024 \arcsec$. A quick look reveals that, at least for the semi-major axis $a = 30~\mathrm{AU}$ and disk mass of  
     $M_{\rm disk} = 0.1 \,\, {\rm M_{\odot}}$, the sensitivity and angular resolution are sufficient for the detection of the density waves (see Fig.~\ref{fig:alma_a}).
     As a next step we quantify the flux difference between the wave maxima and minima in units of 
     sensitivity $\sigma$, employing an algorithm similar to that applied in Sect.~\ref{subsubsec:spat_res}. 
     The results are shown in Figs.~\ref{fig:sigma_10_1} -~\ref{fig:sigma_20_15}.

      We begin the discussion of the results with the highest values of the flux difference, which are measured in 
      units of sensitivity $\sigma$. For all configurations the maximum is reached at $\lambda = 849$ $\rm \mu m$. Similarly, 
      the higher values of the flux difference are concentrated at longer wavelengths and configurations with shorter maximum 
      baselines. However, the resolution at those wavelengths and configurations is not sufficient to resolve the density waves. 
      What we detect is the flux increase with the radius, which corresponds to the inner disk cavity. The combination of 
      wavelength and configuration for which we can detect at least one density maximum are labeled with the appropriate 
      value of flux difference. The detectability of the density waves increases with the semi-major 
      axis $a$. This is not surprising taking into account the results of Tab.~\ref{fig:lam_mag}. The wavelength and
      the magnitude of the density wave increase with the semi-major axis $a$. However, it is surprising, that the  detectability is higher for 
      systems with lower binary mass. From Tab.~\ref{fig:lam_mag} we can infer that the wavelength is increasing 
      for decreasing binary mass, which results in more pronounced maxima. However, the magnitude for a binary mass 
      $M_{\rm B}=1\,\, {\rm M_{\odot}}$ is almost a factor of two greater than for $M_{\rm B}=0.5\,\, {\rm M_{\odot}}$. 
      From this we conclude that the wavelength of the density wave has a greater influence on the detectability than the
      wave magnitude. 
      
      Finally, we want to compare our results to a similar study performed by \cite{Ruge_2015}. One of the major differences 
      of our work compared to the results of \cite{Ruge_2015} is the lack of density waves. There are three major reasons for this. 
      The first reason is the difference in the simulations. \cite{Ruge_2015} conducted a SPH simulation of the inner disk regions. 
      The hydrodynamic simulation performed in this study achieves a better resolution, which allows us to track smaller structures. 
      The second reason is the parameters of the systems. A semi-major axis of $a=2$ AU, considered by \cite{Ruge_2015}, should 
      result in much smaller density wave wavelength than studied here. The last reason are the considered observing wavelengths. 
      The shortest wavelength employed by \cite{Ruge_2015} was 750 $\rm \mu m$. As can be seen in Fig.~\ref{fig:flux_diff_a}, the 
      flux difference between the maximum and minimum decreases with wavelength. Thus, a longer wavelength should decrease the feasibility of detecting the density waves.
      The variability in  flux distribution on the inner disk rim and spiral arm seen in Fig. 7 in the study of
      \cite{Ruge_2015} is caused by the eccentric orbit of the binary ($\varepsilon = 0.3$).

    \subsubsection{European Extremely Large Telescope}
    \label{subsubsec:obs_ELT}
  
  In Sect.~\ref{subsubsec:spat_res} it became clear that the accretion arms are best observed at infrared wavelengths. 
  To simulate an observation with the E-ELT instrument METIS, we convolve the ideal maps (Figs.~\ref{fig:IR_10} -~\ref{fig:IR_20_15})
  with the wavelength-dependent point spread function of an ideal 39 m telescope with a circular aperture. The goal here is to determine 
  whether it is possible to observe the highly variable structures, such as the inner disk rim and accretion arms.

\begin{figure*}
          \resizebox{\hsize}{!}{\includegraphics{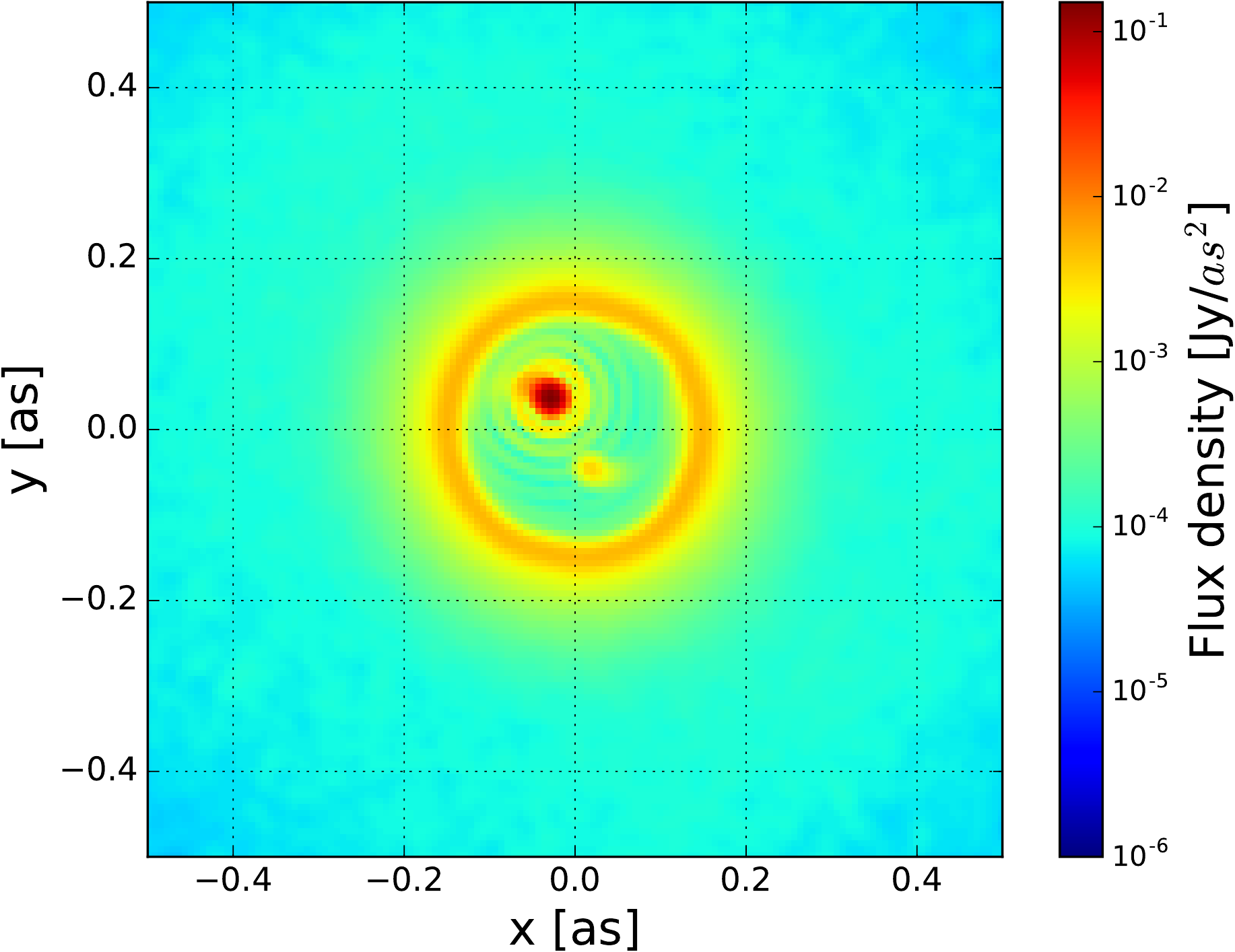} \quad
                                \includegraphics{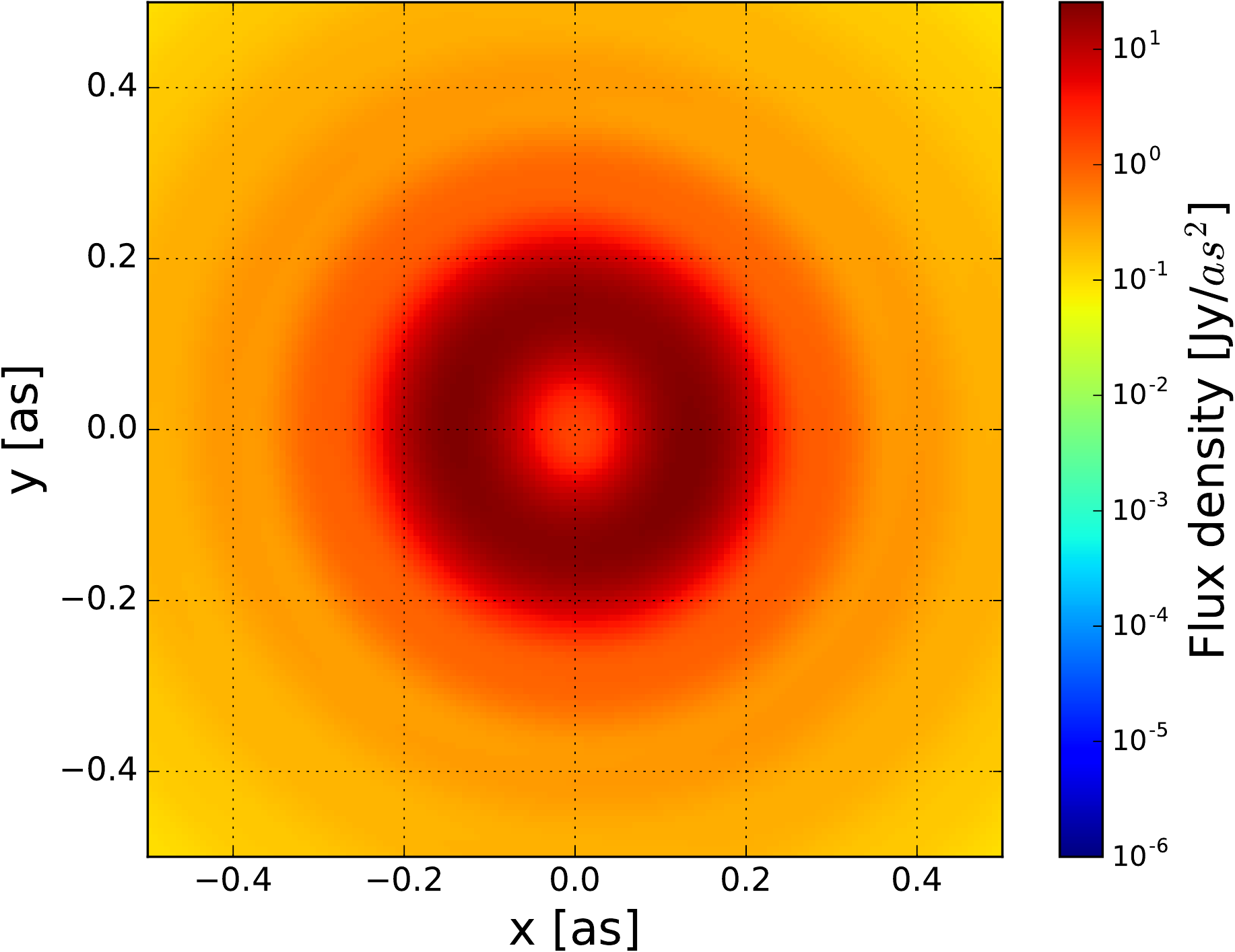}}
          \caption{Flux after convolution at  $\lambda =4.5$ $\rm \mu m$ (\textsl{left}) and $\lambda =20$ $\rm \mu m$ (\textsl{right})  for $M_{\rm B}=1 \,\,{\rm M_{\odot}}$ and  $a = 10$ AU. }
          \label{fig:IR_10_conv}
 \end{figure*}

\begin{figure*}
          \resizebox{\hsize}{!}{\includegraphics{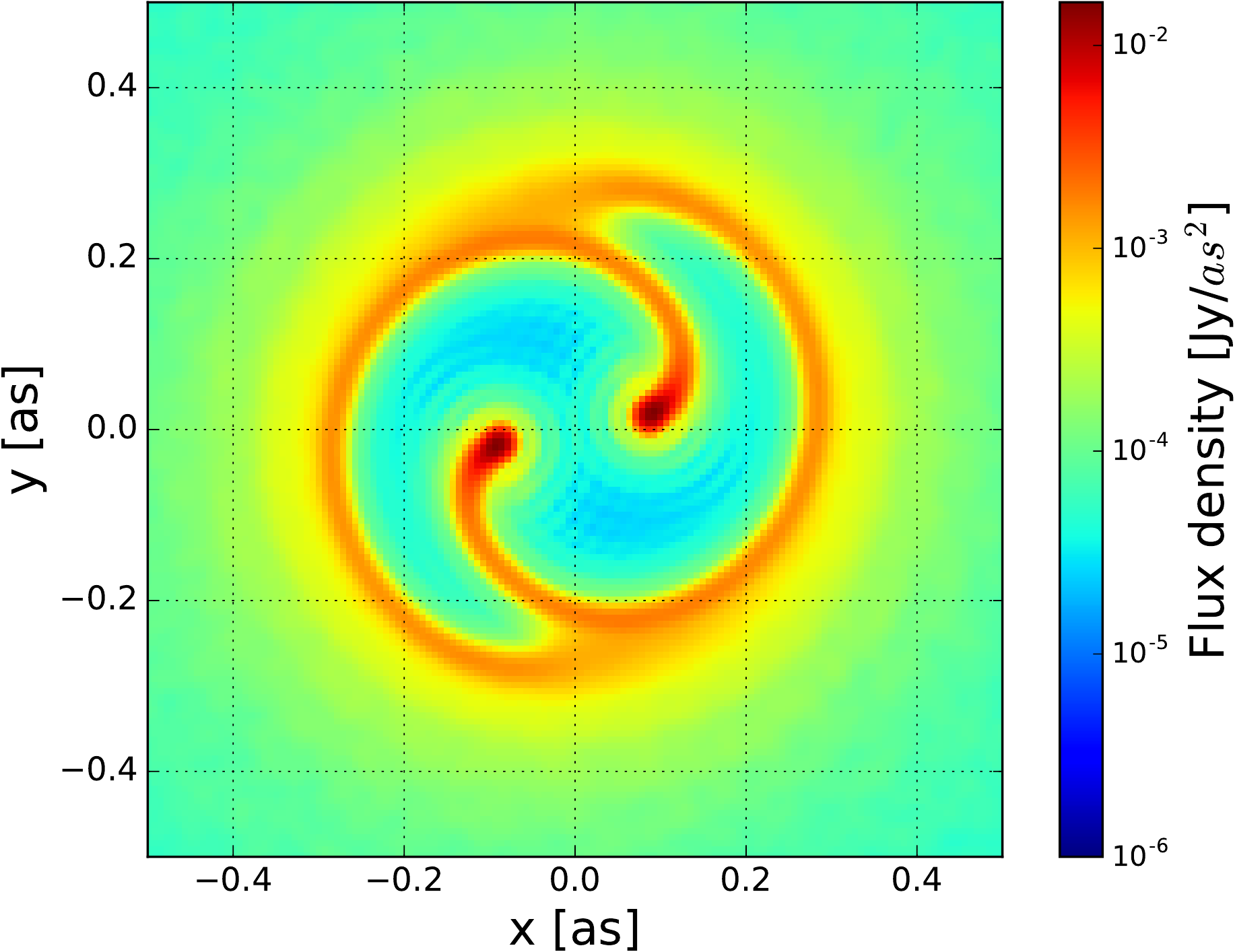} \quad
                                \includegraphics{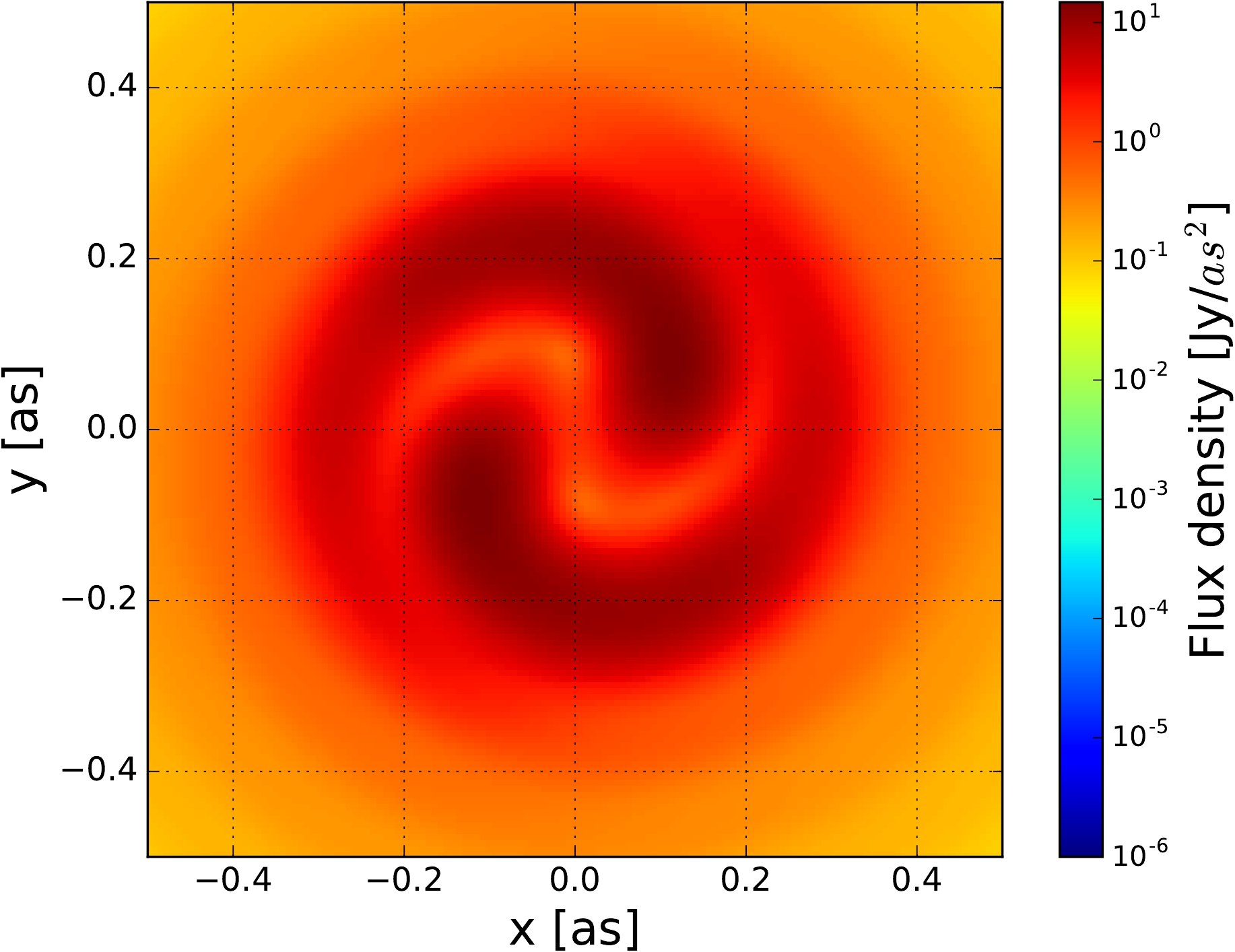}}
          \caption{Flux after convolution at  $\lambda =4.5$ $\rm \mu m$ (\textsl{left}) and $\lambda =20$ $\rm \mu m$ (\textsl{right})  for $M_{\rm B}=1 \,\,{\rm M_{\odot}}$ and  $a = 20$ AU. }
          \label{fig:IR_20_conv}
 \end{figure*}

\begin{figure*}
          \resizebox{\hsize}{!}{\includegraphics{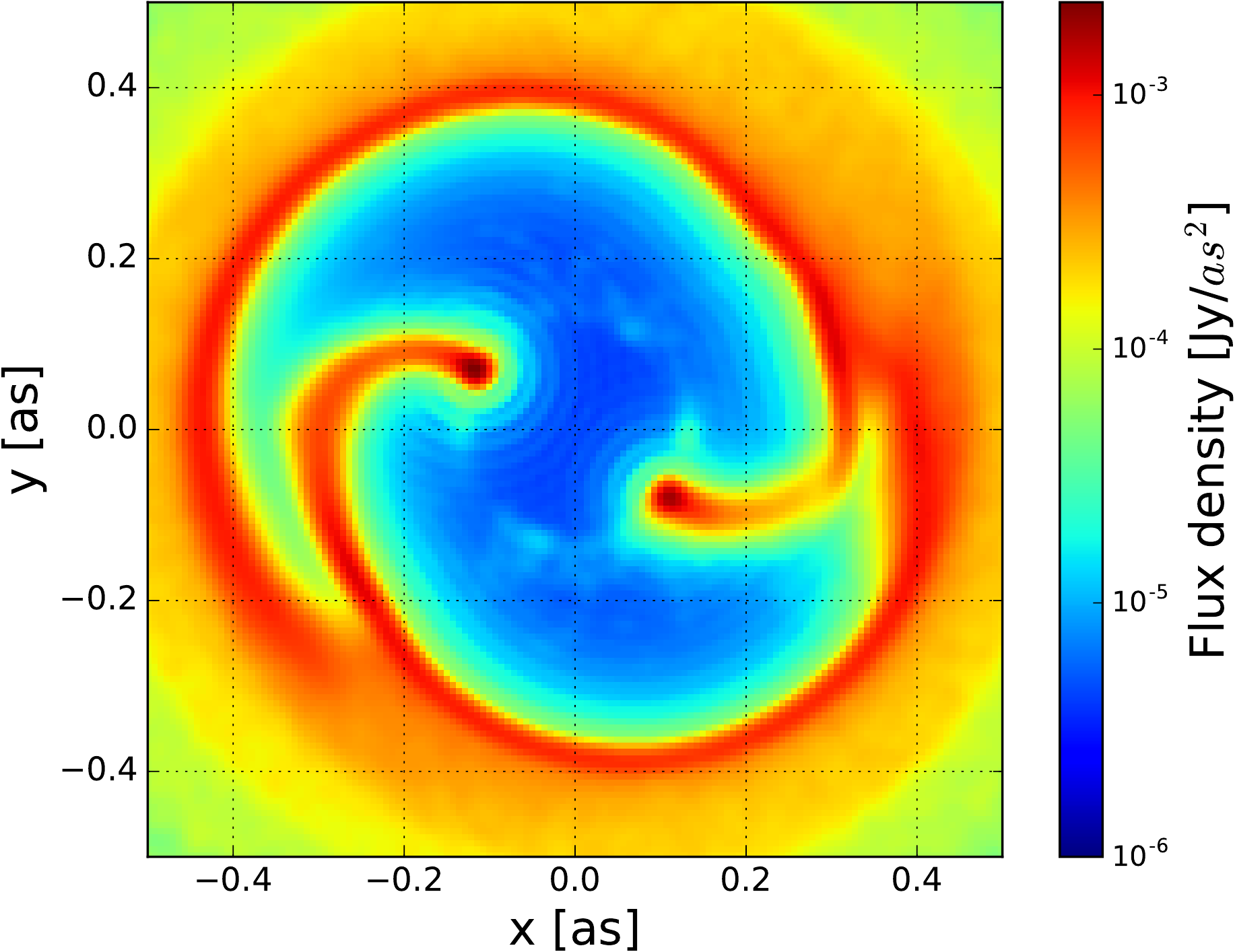} \quad
                                \includegraphics{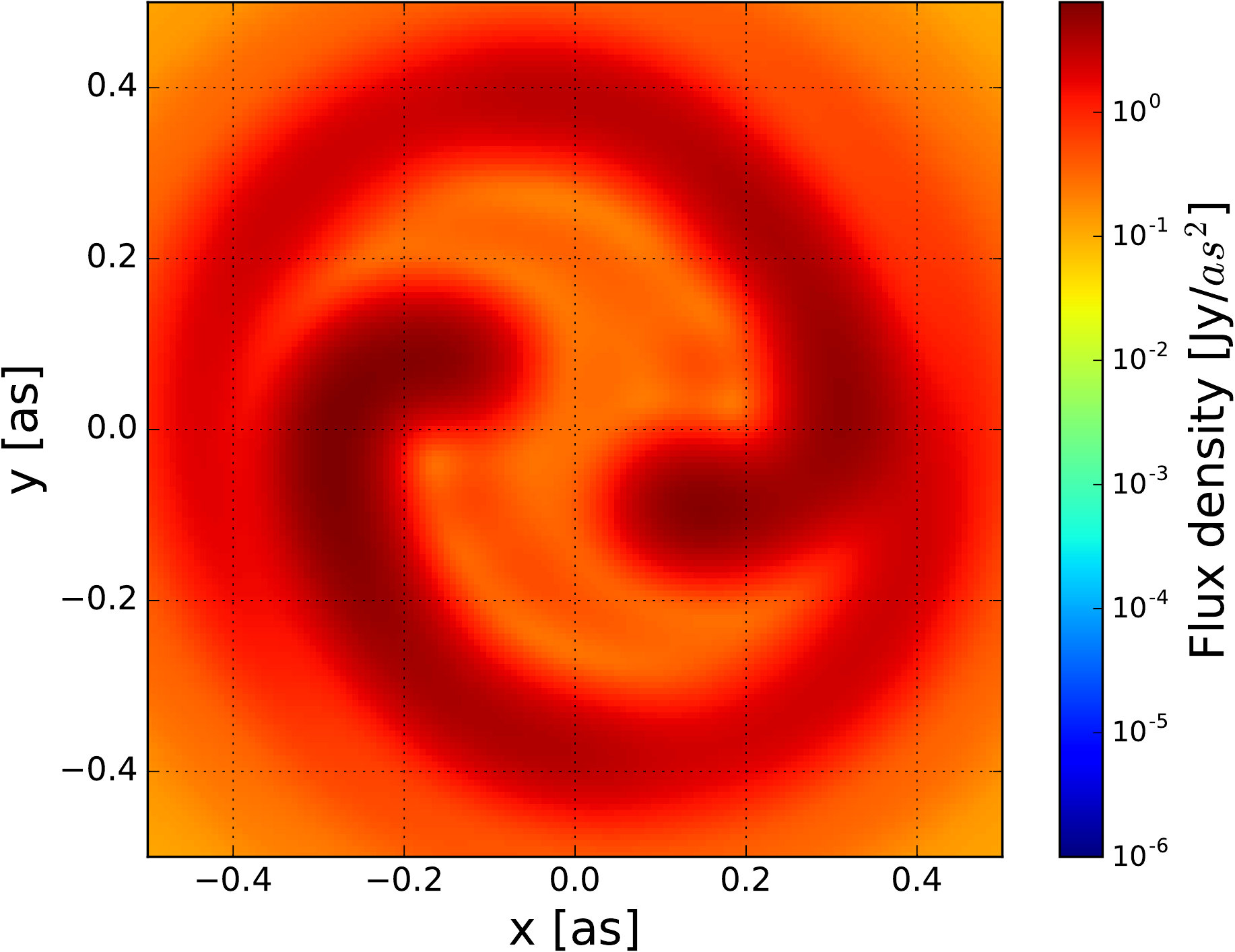}}
          \caption{Flux after convolution at  $\lambda =4.5$ $\rm \mu m$ (\textsl{left}) and $\lambda =20$ $\rm \mu m$ (\textsl{right})  for $M_{\rm B}=1 \,\,{\rm M_{\odot}}$ and  $a = 30$ AU. }
          \label{fig:IR_30_conv}
 \end{figure*}

\begin{figure*}
          \resizebox{\hsize}{!}{\includegraphics{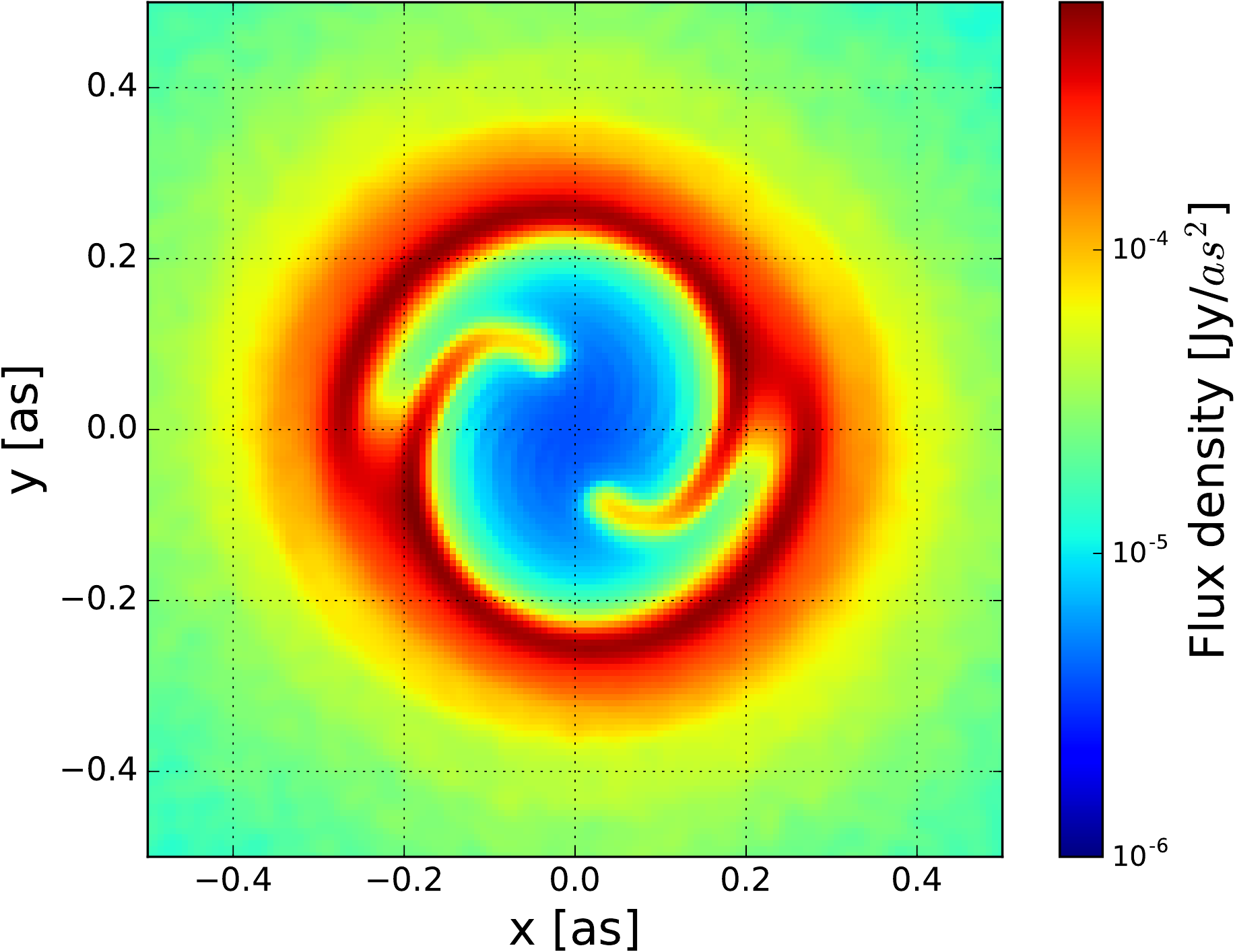} \quad
                                \includegraphics{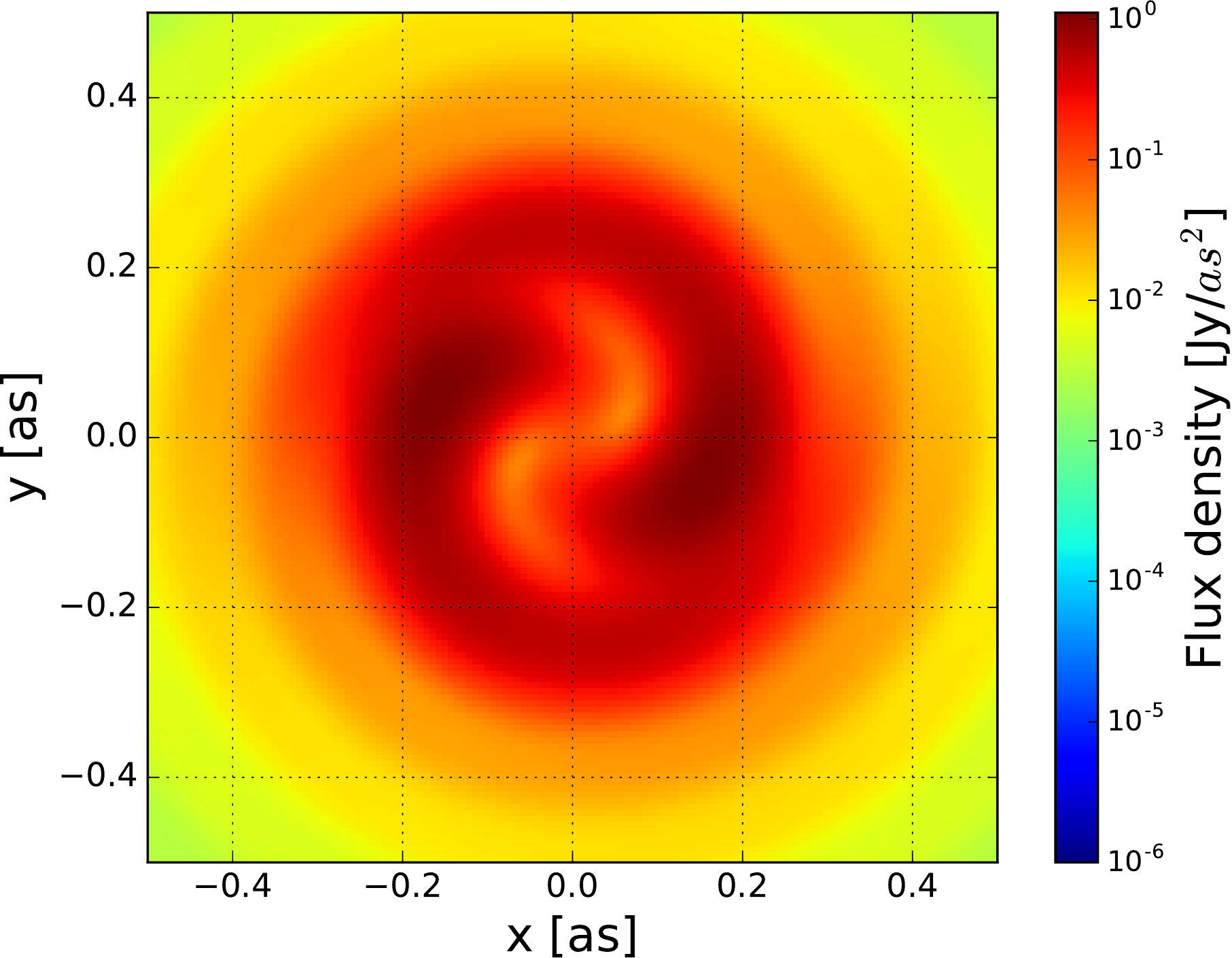}}
          \caption{Flux after convolution at  $\lambda =4.5$ $\rm \mu m$ (\textsl{left} and $\lambda =20$ $\rm \mu m$ (\textsl{right})  for $M_{\rm B}=0.5\,\, {\rm M_{\odot}}$ and  $a = 20$ AU. }
          \label{fig:IR_20_05_conv}
 \end{figure*}

\begin{figure*}
          \resizebox{\hsize}{!}{\includegraphics{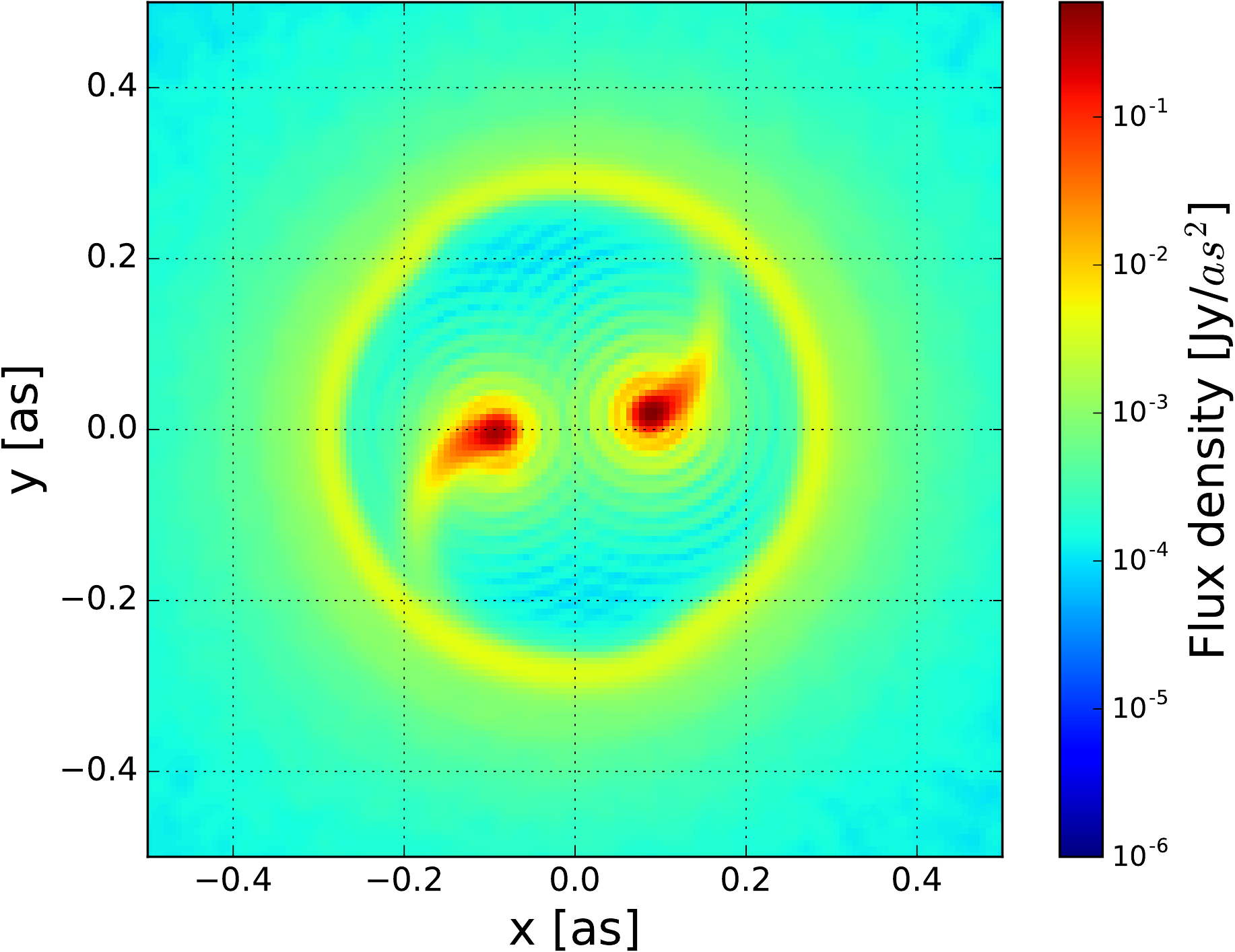} \quad
                                \includegraphics{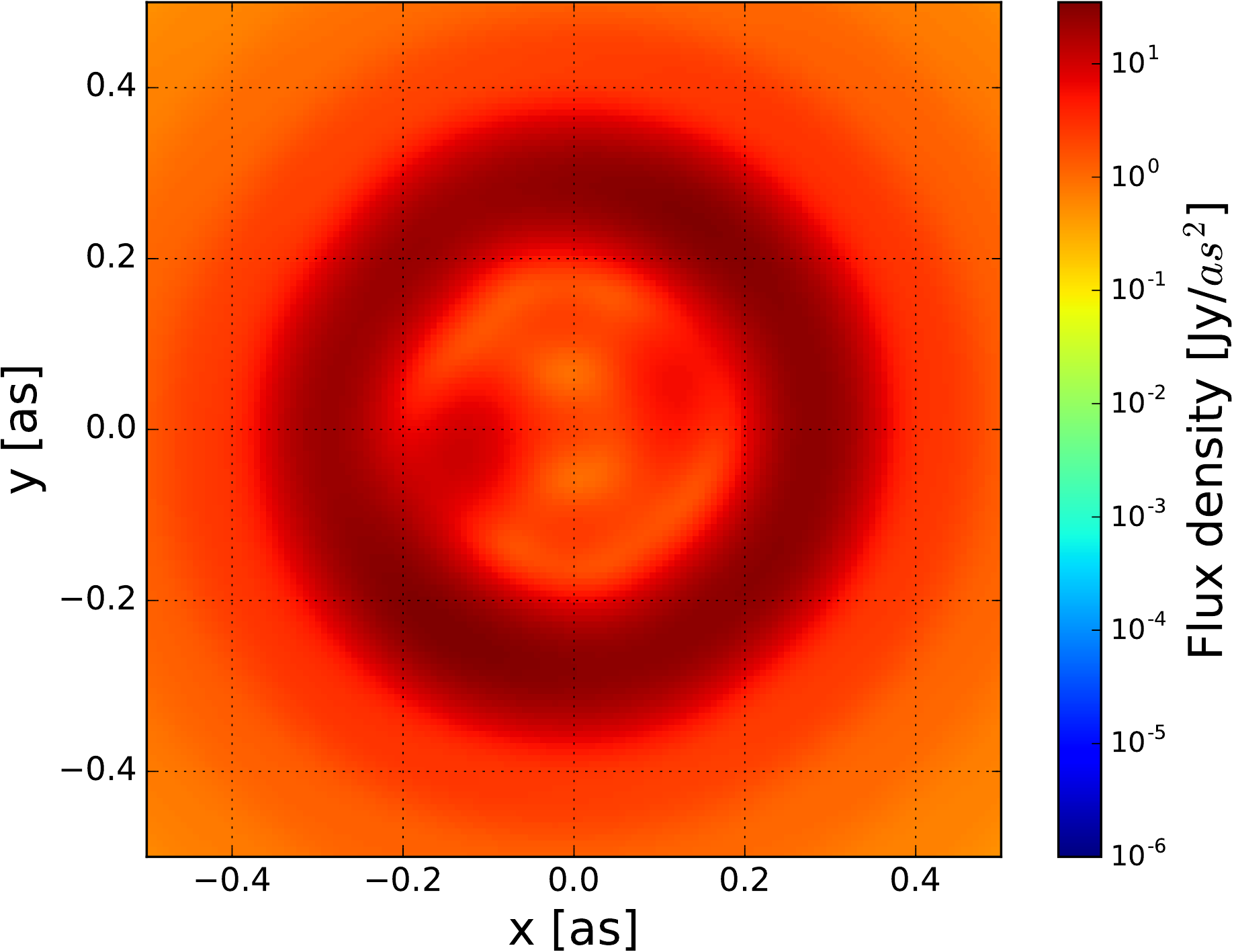}}
          \caption{Flux after convolution at  $\lambda =4.5$ $\rm \mu m$ (\textsl{left}) and $\lambda =20$ $\rm \mu m$ (\textsl{right})  for $M_{\rm B}=1.5 \,\,{\rm M_{\odot}}$ and  $a = 20$ AU. }
          \label{fig:IR_20_15_conv}
 \end{figure*}

   The brightness ratio between the individual binary component and the brightest structures amounts to $\sim 10^2 \, -\sim 10^4$ in the considered cases.
   For this reason, the binaries are removed from the maps before the convolution (see Figs.~\ref{fig:IR_10_conv} -~\ref{fig:IR_20_15_conv}).

   All images at $\lambda =4.5$ $\rm \mu m$ are dominated by the flux from the binary. However, in all systems, with the exception of the case 
   $M_{\rm B}=1 \,\, {\rm M_{\odot}}$ and  $a = 10$ AU, it is still possible to spatially resolve the accretion arms. Similarly, at  $\lambda = 20$ $\rm \mu m$
   one would only detect a bright ring.

%    For a real observation this will be done via a coronagraph.  
   In all but one system  the accretion arms are resolved in both wavelength regimes. In the case of $M_{\rm B}=1 \,\, {\rm M_{\odot}}$ and  $a = 10$ AU, 
   we only detect a bright ring that corresponds to the inner disk rim. For $\lambda =4.5$ $\rm \mu m$ we find two small flux maxima
   that are caused by hot dust in the direct vicinity of the binary at the edge of the computational domain. In Table~\ref{fig:eelt_post_falt} the fluxes originating from 
    inside a circle with radius of $2 \times a$ are compiled. We find  no major differences from the original maps 
   caused by the convolution process (see Tab.~\ref{fig:eelt_pre_falt}). 
   
   \begin{table}
     \caption{Integrated flux of the scattered and re-emitted radiation in units of mJy originating from inside a circle with the radius of $2 \times a$. 
              The direct stellar radiation is not considered. }
\begin{tabular}{c|c|c}
        \multicolumn{3}{c}{}                                                                                                           \\ \hline \hline
                    Parameter                                     & $\lambda = 4.5 {\rm \mu m}$  & $\lambda = 20 {\rm \mu m}$          \\ \hline
         $M_{\rm B} = 1.0 {\rm M_{\odot}} \, ,\,\,\,\, a = 10$ AU & $0.42$                       & $ 2702.1  $                  \\ \hline
         $M_{\rm B} = 1.0 {\rm M_{\odot}} \, ,\,\,\,\, a = 20$ AU & $0.27$                       & $ 2031.33 $                  \\ \hline
         $M_{\rm B} = 1.0 {\rm M_{\odot}} \, ,\,\,\,\, a = 30$ AU & $0.16$                       & $ 1252.79 $                  \\ \hline
         $M_{\rm B} = 0.5 {\rm M_{\odot}} \, ,\,\,\,\, a = 20$ AU & $0.078$                      & $ 144.97  $                  \\ \hline
         $M_{\rm B} = 1.5 {\rm M_{\odot}} \, ,\,\,\,\, a = 20$ AU & $1.66$                       & $ 6610.53 $                  \\ \hline \hline
 \end{tabular}
  \label{fig:eelt_post_falt}
  \end{table}

    \section{Summary}
    
    The goal of this work was to  study the observability of 
    characteristic structures in circumbinary disks, generated by binary-disk interaction. To achieve this goal, a 2D hydrodynamic simulation 
    was performed with the code \texttt{Fosite} to calculate a surface density distribution. The results were applied to a 3D 
    grid-based Monte Carlo code \texttt{Mol3D} and a temperature profile;
    in addition, resulting scattered light and re-emission maps were simulated. These images were used to simulate imaging observations in the 
    near to mid-infrared wavelength range (E-ELT) and at submillimeter/millimetre wavelengths (ALMA).

    We found that the torque exerted on the circumbinary disk by the binary generates  unique 
    features that can be used to distinguish them from 
    protoplanetary disks around single stars. In particular these are the inner cavity, accretion flows
    from disk inner rim to the binary orbit,
    and the density waves on the disk inner edge. We quantified the dependence of those features on binary mass 
    $M_{\rm B}$, semi-major axis $a$, and disk mass $M_{\rm disk}$. 
    Furthermore, we derived a wave equation governing the behaviour of the density waves, which can 
    be transformed  into a standard 1D wave solution in two dimensions.
    
    We have shown that those features alter the thermal and radiation balance of the 
    disk to a sufficiently high degree, which can be seen in the re-emission and scattered light maps. 
    Here we once again studied the dependence of the observable features on the parameters and derived predictions 
    with regard to the successful observation of those features.

    Subsequently synthetic observations of the simulated circumbinary disks with ALMA and E-ELT were generated. 
    We could show that ALMA configurations with the highest angular resolution are capable of observing the density waves
    generated by the binary with $a = 20$ AU and $a = 30$ AU. Furthermore, we made a parameter study that showed that even a configuration with a far 
    smaller maximum baseline is sufficient to observe the inner cavity of those disks.  
    Simulated observations with E-ELT have shown that it will be possible to resolve the accretion arms at the considered wavelengths of
     $4.5$ $\rm \mu m$ and $20$ $\rm \mu m$.

    Further studies of this kind would need to include systems with binary mass ratio $q = M_{\rm sec}/M_{\rm prim} \neq 1$ and 
    an eccentricity value greater than $0$, as they comprise the majority of the known systems \citep{Raghavan_2010} and those parameters have a 
    large impact on the disk structure \citep{Ruge_2015}. Furthermore, this study was focused only 
    on dust emissions. However, the gas of the circumbinary  disk is influenced in the same manner.
    If realised, the observation of binary-induced structures in molecular lines would allow one to observe velocity 
    fields in the disk and in the accretion flows in the disk centre. 
    
\begin{acknowledgements}
      We thank all the members of the Astrophysics Department Kiel for helpful discussions and remarks and for their language corrections. 
   
   This study was funded by the German Science Foundation (DFG), grant: WO 857/12-1.
\end{acknowledgements}  

  \begin{appendix}
  \section{Derivation of the wave equation }
  \label{sec:app}

 In this section we present a detailed derivation of the wave equation mentioned in  Sect.~\ref{subsec:res_fos}.

Since the wave propagates in  $r$ direction we neglect the $\phi$ component and terms containing the derivative with regard to $\phi$, which leaves us
  with the  2D continuity and the $r$ component of the Navier-Stokes equations in polar coordinates,

  \begin{eqnarray} 
    \partial_t \Sigma + \frac{1}{r} \partial_r (r \Sigma v_r) &=& 0 \,  ,\\
     \Sigma \partial_t v_r + \Sigma v_r \partial_r v_r &=& - \partial_r (c_s^2 \Sigma).
  \end{eqnarray}
  
  \noindent
 The parameter $\Sigma$ is the vertically integrated surface density; $v_r$ is the radial drift 
  velocity, which is in this case generated by the disks differential rotation combined with the sheer viscosity; and $c_s$ is the sound speed of
  the gas. By applying the time derivative on A.1 we get

   \begin{eqnarray} 
    \partial_t^2 \Sigma + \frac{1}{r} \partial_r (r \Sigma \partial_t v_r  + r v_r \partial_t \Sigma ) &=& 0 .
  \end{eqnarray}
  
  \noindent
  The time derivation of the surface density can be substituted using A.1 once again and A.2 to substitute $\Sigma \partial_t v_r$,
  
  \begin{eqnarray} 
    \partial_t^2 \Sigma + \frac{1}{r} \partial_r [ - r \Sigma v_r \partial_r v_r - r  \partial_r (c_s^2 \Sigma)   -  v_r \partial_r ( r \Sigma v_r)  ] &=& 0 ,
  \end{eqnarray}
  
  \noindent
  which can be further simplified to
  
  \begin{eqnarray} 
    \partial_t^2 \Sigma - \frac{1}{r} \partial_r \left[r \partial_r \left(( c_s^2 + v_r^2 ) \Sigma \right) + \Sigma  v_r^2 \right] &=& 0 .
  \end{eqnarray}
  
  \noindent
  Assuming the drift velocity $v_r$ and its derivation  $\partial_r v_r$ are much smaller than the sound speed $c_s$ and its derivative
  $\partial_r c_s$ ,  Eq. (A.5) can be reduced to  Eq. (3.1.1)
   A full summary of the solution for the simplified Eq. (3.1.1) can be found in \cite{polyanin_2002}, p. 296-297.

  \end{appendix}

\bibliographystyle{aa} % style aa.bst
\bibliography{literatur} % your references Yourfile.bib

\end{document}